\documentclass[prd,letterpaper,onecolumn,amsmath,amsfonts,amssymb,nofootinbib]{revtex4}
\pdfoutput=1
\usepackage {amsmath}  
\usepackage[dvips]{graphicx}           
\usepackage{float}         
\usepackage{feynmp}     
\usepackage{amssymb}                      
\usepackage[colorlinks=true                                                          
,urlcolor=blue          
,anchorcolor=blue               
,citecolor=blue         
,filecolor=blue          
,linkcolor=blue      
,menucolor=blue      
,pagecolor=blue  
,linktocpage=true
,pdfproducer=medialab                       
,pdfa=true
]{hyperref}  
 \usepackage{csquotes}
\newcommand{\be}{\begin{equation}}
\newcommand{\ee}{\end{equation}}
\newcommand{\bear}{\begin{eqnarray}} 
\newcommand{\eear}{\end{eqnarray}}

\newcommand{\lapproxeq}{\lower .7ex\hbox{$\;\stackrel{\textstyle
<}{\sim}\;$}}
\newcommand{\gapproxeq}{\lower .7ex\hbox{$\;\stackrel{\textstyle
>}{\sim}\;$}}
\newcommand{\stackdown}[2]{\lower 1.4ex\hbox{$\;\stackrel{\textstyle{#1}}
{\scriptstyle{#2}}\;$}}
\newcommand{\beq}{\begin{equation}}
\newcommand{\eeq}{\end{equation}}

\newcommand{\ba}{\begin{eqnarray}}
\newcommand{\ea}{\end{eqnarray}}

\newcommand{\bea}{\begin{eqnarray}}
\newcommand{\eea}{\end{eqnarray}}

\makeatletter
\def\slash{\@ifnextchar[{\fmsl@sh}{\fmsl@sh[0mu]}}
\def\fmsl@sh[#1]#2{%
  \mathchoice
    {\@fmsl@sh\displaystyle{#1}{#2}}%
    {\@fmsl@sh\textstyle{#1}{#2}}%
    {\@fmsl@sh\scriptstyle{#1}{#2}}%
    {\@fmsl@sh\scriptscriptstyle{#1}{#2}}}
\def\@fmsl@sh#1#2#3{\m@th\ooalign{$\hfil#1\mkern#2/\hfil$\crcr$#1#3$}}
\makeatother
\usepackage{color}

\definecolor{orange}{rgb}{0.9,0.2,0}
\definecolor{brown}{rgb}{0.7,0.3,0.2}
\definecolor{fuxia}{rgb}{1,0,1}
\definecolor{skyblue}{rgb}{0,0.1,0.9}
\definecolor{violetred}{rgb}{0.8,0.13,0.56}
\definecolor{deeppink}{rgb}{1.00,0.08,0.5}
\definecolor{pink}{rgb}{1.00,0.75,0.80}
\definecolor{orchid}{rgb}{0.85,0.44,0.84}
\definecolor{lightpink}{rgb}{1.00,0.71,0.76}
\definecolor{bluish}{rgb}{0,0.6,0.8}  
\begin{document}

\title{Reheating in $R^2$  Palatini inflationary models   } 

\author{Ioannis D. Gialamas$^{a}$}
\email{i.gialamas@phys.uoa.gr} 
\author{A. B. Lahanas$^{a}$ }
\email{alahanas@phys.uoa.gr}

\vspace{0.1cm}
\affiliation{
${}^a$ National and Kapodistrian University of Athens, Department of Physics,
Nuclear and Particle Physics Section,   GR--157 71   Athens,~Greece 
\\
} 
  
\vspace*{2cm}    
\begin{abstract}
We consider $R^2$ inflation in the Palatini gravity assuming the existence of scalar fields, coupled to gravity in the most general manner. These theories,  in the Einstein frame, and  for one scalar  field $h$, share common features with $K$ - inflation models.  We apply this formalism for the study of popular inflationary models, whose  potentials  are monomials, 
$ V \sim h^{n} $,  with $ n $  a positive  even integer. We also study the Higgs model  non-minimally coupled to gravity. Although these have been recently studied, in the framework of the Palatini approach, we show that the scalar power spectrum  severely constrains these models.  Although we do not propose a particular reheating mechanism, we show that the quadratic 
$ \sim h^2$ and the Higgs model can survive  these constraints  with a  maximum reheating temperature as large as   $ \sim  10^{15} \, GeV$, when reheating is instantaneous.  However, this can be only attained at the cost of a delicate fine-tuning of couplings.   Deviations from this fine-tuned values can still yield predictions compatible with the cosmological data, for  couplings that lie in  very tight range, giving lower reheating temperatures. 
\end{abstract} 
\maketitle


\section{Introduction}

It has been long known that the Palatini formulation of General Relativity (GR), or first-order formalism, is an alternative to the well-known  metric formulation, or  second-order formalism.  In the latter the space time connection is determined by the metric while in the Palatini approach the connection $\Gamma^\mu_{\lambda \sigma}$ is treated as  an independent variable
\cite{Sotiriou:2006hs,Sotiriou:2006qn,Sotiriou:2008rp,Borunda:2008kf,DeFelice:2010aj,Olmo:2011uz,Capozziello:2011et,Clifton:2011jh}, ( see also \cite{Nojiri:2017ncd}, and references  therein).
It is through the use of the equations of motion that $\Gamma^\mu_{\lambda \sigma}$ receive the well known form of the Christoffel symbols, describing thus a metric connection.  Within the context of GR the two formulations are  equivalent.  However in the presence of fields that are coupled in a non-minimal manner to gravity this no longer holds, \cite{Sotiriou:2006hs,Sotiriou:2006qn,Sotiriou:2008rp}.  In that case the two formulations describe  different physical theories.    

Encompassing the popular inflation models into  Palatini Gravity, in an effort to  describe the cosmological evolution of the Universe, leads to different cosmological predictions, from the metric formulation,  due to the fact that the dynamics of the two approaches differ. A notable example is the Starobinsky model, for instance, where except the graviton there exists an additional propagating scalar degree of freedom,  the scalaron, whose mass is related to the coefficient of   the $R^2$ term. In the Einstein frame this shows up as a dynamical scalar field, the inflaton, moving in a potential, the celebrated Starobinsky potential, 
\cite{Starobinsky:1980te,Mukhanov:1981xt,Starobinsky:1983zz}.  
Within the framework  of the Palatini Gravity,  in any  $f(R)$ theory \cite{Sotiriou:2008rp}, there are no  extra  propagating degrees of freedom, that can play the role of the inflaton, and hence the inflaton  has to be put in by hand as an additional field coupled to $f(R)$ gravity.  

The differences between metric and Palatini formulation in the cosmological  predictions, as far as inflation is concerned, arise from the non-minimal couplings of the scalars,  that take-up the role of the inflaton. These couplings  are different in the two approaches. This has been first  pointed out in  \cite{Bauer:2008zj} and has attracted the interest of many authors since,  
\cite{Koivisto:2005yc,Tamanini:2010uq,Bauer:2010jg,Enqvist:2011qm,Borowiec:2011wd,Stachowski:2016zio,Fu:2017iqg,Rasanen:2017ivk,Tenkanen:2017jih,Racioppi:2017spw,Markkanen:2017tun,Jarv:2017azx,Rasanen:2018ihz,Racioppi:2018zoy,Carrilho:2018ffi,Enckell:2018kkc,Bombacigno:2018tyw,Enckell:2018hmo,Antoniadis:2018ywb,Antoniadis:2018yfq,Rasanen:2018fom,Almeida:2018oid,Takahashi:2018brt,Kannike:2018zwn,Tenkanen:2019jiq,Shimada:2018lnm,Wu:2018idg,Kozak:2018vlp,Jinno:2018jei}, with still continuing activity, \cite{Edery:2019txq,Rubio:2019ypq,Jinno:2019und,Giovannini:2019mgk,Tenkanen:2019xzn,Bostan:2019wsd,Tenkanen:2019wsd}.

The measurements of the cosmological parameters, by various collaborations,  has tighten the allowed limits of these observables which in turn constrain severely, or even exclude, particular  inflationary models, \cite{Akrami:2018odb,Aghanim:2018eyx,Ade:2018gkx}.  In particular, the  spectral index $n_s$ and the bounds on the tensor-to-scalar ratio $r$ impose severe restrictions and not all models can be compatible with the observational data
{\footnote{ In this work, standard assumptions are made for neutrino masses and their effective number. Relaxing these it induces substantial  shifts in $n_s$  \cite{Gerbino:2016sgw}.
}}.
Within the class of $f(R)$ theories the Starobinsky model, which is an $R^2$- theory, is singled out, although other popular  models can also successfully pass the tests provided by the recent cosmological observations.  The measurements of the primordial scalar perturbations, and of the associated power spectrum amplitude $A_s$,  constrains the scale of inflation in models  encompassed in the framework of the metric formulation. We shall show that in the Palatini formalism this imposes restrictions that are more stringent, at least in some cases, than the ones arising from the observables $n_s, r$ and should be duly taken into account.   
 In this work we shall consider ${\cal{R}}^2$ theories, in the framework of the Palatini Gravity, and study the cosmological predictions of some of the popular models existing in  literature.  We will show that these do not comfortably stand, unless the parameters describing the models are fine-tuned, the main source of this fine-tuning being  the power spectrum amplitude. 
 
 This paper is organized as follows :
 
 In section II, we present the salient features and give a general setup of   $f({\cal{R}})$ - theory
 {\footnote{
 Throughout this paper we use different symbols for the Ricci scalar which  in the metric formulation we denote by $R$, and in the Palatini approach denoted by   ${\cal{R}}$. 
 }}
 ,
  in the presence of  an arbitrary number of scalar fields, coupled to gravity in a non-minimal manner, in general.  Although this is not new, as this effort has been undertaken by other authors as well,  we think that the general, and model-independent, expressions  we arrive at, are worth being discussed.
 We shall then  focus on the case of ${\cal{R}}^2$ theories for which the passage to the Einstein frame is easily implemented. 
 These theories have a gravity sector, specified by two arbitrary functions, sourcing in general  non-minimal couplings of the scalars involved in Palatini Gravity and a third function which is the scalar potential. 
 In the Einstein  frame, and when a single field is present, these models have much in common with the $K$ - inflation models
 \cite{ArmendarizPicon:1999rj}. 
 
 In section III, we discuss the arising equations of motions and the slow-roll mechanism, and give the pertinent slow-roll parameters, adapted to the particular setup. This is necessary since it is our aim to employ a scheme in which the passage to canonically normalized fields is not mandatory.  This we find it more convenient especially because the use of  canonically normalized fields results, in most of the cases, to expressions that cannot be cast in closed forms. 
 
 The discussion of the cosmological observables is the subject of section IV.  We focus, in particular, to the power spectrum amplitude which, as already advertised, puts severe constraints on the inflation models that we are going to discuss. We find it necessary to include  higher order corrections,  in the slow-roll parameters, of the power spectrum, since these may account for contributions 
 comparable, in magnitude, to the errors associated with the power spectrum. 
 Although we shall not adopt  a particular reheating mechanism,  the dependence of the  number of e-folds 
 on the reheating temperature  is of paramount importance,   for the study of the cosmological predictions. This is, also,  reviewed in section IV.
  
 In section V, we consider particular inflation models, namely the class of models in which the scalar field $h$, is characterized by monomial potentials  $\sim h^{n}$, with $ n $ a positive  even integer, and the Higgs model. 
 Although these have been much studied,  we shall show that the cosmological data put severe restrictions on the associated couplings leading to fine-tuned adjustments of the parameters involved, when the power spectrum data are taken into account.  The reheating mechanism can be instantaneous, at the cost of an unnatural  fine-tuning of the couplings pertinent to the potential, describing the aforementioned models. 
 For the models discussed,  the instantaneous reheating temperature $T_{ins}$, which sets the maximum temperature,  can be as large as $ \sim 10^{15} \, GeV$. 
 Departing from these fine-tune values,  we can still be in agreement with all data, with   temperatures that are significantly lower than  the instantaneous reheating temperature. This requires that the  coupling of the potential lies within a very tight range. Outside this range  these models cannot be made compatible with the power spectrum data for any value of the equation of state parameter $w$ in the range $ -1/3 < w < 1$.
 
 In sections VI,  we end up with our conclusions.

\section{The model}

We shall consider an action where scalar fields $h^J$ are coupled to gravity in the following manner
\bea
S \, = \, \int \, d^4 x \, \sqrt{ - g} \, \left( \, f({\cal{R}}, h) + \dfrac{1}{2} G_{IJ}(h) \, \partial h^I \partial h^J - V(h)  \, \right)\, .
\label{act1}
\eea 
In it ${\cal{R}}$ is the scalar curvature in the Palatini formalism and $f({\cal{R}}, h)$ and arbitrary function of the scalars $h^J$ and 
${\cal{R}}$.  This action is reminiscent of an $f(R)$ theory  in which scalar fields are involved with kinetic terms  written in the most general way resembling  $\sigma$ - models.  Following standard procedure we write this action in the following manner,
introducing the auxiliary field $\Phi$. 
\bea
S \, = \, \int \, d^4 x \, \sqrt{ - g} \,  \left( \, f(\Phi, h) + f^\prime(\Phi, h) \, (  {\cal{R}} - \Phi) + \dfrac{1}{2} G_{IJ}(h) \, \partial h^I \partial h^J - V(h) \, \right) \, .
\label{act2}
\eea
In this $  f^\prime(\Phi, h)$ denotes the derivative with respect $\Phi$. 
One can define $\psi$ in the following way 
\bea
\psi = \dfrac{ \partial f ( \Phi, h)}{ \partial \Phi} , \quad \text{with \, inverse} \quad \Phi = \Phi ( \psi, h)   \, ,
\label{psiphi}
\eea 
so the action is written as follows,
\bea
S \, = \, \int \, d^4 x \, \sqrt{ - g} \,  \left( \,  \;
\psi \, {\cal{R}} +  \dfrac{1}{2} G_{IJ}(h) \, \partial h^I \partial h^J  - \psi \Phi + f(\Phi, h) - V(h) \,  \right) \, .
\label{act3}
\eea
One can go to the Einstein frame by performing a Weyl transformation of the metric
\bea
g_{\mu \nu} = \lambda \, {\bar g}_{\mu \nu} , \quad  \text{with} \quad \lambda \, \psi=\dfrac{1}{2} \, .
\eea 
That done the theory  in the Einstein frame receives the following form,
\bea
S \, = \, \int \, d^4 x \, \sqrt{ - \bar{g}} \, \,   \left( \;
 \, \dfrac{\overline{\cal{R}}}{2} +  \dfrac{1}{4  \psi} \, G_{IJ}(h) \, \partial h^I \partial h^J 
 - \dfrac{1}{4  \psi^2} \,  ( \psi \Phi - \, f(\Phi, h) + V(h))  \, \right) \,  .
\label{ein1}
\eea
The last step is to eliminate the field $\psi$ whose equation of motion is trivially found to be
\bea
\psi \, (\partial h )^2 \, = \, \psi \, \ \Phi   - 2 \, f(\Phi, h)  + 2 V(h) \, , 
\label{solve1}
\eea
where, in order  to speed up notation,  we have  denoted $ G_{IJ}(h) \, \partial h^I \partial h^J =  (\partial h )^2$. 

Note that (\ref{solve1})  is not solvable, in general, but we shall exemplify it in ${\cal{R}}^2$-theories where this can be analytically solved.  In the following we shall focus on such theories which can be considered as generalizations of the Starobinsky action. However there are two major differences, first the coefficients of the linear and quadratic, in the curvature $  {\cal{R}}$, terms are not in general constants, and second the framework  is the Palatini formalism in which the  connection is not the well-known Christoffel  connection but it is treated as an independent field.

We shall apply the previous formalism when only a single scalar,  $h$,  is present and $ f(h , {\cal{R}})$ is quadratic in the curvature having the form
\bea
f({\cal{R}}, h) \, = \, \dfrac{g(h)}{2} \,  {\cal{R}} \, + \,\dfrac{  {\cal{R}}^2}{ 12 M^2(h) } \,   .
\label{fr}
\eea
Since a single scalar field is assumed its kinetic term can be always brought to the form $ \, (\partial h )^2  / 2 $, that is in the action  (\ref{act1})  the field can be taken  canonically normalized.  Therefore in this theory there are three arbitrary functions, namely $g(h), M^2(h), V(h)  $, and any choice of them specifies a particular model. We have set the reduced Planck mass 
$ m_P = (8 \pi G_N )^{-1/2}$ dimensionless and equal to unity and thus  all quantities in (\ref{fr}) are dimensionless. When we reinstate dimensions the functions $g, V$ have dimensions $mass^2, mass^4$, respectively, while $M^2$ is dimensionless. 
Note that a non-trivial field dependence of the functions $g(h)$ and / or $ M^2(h)$ is a manifestation of non-minimal coupling of the scalar $h$ to Palatini Gravity.
Note that since we employ Palatini formalism,  there is no a scalaron field,  associated with an additional propagating degree of freedom,  which in the Einstein frame of the metric formulation  play the role of the inflaton.

With the function $ f({\cal{R}}, h) $ as given by (\ref{fr}) we get from  Eq. (\ref{psiphi}), 
\bea
\psi = \dfrac{g(h)}{2} \, + \, \dfrac{\Phi}{ 6 M^2(h) } \, , 
\label{ppp1}
\eea 
whose inverse is, 
\bea
\Phi = 6 M^2(h)  \, \left( \psi - \dfrac{ g(h)}{ 2}  \right) \, . 
\label{sigma}
\eea
Using these we can  solve  (\ref{solve1}) in terms of $\psi$  in a trivial manner, 
\bea
\psi = \dfrac{4 \, V + 3 M^2 g^2 }{ 2 (\partial h )^2  + 6 M^2 g }  \, , 
\label{psisol}
\eea
that is $\psi$, an hence $\Phi$ from (\ref{sigma}), are expressed in terms of $ h, (\partial h )^2 $. 
Plugging $\psi, \Phi$ into (\ref{ein1}) we get, in a straightforward manner, 
\bea
S \, = \, \int \, d^4 x \, \sqrt{ - {g}} \, \,  \left(  \;
 \, \dfrac{{\cal{R}}}{2} +  \dfrac{K(h)}{2} \,  (\partial  h )^2 \, + \, \dfrac{L(h)}{4} \,  (\partial h )^4 \, - \, V_{eff}(h) \, \right) \,  .
\label{einfinal}
\eea
In this action we have suppressed the bar in the the scalar curvature and, also,  $ \sqrt{ -g}$, and in order to simplify notation   we have denoted $  \partial_\mu h \partial^\mu h$  by $  (\partial h )^2$ and 
$(  \partial_\mu h \partial^\mu h)^2$ by $ (\partial h )^4 $.  
Note the appearance of quartic terms $  (\partial h )^4$ in the action. As for the functions $K,L, V_{eff}$, appearing in (\ref{einfinal}), they  are analytically given by
\bea
L(h) \, = \, ( 3 M^2 g^2 + 4 V)^{-1} \, , \, K(h)  \, = \, 3 M^2 g L \, , \, V_{eff} = 3 M^2 V L  \, .
\label{functs}
\eea
Observe that since terms  up to  ${\cal{R}}^2$  have been considered,   in the $ f({\cal{R}}) $ - gravity,   higher than $  (\partial  h )^4$  terms do not appear in the action. 

 The above Lagrangean may  feature, under conditions,  K - inflation models \cite{ArmendarizPicon:1999rj}, which involve a single field,  described by an  action whose general form is
\bea
S \, = \, \int \sqrt{-g} \;  \left(  \, \dfrac{{\cal{R}}}{2} + p (h, X) \, \right) \, d^4 x   \, .
\label{ssss}
\eea
where $X \equiv (1/2) \partial_\mu h \partial^\mu h$. 
The comological perturbations of such models were considered in \cite{Garriga:1999vw}.
However the importance of a time-dependend speed of sound $c_s$ in K - inflations models was emphasized in \cite{Lorenz:2008je} and cosmological constraints were derived,  using improved expressions for the density perturbations power spectra.  
Specific  models with  $ p (h, X) = F(X) - V(h)$ were considered  in  \cite{Li:2012vta}.   In (\ref{einfinal})  the Lagrangean density involving the scalar field is identified with $p(h, X)$,  but  the function $F(X)$ is now replaced by  $  {K(h)}\,  X\, + \, {L(h)} \,  X^2 $, which  depends,  in addition to  $X$, on the field $h$, as well, through  $K(h), L(h)  $. 

In a flat Robertson-Walker metric, where the background field $h$ is only time dependent, the energy density and pressure are given by
\bea
\rho(h, X) \, = \, {K(h)} X + 3 \,L(h) X^2 + V_{eff}(h)
\quad , \quad 
p(h, X) \, = \, {K(h)} X +  \,L(h) X^2 - V_{eff}(h)  \, ,
\label{enpre}
\eea
with $X$ being, in this case, half of  the velocity squared,   $ X = {\dot h}^2 / 2$. 
  
We shall assume that the function $L(h)$ is always positive to avoid  phantoms, which may lead to  an equation of state with  
$w < -1$. This may occur  when $ L < 0$ and $ X $  becomes sufficiently  large. However, there is no restriction for  the sign of $K(h)$ which  may be negative in some regions of the field space, signaling that the kinetic term has the wrong sign in those regions.  Obviously  the sign of $K(h)$ should be positive at the minimum of the potential.  Options where $K$ is negative in some regions, although interesting,  will not  be pursued in this work.   Besides,  we shall assume that the potential is positive $V_{eff}(h) \geq 0$ and it  has  a Minkowski vacuum. This ensures that the  energy density is positive definite.  When inflation models are considered,  the inflaton  will roll down towards  this minimum signaling the end of inflation and beginning of Universe thermalization. These are rather mild conditions.   

Concerning  the potential $V_{eff}$,  appearing in the Lagrangian (\ref{einfinal}) in the Einstein frame, from  the last of (\ref{functs}) we see that due to the fact that  we have assumed  $L, M^2 > 0 $,  positivity of  $V_{eff} \geq 0$ entails $ V \geq 0$.   Moreover one  can trivially show, from  (\ref{functs}), that $V_{eff}$  be cast in the following form,  
\bea
 V_{eff} = \dfrac{3 \, M^2 }{4} \, \left(  1 - \dfrac{K^2}{3 M^2 L} \right)  \, .
 \label{potspe}
\eea
From this it is seen that besides being positive  the potential is bounded from above by, 
\bea
V_{eff} \leq   \dfrac{3 M^2}{4 }   \, .
\label{bbb}
\eea
This upper bound can be easily saturated, for large $h$, by choosing appropriately the functions involved, namely $g , M^2$ and $V$.
Actually the asymptotic values of these functions, for large $h$, control the behavior of the potential in this regime
{\footnote{
It is fairly easy to see that   saturation of the bound (\ref{bbb}), for large field values,  is  easily obtained if  \\
$  \dfrac{g^2 \, }{V}  \longrightarrow \, 0 \, \;  ,  \; \text{and} \; \, \;   \dfrac{g^2 \, M^2 }{V}  \ll \, 1 \; \, , \quad \text{as} \quad h \longrightarrow large $.
}} .
 If we opt that the function $M^2$ approaches a plateau, or is constant, so does the potential  which may therefore drive  successful inflation.  The requirement to have a Minkowski vacuum can be, also, easily satisfied, and therefore many options are available for potentials bearing the characteristics demanded for the inflationary slow-roll mechanism to be implemented. This will be exemplified in specific models, to be discussed later.

Concluding this section,  we presented a general, and  model independent, framework of ${\cal{R}}^2 $ - theories, in the Palatini formulation of Gravity, which may be useful for the study of inflation models and may support slow-roll inflation.  In the Einstein frame these theories have much in common with the $K$-inflation models.  This formalism will be implemented,  for the study of various models of inflation in the following sections.

\section{The equations of motion and the slow-roll}

When non canonical kinetic terms are present the equations of motions for the would be inflaton scalar field $h$ differ from their standard form. As a result, the  cosmological parameters describing the  slow-roll evolution should be modified appropriately. Certainly one can normalize the kinetic term of the scalar field appropriately but this is not always very convenient. Actually  the integrations  needed, in order  to pass from the non canonical to the canonical field, are not easy, in most of the cases,  to be carried and the results cannot be presented in a closed form. Therefore it proves easier to work directly with the non canonical fields and express the pertinent  cosmological observables in a manner that is appropriate for this treatment.

It is not hard to see that the field $h$ satisfies the equation of motion given by
\bea
( K + 3 L \, \dot{h}^2 ) \ddot{h} +3 H(  K +  L \, \dot{h}^2  ) \, \dot{h} + V^\prime_{eff}(h) 
+\dfrac{1}{4} \, ( 2 K^\prime + 3 L^\prime \, \dot{h}^2 ) \,\dot{h}^2 =0  \, ,
\label{eomun}
\eea
in it all primes denote derivatives with respect $h$. If the field were canonical, $K = 1$,  and there were no quartic in the velocity terms, that is $L = 0$, then the equation above receives its   well-known form.  In this, 
the effect of using a non canonical, in general, field $h$ is encoded in the function $K$. The effect of the presence of terms 
 $({\partial h})^4$ in the  action is encoded within the function $L$.   The terms that depend on $L$ are  multiplied by an extra power of the velocity squared, as compared to the $K$-terms. These cannot be neglected  although, as we discuss below, they are  small in particular models, during inflation. 

We can gain more insight is we momentarily   use a canonically normalized field, say $\phi$, defined by
 \bea
 \phi = \int{ \sqrt{K(h)}} \, dh   \, .
\label{canon}
 \eea
 To avoid ghosts we shall assume that $ K > 0$, so that the integration above makes sense.  Actually if  $K$ is negative the kinetic term of the field $\phi$ will have the wrong sign, i.e. $ - ( {\partial \phi} )^2 / 2$.  It could happen however that this function is negative in some region but at the Minkowski vacuum is strictly  positive.  In this way ghosts are also avoided.  This case, interesting as might be, is not discussed and we prefer to take a rather conservative view point of having $ K > 0$ in the whole region. 
Then in terms of the field $\phi$  the equation of motion (\ref{eomun}) takes up the form
 \bea
 \left( 1 +   \dfrac{3 L}{K^2}  \dot{\phi}^2 \right) \, \ddot{\phi} +3 H \left(  1 +   \dfrac{ L}{K^2}  \dot{\phi}^2 \right) \, \dot{\phi} + 
  \dfrac{d V_{eff}}{d \phi} +
 \, \dfrac{3 L}{ 4 K^2} \,  \dfrac{d \,ln \, ({L}/{K^2} )}{ d \phi }  
 {\dot{\phi}}^4
 = 0   \, .
 \label{eomnum2}  
 \eea 
  From this form it appears that the smallness of the  $ {\partial h}^4$ terms in the action is quantified by the smallness of the ratio 
  $ \frac{ L}{K^2}  \dot{\phi}^2 \ll 1 $, which is equivalent to  $ \frac{ L}{K}  \dot{h}^2 \ll 1 $. 
  Neglecting this in the equation above we recover the well-known  form of the equation of motion for the canonical field $\phi$. 
  
\begin{figure}
\centering
  \centering  
  \includegraphics[width=0.45\linewidth]{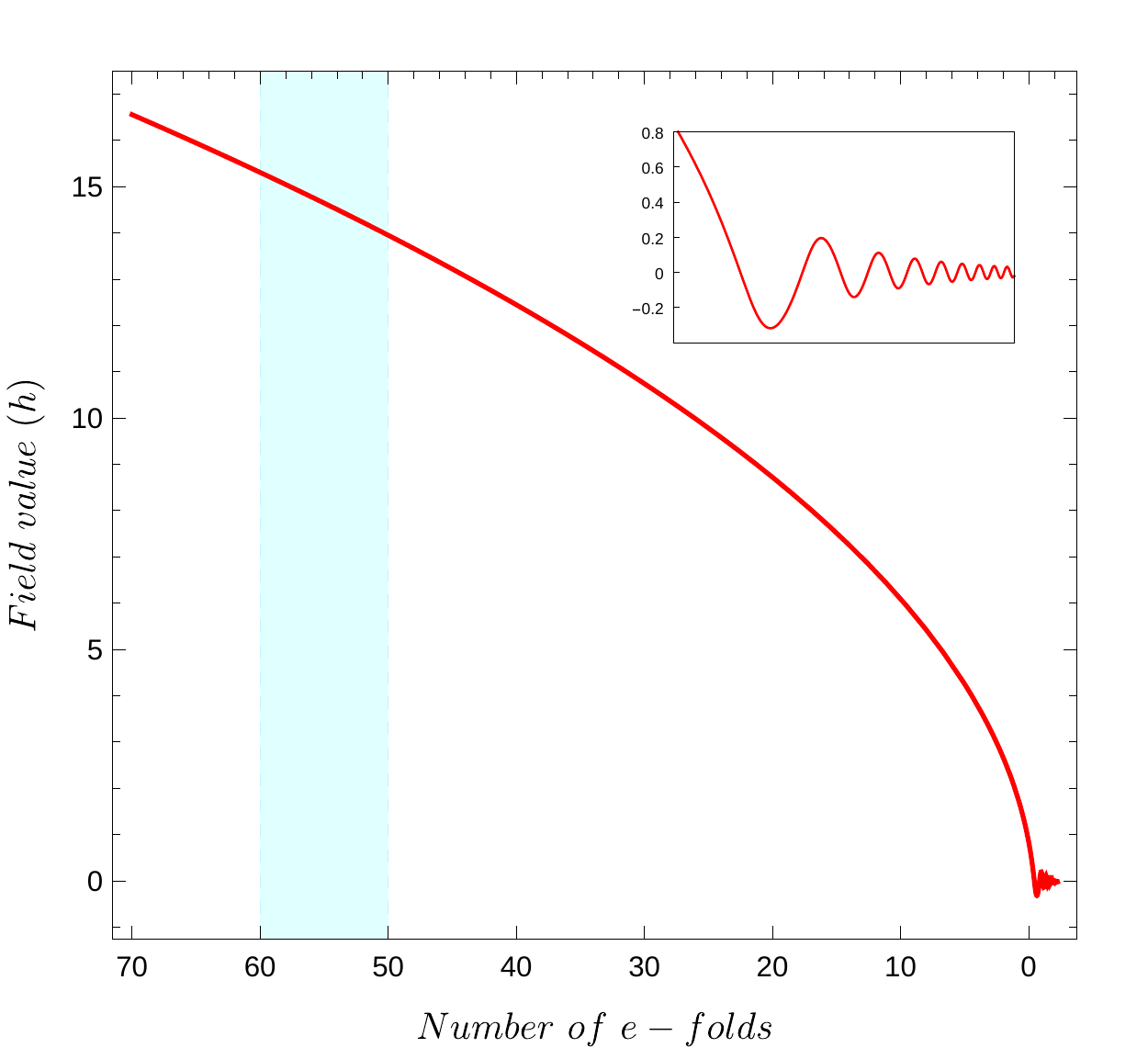}     
   \includegraphics[width=0.45\linewidth]{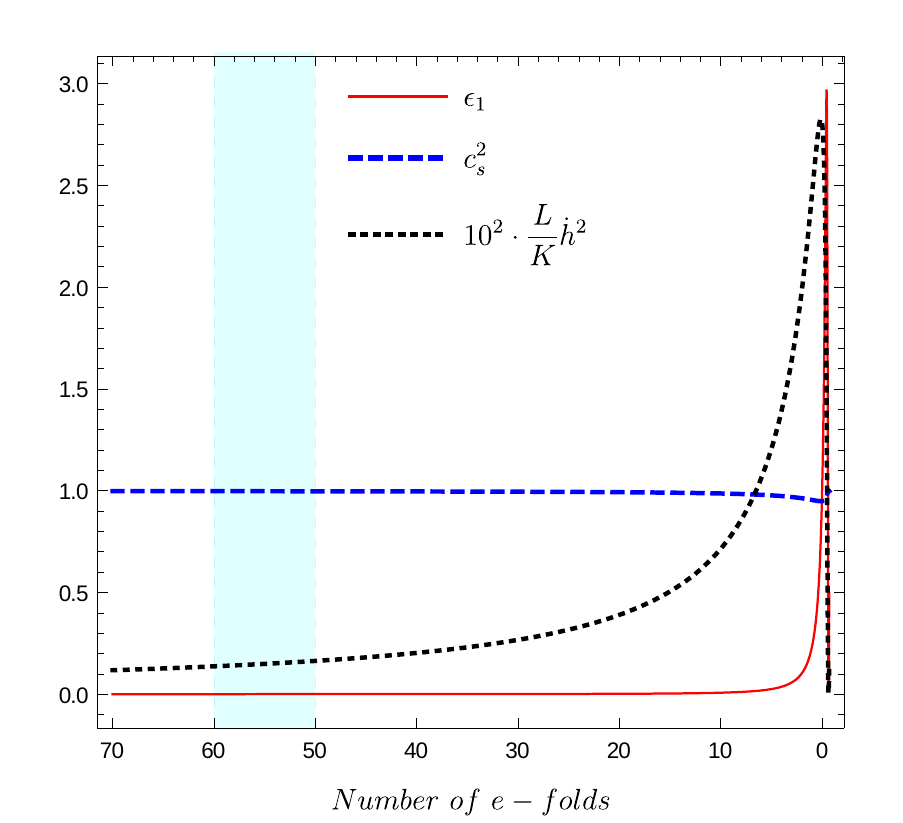} 
\caption{
On the left pane the evolution of the field $h$ with number of e-folds is shown.
 In the small inset figure, the rapid oscillations of $h$, as it approaches  the minimum of the potential, are shown magnified.  
 On the right we display the parameter $\epsilon_1$, the sound speed squared $c_s^2$ and the quantity $  10^{2} \,{ L}\, \dot{h}^2 / K$ .    }
   \label{fig00}
\end{figure}

  However, it should be stressed that our numerical study duly takes  into account the contribution of these terms and no approximation, whatsoever,  is made, although we have found numerically  that they are small, at least in the models under consideration  in this work. 
This has been also pointed out in   \cite{Enckell:2018hmo,Antoniadis:2018ywb,Antoniadis:2018yfq,Tenkanen:2019jiq}. 
In order to show this, and anticipating conclusions based on our numerical treatment that are  presented in forthcoming sections,  in Figure \ref{fig00} we plot on the left panel the evolution of the field $h$ versus the number of e-folds 
$ N = ln (a_{end}/a(t) )$,  from the  time the number of e-folds reached $N = 70$ to  the end of inflation corresponding  to $N = 0$.  
 On the right panel, we display the parameter $\epsilon_1 =  - \frac{\dot{H} }{ H^2}$, the sound of speed squared $ c_s^2$ and the evolution of  $  \frac{ L}{K}  \dot{h}^2 $. For reference, we have also included a vertical band to mark the region $ N = 50 - 60$, usually quoted in literature.  
 These plots regard  the   minimally coupled model, for which $g =1, M^2 = \frac{1}{3 a}$ and $ V = \frac{m^2}{2} h^2$. The  values of the parameters $a, m^2$, used in producing Figure  \ref{fig00},  correspond to Model I ( C - case ), discussed later in section V. for which $ a =  2.0 \times 10^9 $ and $ m = 6.32 \times 10^{-6}$. However similar findings hold for the other models studied  in  this work, as well.
 
  On the left pane of this figure one notices  the rapid damped oscillations of $h$,  after the end of inflation,  when  it starts falling to the minimum of the potential. These are clearly visible in the  magnified inserted  small figure.  From  the right pane,  one can see that 
  $\epsilon_1 << 1.0$,  $ c_s^2 \simeq 1.0$, and $  \frac{ L}{K}  \dot{h}^2 \sim  {\cal{O}} ( 10^{-2} )$, for any number of e-folds that is 
 larger than about $5$, or even smaller.  For $ N \lesssim 5$ the function    $\epsilon_1$ starts growing and  $  \frac{ L}{K}  \dot{h}^2 $ increases significantly, however its magnitude stays small till the end of inflation.
   
 On the other hand, the scales of interest,  from CMB observations, are within the range $10^{-4} Mpc^{-1} \lesssim k \lesssim 10^{-1} \, Mpc^{-1}$ and the number of e-folds that are left to the end of inflation, from time  $t_k$ a scale $k$ crossed the sound horizon,  is $ N_k = ln (a_{end}/a(t_k) )$.  Even  for the largest scale, in the aforementioned range,  the number of e-folds cannot be less than about  $\simeq 20$, as we have found numerically. Therefore,   any scale $k$ in the range of interest,  crossed the sound horizon long before  the end of inflation,  when 
$\epsilon_1 << 1.0$,  $ c_s^2 \simeq 1.0$, and $  \frac{ L}{K}  \dot{h}^2 $ was  small  $ {\cal{O}} ( 10^{-2} )$.  Therefore,  for the cosmological scales 
of interest the contribution of the $L$ - terms is small. 

Although small, for a broad range of the parameters and for the class of models studied here, the role of $ \frac{ L}{K}  \dot{h}^2 $ is important in the determination of the energy density $ \rho_{end} $  at the end of inflation, which in turn  affects the instantaneous reheating temperature $T_{ins}$.
This delicate issue will be discussed later in Section V. Even in this case, however,  we have found that the values of $ \rho_{end} $  deviate from those obtained approximately by factors of order  ${\cal{O}}(1)  $. 

The previous arguments state that, for the cases of interest to us, one can use the slow-roll approximation and at the same time  neglect the  $ L  $ - terms,  provided their omission is adequately  justified.  We repeat that, our results are based on a numerical study and no such an approximation is made. However,  this does not deprive us from the right, and for an analytic treatment of the models under consideration, to present qualitative  arguments, based on this approximate scheme,  aiming at a better understanding of the results  that are reached based on a numerical study  in which all terms are included and no  approximation is made. 
  
Provided that the contribution of  the  $L$ - term is small,  and with $ K > 0$, the first  slow-roll parameters, as defined in terms of the potential are given by,  in terms of the non canonical field $h$,  
\bea
\epsilon_V \, = \, \dfrac{1}{ 2 K(h) } \, 
\left( \dfrac{V_{eff}^\prime }{ V_{eff} } \right)^2
\quad , \quad
\eta_V \, = \, 
  \dfrac{  \left(  K^{-1/2} V_{eff} ^\prime \right)^\prime}{ K^{1/2} \,V_{eff} }   \, .
\label{epv}
\eea 
In these equations primes denote derivatives with respect the field $h$. 
It is trivial to show that these definitions  indeed coincide with the well-known definitions if the canonically normalized field $\phi$ of Eq. (\ref{canon}),  is used.  
As for the  the number of e-folds, left to the end of inflation, this is given by  
\bea
N_* \, = \, \int^{h_*}_{h_{end}}  K(h) \, \dfrac{ V_{eff}(h) }{  V_{eff}^\prime(h) } \, dh  \, .
\label{nleft}
\eea
In this  $h_*$ is the pivot value and $h_{end}$  the value of the field at the  end of inflation.

\section{Cosmological observables}  

Concerning the cosmological observables, we start by discussing the scalar and tensor power spectra which play an important role in 
inflationary cosmology.   The CMB observations restrict considerably  the predictions of inflationary  models, in general,  and also put severe constraints on specific models, encompassed in the  framework of the Palatini Gravity, as we shall analyze in forthcoming sections.  

There is a long history on studies towards this direction  the first calculations were performed in \cite{Starobinsky:1979ty,Mukhanov:1985rz,Mukhanov:1988jd}. Since then there has been an intense activity in the field using different methods and improvement of  the calculations,  by considering higher order corrections, demanded by the precise measurements  of the cosmological parameters, 
or  tackle theories with a variable speed of sound, 
\cite{Lucchin:1984yf,Stewart:1993bc,Gong:2001he,Schwarz:2001vv,Martin:2002vn,Habib:2002yi,Leach:2002ar,Habib:2004kc,Casadio:2004ru,Wei:2004xx,Casadio:2005xv,Kinney:2007ag,Lorenz:2008je,Lorenz:2008et,Agarwal:2008ah,Martin:2013uma,Jimenez:2013xwa,Alinea:2015gpa}

Choosing an arbitrary pivot scale,  $k^*$,  that exited the sound horizon at $t^*$, that is $k^* c_s(t^*) = a(t^*) H(t^*)   $,  the scalar and tensor power spectra can be expanded about this pivot. keeping  the first order terms in the slow-roll parameters, one has 
\bea
{\cal{P}}_\zeta (k) \, = \, \dfrac{H_*^2}{8 \pi^2 m_P^2 \, \epsilon_1^* c_s^*}  \,
A \left( 1 - 2 ( D+1) \, \epsilon_1^* - D \epsilon_2^* - (2+D) \, s_1^*
+( -2 \epsilon_1^* - \epsilon_2^*-s_1^* ) \, ln \dfrac{k}{k^*}
 \right)
\label{power}
\eea
and 
\bea
{\cal{P}}_h (k) \, = \, \dfrac{2 \, H_*^2}{ \pi^2 m_P^2 }  \,
A \left( 1 - 2 ( D+1 - ln c_s^*) \, \epsilon_1^* 
+( -2 \epsilon_1^*  ) \, ln \dfrac{k}{k^*}
 \right)  \quad . 
 \label{ten1}
\eea
A subtle point concerns the dependence of the tensor power spectrum on the sound velocity $c_s$. Usually this is calculated evaluating all quantities at the time of  Hubble horizon, which differs, in general,  from the time of sound horizon. However if we want to compare the scalar and tensor spectra using the results of cosmological measurements the same pivot should be used, for consistency, as has been emphasized in  \cite{Lorenz:2008je,Agarwal:2008ah,Martin:2013uma,Jimenez:2013xwa}.  Using a pivot scale $k^* c_s(t^*) = a(t^*) H(t^*)   $ in the scalar spectrum a dependence on $c_s$ emerges in the tensor spectrum, as well. 

In these equations the Hubble flow functions (HFF), usually referred to as  slow-roll parameters,   are defined in the usual manner, in terms of the Hubble rate, 
\bea
\epsilon_1 \equiv - \dfrac{d ln H}{dN} = - \dfrac{\dot{H} }{ H^2 } \quad , \quad \epsilon_2 \equiv \dfrac{dln \epsilon_1}{dN} = \dfrac{\dot{\epsilon_1} }{ \epsilon_1 H } 
 \quad , \quad  s_1 \equiv \dfrac{d ln c_s }{ dN} = \dfrac{\dot{c}_s }{ c_s H }  \, , 
 \nonumber
\eea
where  $ dN = H dt$.  A star in the HFF in the expressions above means that these  are evaluated at  $t^*$. 
  The equations (\ref{power},\ref{ten1}) can be found in references \cite{Lorenz:2008je,Lorenz:2008et,Martin:2013uma,Jimenez:2013xwa}. In those references a slightly different pivot scale is usually quoted, $k_\diamond c_{s \diamond} (\eta_\diamond)= - 1/\eta_\diamond $, 
  where $  \eta_\diamond $ is the conformal time, 
  and the corresponding expressions are given in terms of  $k_\diamond $. However to first order, in  HFF,  they  have the same form when  the  quanities denoted by  a diamond symbol  are  replaced by the corresponding  starred ones. It is only the second order terms that are affected and the corresponding coefficients differ. Note that higher order corrections have been calculated but here we retain the next to leading corrections, that is first order in the slow-roll parameters.  In \cite{Martin:2013uma}    the constants $A, D$ are analytically given. In fact their values are $ A = 18 e^{-3}$ and $ D = 1/3 - ln 3$, as they follow using the 
  {\it{uniform approximation}},  a method that is suitable for theories with varying speed of sound $c_s$,  which resembles the WKB method. Taking next order corrections  in the adiabatic approximation these constants are dressed ( renormalized ) and $A$ turns out to be very close to unity while $D$ becomes $ D = 7 /19 - ln 3$. For our numerical treatment we shall therefore use the renormalized values. For details we point the reader to reference \cite{Martin:2013uma} where a detailed study is presented, encompassing also higher order corrections,  and comparison with other calculations is made.

The corresponding amplitudes, for scalar and tensor perturbations, are then given by,
\bea
A_s(k^*) = {\cal{P}}_\zeta (k^*) \quad , \quad A_t(k^*) = {\cal{P}}_h (k^*)   \quad .
\label{amplitude}
\eea

Concerning the spectral index of the scalar power spectrum, following standard definitions, this is given by
\bea
n_s  \, = \, 1  -2 \, \epsilon_1^* - \epsilon_2^* -  s_1^* 
\label{speind}
\eea
while the tensor-to- scalar ratio is  given by,
\bea
r \equiv \dfrac{{\cal{P}}_h (k^*) }{{\cal{P}}_\zeta (k^*)} = 16 \,  \epsilon_1^* \,  c_s^* \, ( 1 + 2 \, ln c_s ^*\,  \epsilon_1^*  + D \epsilon_2^* 
+ (2 + D) \, s_1^*)   \quad ,
\label{ratiorcor}
\eea
to the  same order of approximation.

As for the number of e-folds left,   $\, N_k = ln \, \dfrac{ a_{end}}{ a(t) }$,     from the time $t$ some  scale $k$ crossed the sound horizon to the end of inflation,  this is given by,   
 \bea
 N_k \, =&& \, ln \left[ \left(\dfrac{\pi^2}{30} \right)^{\frac{1}{4}} \dfrac{( g^{*,0}_s)^{ \frac{1}{3}}}{ \sqrt{3} } \dfrac{T_0}{H_0} \right]
 - ln\left( \dfrac{k}{  a_0 H_0} \right) - ln c_s
 + \dfrac{1}{4} \left( ln \, \dfrac{ \,  3 H^2}{ m_P^2} + ln \, \dfrac{ 3 H^2 m_P^2}{  \rho_{end}}  
\right)  \nonumber \\
&& \, + \, \dfrac{1- 3 w}{12 ( 1+w)} \, ln \, \dfrac{\rho_{reh} }{ \rho_{end }} 
- \dfrac{1 }{12} \, ln \, g_s^{* (reh)}  \, + \dfrac{1}{4} \, ln \left( \dfrac{ g^{* (reh)} }{ g_s^{* (reh)}  } \right)
\, .
  \label{nfold1}  
 \eea   
 The Hubble rate, as well as $c_s$ on the right hand side of it,  are evaluated at the crossing time $t$. 
 See  \cite{Liddle:2003as,Dodelson:2003vq},  and  also \cite{Martin:2010kz,Lozanov:2017hjm}.  Note that, a  
$ - ln c_s$ is included in the expression for $N_k$,   due to the fact that the speed of sound may not be unity, as is the case in $K$-inflation models. All other terms are well-known. 
 In this equation  $g^*, g_s^*$ are the energy density and entropy density degrees of freedom. In all quantities superscripts or  subscripts labeled  by $0$ or $reh$ denote evaluations at present epoch  and end of reheating  respectively. 
 In this the parameter $w$ is the effective equation of state parameter in reheating period.  The last line depends on the reheating values.  The first term in the equation for $N_k$,  takes the well-known value $66.89$ when the reduced Hubble constant $h$ equals to  $h = 0.676$. In the expression above we prefer to present it analytically as  in \cite{Lozanov:2017hjm}. 
 Taking a  different value for $h$, always  within observational limits, the first term  will be replaced by $ 66.89 - ln(h/0.676)$.  Concerning the last term, this is very small and can be omitted, due to the fact that entropy and energy degrees of freedom are very close to each other for temperatures $ T > 500 \, keV$.  Therefore we shall drop it in the discussion that follows.  The term before the last is also small but we shall retain it.  In fact  assuming the $SM$ content this takes values  $g_s^{* (reh)} = 106.75$, for temperatures above $ \sim 1 \, TeV$, being smaller for lower temperatures.  Therefore  this term contributes little, less that ${ \cal{O}} ( 1 \% )$.  The  term,  $ \sim ln ( { 3 H^2 m_P^2}/{  \rho_{end}}  )$ ,  is also not expected to be large. The largest contributions stem from the term  $ \sim ln ( { 3 H^2 }/{ m_P^2}  )$ and  the reheating term, in most of the models, in which the speed of sound is unity.  With varying speed of sound the contribution from the $ - ln c_s$ term is positive and may also give significant contribution, in general.    The constant  $w$ characterizes an effective equation of state parameter.  
 In the following, whenever   the pivot scale $k^*$  is used in (\ref{nfold1}),  $N_k$ shall be denoted by $N_*$.  
 
 Given an inflationary model, the largest uncertainties in $N_k$ are mainly due to the period of Universe  reheat after this exited from inflation. For a review see for instance \cite{Allahverdi:2010xz}.  For the scales of interest, these uncertainties yield values of $N_k $ in the range $50 - 60$, usually quoted in the literature.
 The reheating  temperature the Universe reached after its thermalization has been extensively studied and various mechanisms and models have been put under theoretical scrutiny,  \cite{Allahverdi:2010xz,Podolsky:2005bw,Martin:2010kz,Adshead:2010mc,Mielczarek:2010ag,Easther:2011yq,Dai:2014jja,Munoz:2014eqa,Cook:2015vqa,Gong:2015qha,Rehagen:2015zma,Lozanov:2017hjm}. 
The number of e-folds accrued during the  reheating period, 
 $   \Delta N_{reh} $,  is given by  
 \bea
  \Delta N_{reh} \equiv ln \, \dfrac{a_{reh}}{a_{end}} = - \dfrac{1}{3 ( 1+w)} \, ln \, \dfrac{\rho_{reh} }{ \rho_{end }}  \, .
  \label{nfoldreh}
 \eea
 The subscripts  (reh), (end) in the cosmic scale factor and the energy densities denote that these quantities are evaluated at the end of the reheating period and inflation respectively. 
 The  effective  equation of state parameter $w$ in the reheating period we consider as a free parameter. At the end of inflation 
 $w = - 1/3$ while the value $ w = 1 / 3$ corresponds to the onset of radiation dominance. In the canonical reheating scenario $w =0$, but  values in the range  $ \simeq 0.0 - 0.25$, or larger, right after inflation,  are also possible in some models
 \cite{Lozanov:2016hid,Lozanov:2017hjm}.

For any model given a  value of $N_k$,  we have a prediction of  
$\Delta N_{reh}$ and  in this sense  (\ref{nfold1})  serves as a probe of the reheating process. Inversely, given a reheating mechanism, within the context of any particular inflationary model,  the value of  $\Delta N_{reh}$ is fixed,  and hence $N_*$ is predicted. 
In terms of $ \Delta N_{reh}  $,  for given $w$,  one has for the reheating temperature,, see for instance  \cite{Munoz:2014eqa}, 
 \bea
 T_{reh} \, = \, \left(  \dfrac{ 30}{\pi^2 \, } \dfrac{      \, \rho_{end} }{   \, g^{* (reh)} \,} \, \, \right)^{1/4} \,
 exp \left({ - \dfrac{ 3(1+w) \Delta N_{reh}  } {  4}  } \right) \,  .
\label{treh}
 \eea  
 In our numerical studies we shall adopt the common  values $ g^{* (reh)}  = g_s^{* (reh)} = 106.75 $, corresponding to the SM content, as discussed before, for temperatures above $\sim 1 \, TeV$. 
 {\footnote{  
 With 
 $  g^{* (reh)}  = 100 $  Eq. (\ref{treh}) coincides with that given in  \cite{Munoz:2014eqa}.
 }} .
 Note  that since $  {a_{reh}} > {a_{end}} $  we have that $ \Delta N_{reh} \geq 0$, and therefore due to  $w > -1$ the reheating temperature $  T_{reh}$ is bounded from above 
 \bea
 T_{reh} \leq   \left( \dfrac{ 30}{\pi^2 \, } \dfrac{      \, \rho_{end} }{   \, g^{* (reh)} \,} \, \, \right)^{1/4}  \,  .
\label{tins}
 \eea
 The bound on the right hand side  of this  defines  the instantaneous reheating temperature, $  T_{ins}$. The  temperature 
 $  T_{reh}$ reaches this upper  bound when  the reheating process  is instantaneous, in which case $ \Delta N_{reh} = 0$.  Note that 
 for rapid thermalization we have $ \rho_{end} = \rho_{reh} $, from Eq. (\ref{nfoldreh}). 
 The reheating temperature should be  larger than  $  \sim 1 \,  MeV$ so that  Big Bang Nucleosynthesis (BBN) is not upset.  Lower values on $T_{reh}$ have been  established in   \cite{Kawasaki:1999na} and  recently in  \cite{Hasegawa:2019jsa}   .
 
 In terms of the reheating temperature the number of e-folds $N_*$, corresponding to a pivot scale say $k^*$,  is written as 
  \bea
 N_* \, = && \, 66.89 - ln c_s^* - ln\left( \dfrac{k^* }{  a_0 H_0} \right) + \dfrac{1}{4} \left( ln \, \dfrac{ \,  3 H_*^2}{ m_P^2} + ln \, \dfrac{ 3 H_*^2 m_P^2}{  \rho_{end}}  
\right)  - \dfrac{1}{12} \, ln \, g_s^{* (reh)}  \, 
\nonumber \\
&& + \dfrac{1 - 3 w }{ 3 (1 + w)} \, \left(   ln \dfrac{T_{reh}}{m_P}  - \dfrac{1}{4} \, ln \dfrac{\rho_{end}}{m_P^4}  
- \dfrac{1}{4} \, ln \dfrac{30 }{ \, \pi^2} +  \dfrac{ln \,  {g^{* (reh)}  }}{4} \,   
\right)  \, .
 \label{nfold2}  
 \eea   
 which we shall use in the following. 
 On the right hand side of it $H_*,  c_s^*$ are  the Hubble rate and the speed of sound are, respectively,  evaluated at $t^*$, the time of sound horizon crossing of the mode $k^*$. 
  In this we have taken $h = 0.676$. Uncertainties in the value of $h$ little affect $N_*$, and for this reason we have also dropped the dependence of  $N_*$ on  $ln(h/0.676)$.

The appearance of the sound of speed parameter $c_s$  is due to the fact that in the Palatini formulation of ${\cal{R}}^2$ gravity  higher in the  velocity $ \dot{h}$  terms unavoidably appear, and its value deviates from unity. In fact $c_s$ is defined by
\bea
c_s^2 \, = \, \dfrac{ \partial p / \partial X}{ \partial \rho / \partial X}  \, ,
\label{sound}
\eea
where $X$, defined after Eq. (\ref{ssss}), is half the velocity squared.  In terms of the field $h$ and its velocity $\dot{h}$ this receives   the form
\bea
c_s^2 \, = \, \dfrac{ 1 + L \, {\dot{h}}^2  / K}{ 1 + 3 L \, {\dot{h}}^2 / K}  \,   .
\label{sound2}
\eea
$ c_s $ is controlled by $ L \, {\dot{h}}^2 /  K   $,  the same combination  that appears in the equation of motion for the field $h$, and approaches unity when $ L \, {\dot{h}}^2 /  K \ll 1 \,$.

The Planck 2018 data  \cite{Akrami:2018odb,Aghanim:2018eyx},    yield a value for $A_s$ 
\bea
Log( 10^{10} \, A_s ) \, \simeq \, 3.04  \, ,
\label{aobs}
\eea
at a  pivot scale  $k^* = 0.05 Mpc^{-1}$.  In slow-roll inflation, in models with $c_s=1$, this pivot  crossed the horizon at times $t^*$ that  are well within the slow-roll regime.  Then  to lowest order  in HFF the scalar amplitude of  (\ref{amplitude}) is written as,  
\bea
A_s \, \simeq \,  \dfrac{1}{24 \pi^2 m_P^4 } \,  \dfrac{V_{eff}*}{\epsilon_1^*}    \, ,
\label{ampl3}
\eea
 Then in this approximation, and using (\ref{aobs}), we get
\bea
\dfrac{V^*_{eff}}{m_P^4} = 4.97 \times 10^{-7} \, \epsilon_1^*  \, .
\label{potbound}
\eea

 The corresponding amplitude for the tensor perturbations, in leading order, is found to be
\bea
A_t \, \simeq \,  \dfrac{ {2 \, V_{eff}*}  }{3 \pi^2 m_P^4 } \,   ,
\label{tensor}
\eea
 resulting to a tensor to scalar ratio
 \bea
 r = \dfrac{ A_t}{ A_s } = 16 \,  \epsilon_1^*  \, , 
 \label{ratior}
 \eea
 which yields, on account of (\ref{ampl3}) 
 \bea
 \dfrac{V^*_{eff}}{m_P^4} = \dfrac{3 \pi^2}{2} \, A_s r      \, , 
 \eea
 a well-known result. The Planck 2018 data, when combined with the BICEP2/Keck Array BK15 data, see 
 \cite{Akrami:2018odb,Ade:2018gkx}, yield an upper bound $\,r  < 0.063$,  which when the  pivot scale $ k^* = 0.002 \, Mpc^{-1}$ is used,  decreases   to $\,r_{0.002}  < 0.058$. With   $\, r  < 0.063$ we have an upper bound on the value of the potential  given by
 \bea
 \dfrac{V^*_{eff}}{   m_P^4}  \lesssim 2.0 \times 10^{-9}   \, , 
 \label{boundr}
 \eea
 which constrains the scale of inflation. 
 In terms of the Hubble rate this is actually the bound $ H_* / m_p < 2.5 \times 10^{-5} $ quoted in \cite{Aghanim:2018eyx,Akrami:2018odb}.     
 
 The  bound  $\,r < 0.063$ translates to a bound on  $  \epsilon_1^*$, which is actually follows from (\ref{ratior}), given by
 \bea
  \epsilon_1^*  < 0.004  \, .
 \eea
 In models with varying speed of sound $c_s$ the previous arguments do not hold, in general, and the role of $c_s$ has to be taken duly into account. This is done in our numerical analysis which will be presented later on.  
 However, as already discussed in the previous section,  we have found numerically  that $c_s$ is constant and very close to unity  for times until the time $t^*$ of the horizon crossing and only at the late stages of inflation deviation of $c_s$ from unity starts to show up. 
  Therefore on these grounds the arguments given before can be used to have a first  estimate of the bounds put in the parameters of  the models that we will  study in the forthcoming sections. However,  our final results do not rely on these qualitative arguments but rather  on a numerical study where the dependence  on $c_s$ is duly considered. 
 Note that separate lower bounds on $c_s$ are obtained from absence of non gaussianities, which are however satisfied in all models considered in this work,  due to the fact that $c_s^*$ is very close to unity,  as we have  already remarked.

  \section{Models}  
  
  Before embarking on considering specific models and presenting our results,  we find it appropriate to briefly  outline  the procedure followed in this section. 
  As already stated, towards the end of section III, our predictions are based on a study which solves Friedmann equations and the evolution equation  (\ref{eomun}) numerically with no-approximations made.  However before doing that we find it useful to first employ  the  slow-roll  approximation,  neglecting the contribution of  $L$ - terms. This is made   for comparison with the numerical results which are the only reliable source to reach physics conclusions. 
   In the models under consideration in this work,  the numerical study reveals that this approximate scheme makes sense since it is justified by the results of the numerical treatment.  It is  for this reason it  explains at a very satisfactory level the results that are derived  by our numerical treatment. However it should be remarked that this may not be a generic feature and may not be   valid  for other models encompassed within the framework of the Palatini $ {\cal{R}}^2$ Gravity. 
   
   Concerning our numerical analysis, the approximate scheme employed  is also useful towards having a first estimate  of  the magnitudes  of the parameters involved, which are  consistent with the  limits imposed  by the measurements of the cosmological parameters. In our numerical approach we scan the parameter space starting  from initial  values of the parameters that fall within the range suggested by this analysis. 
  
   In our procedure the time corresponding to the end of inflation, $t_{end}$,  is determined, as usual,  by the condition 
$\epsilon_1(t_{end})=1$, or same  $\, \ddot a{(t_{end})} = 0$.  The time $t^*$, corresponding to the sound horizon crossing, for a chosen pivot scale $k^*$, for any given reheating temperature $T_{reh}$, and effective equation of state parameter $w$, is then found by solving Eq.  (\ref{nfold2}). This  is a fairly easy task to implement numerically. That done all quantities at $t^*$,  which refer to the pivot scale $k^*$, are easily calculated.

  \subsection{ Minimally coupled models with potentials $ \sim h^n$}
  
  In this section we consider specific models using the formalism presented in previous sections, and discuss their predictions. 
 An interesting  class of models is the one in which the potential $V$  is a monomial   in the field $h$,  $V \sim h^n $, with $n$ even integer, and $g, M^2$ are constants, that is the scalar $h$  couples to gravity in a minimal manner. We set $g=1$
 {\footnote{
 When Planck mass is reinstated in the action  this corresponds to $ g = m_P^2$. 
 }} 
 and hence these models are described by 
 \bea
g(h) =1\quad , \; M^2(h) = \dfrac{1}{3 a} \quad , \quad V(h) \, = \, \dfrac{\lambda }{n} \, h^n \quad \text{with} \;  \; n = 
\text{positive even integer   }
\, .
\label{xx1}
\eea
Therefore two parameters, $a$ and $\lambda$ are involved which are in principle unknown.  Cosmological data will constrain their allowed values as we shall see shortly. In order to facilitate the analysis we define the parameter $c$ defined by the combination, 
\bea
c \, = \,  \dfrac{4 \,  \lambda  \, a }{n} \,   .
\label{xx2}
\eea
Then the  functions $K, L$ are given by
\bea
K(h) \, = \, ( 1 + c h^n )^{-1}  \; , \;  L(h) \, = \, a \,( 1 + c h^n )^{-1} \, ,
\label{xx3}
\eea
while the potential $V_{eff}$ receives the form
\bea
V_{eff}(h) \, = \, \dfrac{1}{4 a } \, \dfrac{c h^n }{ 1 + c h^n }   \, .
\label{xx4}
\eea
For large values of $h$ this is $\simeq 1 / 4 a$ therefore  $1/a$, which is proportional to  $M^2$,  sets actually the inflation scale. 

In order to find the region of the parameters $a , \lambda$, or equivalently $a, c$, which are consistent with cosmological data, we shall first consider the amplitude of the power spectrum $As$. It suffices, for this purpose,  to consider  the simplified form given by (\ref{ampl3}),  take $c_s^* \simeq 1$ and  replace $\epsilon_1^*$ by $\epsilon_V$ as given by (\ref{epv}). Then from the analytic form of the potential, given before, and from (\ref{epv}) the amplitude $A_s$ of Eq. 
(\ref{ampl3}) takes the form, putting  $ m_P = 1 $, 
\bea
A_s \simeq \dfrac{1}{24 \pi^2}   \dfrac{1}{ 2 n^2} \, \left( \dfrac{c}{ a}  \right)\, h_*^{n+2} \, = 
\, \dfrac{1}{12 \pi^2}  \, \dfrac{ \,  \lambda}{ n^3} \,  \, h_*^{n+2} \, ,
\label{xx5}
\eea
where $h_*$ is the  value of the field at $t^*$. 
One sees immediately that it  is the ratio $c/a$, or equivalently the parameter $\lambda $, that controls the magnitude of the amplitude $A_s$. For the central value of $A_s$, which is $ A_s \simeq  2.1  \times 10^{-9}$, on account of (\ref{xx5}) , we have 
\bea
\lambda  \, h_*^{n+2}  \simeq ( 2.49  \times 10^{-7} ) \; n^3
\quad \text{or} \quad 
 \left( \dfrac{c}{a} \right) \, h_*^{n+2} \simeq ( 9.95   \times 10^{-7} ) \; n^2  \, .
\label{xx6}
\eea
To further quantify the allowed range of the parameters we also need have an estimate for $h_*$. To this goal we use (\ref{nleft}) from which it follows  that 
\bea
N_* = \dfrac{1}{2 n} \, (  h_*^2 - h_{end}^2 )    \, ,
\label{xx7}  
\eea
which yields
\bea
h_*^2 = 2 n N_* +  h_{end}^2  \, .
\label{xx8}  
\eea
$h_{end}$ is defined as the value for which  $\epsilon_V = 1$. For the specific models
\bea
\epsilon_V  = \dfrac{n^2}{2} \; \dfrac{1}{ h^2 ( \,1+ c \, h^n)}  \, , 
\label{xx9}
\eea
therefore  $ h_{end}^2 $ is solution of the equation
\bea
c \, h_{end}^{n+2} + h_{end}^2 -\dfrac{n^2}{2} = 0  \, .
\label{xx10}
\eea
For $c=0$ the solution is exactly $ h_{end}^2  = n^2/2$ while  for any  $ c > 0 $ the only real and positive solution for $ h_{end}^2  $ is easily found to  be bounded by $ n^2 / 2$.  From this bound on $ h_{end}^2  $ and using the fact that $N_*$   is $ \sim 50$, or so,   it follows from (\ref{xx8}) that $h_*$ is well approximated by 
\bea
h_* = \sqrt{ 2 n N_* }  \, ,
\label{xx11}  
\eea
provided that $ n << 4 N_*$.  This covers a large class of models ranging from $n=2$ up to $ n = 10$ or even larger.  
Using $h_*$, given above, $A_s$ of Eq. (\ref{xx5}) is written, in terms of $N_*$,  as 
\bea
A_s \simeq  
\, \dfrac{1}{12 \pi^2}  \, \dfrac{ \,  \lambda}{ n^3} \,  \, { ( 2 n N_*) } ^{(n/2+1) }   \, .
\label{yy5}
\eea
For $ A_s \simeq 2.1 \times 10^{-9}$ we have that  the coupling $\lambda$ is constrained  to be 
\bea
\lambda \simeq  ( 4.97 \times 10^{-7} ) \; \;  \dfrac{k^2}{{(4k)}^{k}} \; \dfrac{1}{N_*^{k+1}} 
\quad \text{where} \; \;  n = 2 k  \, .
\label{xx12}
\eea  
Note that this is inverse proportional to $ N_*^{k+1} $.  For $ N_* = 55$ and 
for $n=2$, that is $ V \sim h^2$, this yields $\lambda \simeq  4.11 \times 10^{-11} $ while for $k=2$,  that is $ V \sim h^4$ we get
$\lambda \simeq  1.87 \times 10^{-13} $. Note that for the $n=4$ case Eq. (\ref{yy5}) coincides with that given in 
\cite{Tenkanen:2019jiq}. In that work a small value of the quartic coupling,  $\lambda \simeq  2.0 \times 10^{-13} $,  is also quoted,  quite close to ours given before. 

As for the parameter $a$  a lower bound can be established from the bound  (\ref{boundr}), that is  from the observational bound on the tensor to scalar ratio $r$. Using the analytic form of the potential one finds
 \bea
 \dfrac{1}{4 a} \, \dfrac{c h_*^n}{1 + c h_*^n} \, < \, 2.0  \times 10^{-9}    \, .
 \label{xx13}
 \eea
 Replacing $c$ in terms of $a$ from (\ref{xx2}), and using the value of $h_*$ given before in (\ref{xx11}), we have  from (\ref{xx13}), after some trivial manipulations, 
 \bea
 a \gtrsim 10^8 \left( 1.25 - \dfrac{   N_*}{50 \, n} \right)  \, .
 \label{xx14}
 \eea
For instance,  for the quartic potential $ V \sim h^4$ and for $ N_* = 55 $ this yields $ a \geq 0.97 \times 10^8$, resulting to an inflationary scale, lower than $ \sim 10^{-5}$, or so. 
Note that  (\ref{xx14}) is the lowest allowed value of $a$ consistent with the power spectrum and the bound on the potential imposed  by  the tensor to scalar ratio   $r <  0.063$.

The constraints on the parameters given before  arise from the amplitude of the power spectrum, in combination with the bound on $r$,  and set the range where acceptable values for $A_s$ can be obtained,  However the primordial tilt  $n_s$ puts additional  constraints and in order to have an estimate of it we use the approximate formula given by
\bea
n_s \simeq 1 - 6 \epsilon_V + 2 \eta_V  \, .  
\label{xx15}
\eea
The parameter $ \epsilon_V $ is given by (\ref{xx9}) and for $ \eta_V$ we employ (\ref{epv}) from which it follows that 
\bea
\eta_V =  \dfrac{n \, \left( n - 1 -( n/2+1) c h^n \right)  }{ h^2 ( 1 + c h^n)}  \, .
\label{xx16f}
\eea
From this, and $\epsilon_V$ of Eq. (\ref{xx9}),  we get, on account of (\ref{xx15}),   
\bea
n_s = 1 - \dfrac{n^2 + 2 n}{h^2}   \, .
\label{zz15}
\eea
Replacing $h$ by $h_* = \sqrt{ 2 n N_*} $  a rather simple expression for $n_s$ is obtained  given by
\bea
n_s =1 - \dfrac{n +2 }{2 \, N_*}   \, .  
\label{xx16}
\eea
Note that for $n=2$ and $N_* = 55$ the above formula yields $ n_s = 0.9636$ which is well within observational limits but for $n=4$ a rather large value of $N_*$ is needed to have an acceptable value for $n_s$.  
In fact $ N_* > 76 $ is required to have $ n_s =0.9607$, the lowest allowed if the data $ n_s = 0.9649 \pm 0.0042$  is used.  This is a rather large value for the number of e-folds $N_*$. The situation becomes even worse for models with  $ n > 4$. 

It is important, in the framework of this qualitative discussion, to have estimates of the variations of the quantities of interest with varying the parameters of the models at hand. Starting from the power spectrum amplitude, given by (\ref{yy5}), it is a trivial task to see that such  a  variation yields 
\bea
\delta A_s \, = \, \left(\dfrac{ \delta \lambda}{\lambda} +  \dfrac{n + 2}{2}  \,  \dfrac{\delta N_* }{ N_* } \right)\, A_s \, .
\label{varas}
\eea  
The first term stems from the explicit dependence of $A_s$ on $\lambda$.  
For fixed $\lambda$, and varying  only $a$, it is only  the second term that contributes.  In this case it can be seen that,  if the  variation of e-folds is of order unity or so,  it may produce a substantial change in $A_s$,  of the same order of  the errors accompanying the measurements of $A_s$.  Due to the prefactor $( n + 2 )/2$,  on the right hand side of (\ref{varas}), this is larger for models with larger $n$.  

On the other hand, the corresponding variation of the spectral index $n_s$ is found, from (\ref{xx16}),
\bea
\delta n_s \, = \, \dfrac{ n + 2}{ 2 \, N_*^2 } \, \delta N_*  \, .
\label{varns}
\eea 
This is proportional  to the relative change  $  \delta N_*  / N_* $ but is accompanied by an extra $N_*$ in the denominator. 
Due to that one expects that $n_s$ little varies with changing the number of e-folds. 

In order to estimate  the variations $  \delta N_* $, and hence   $\delta A_s, \delta n_s$,  with varying the couplings involved, namely $a$ and $\lambda$ for the models under investigation, one should start  from Eq (\ref{nfold2}), and for a fixed value of the reheating temperature,  vary $N_*$ with respect $a$, $\lambda$. The only  dependence on these is through the  logarithm of $3 H_*^2 $, which in the slow-roll regime equals to $V_{eff}(h_*)$, and the logarithm involving $\rho_{end}$.  We skip the details of such an analysis. We merely state that the final result  is of the form 
\bea
\delta N_* \, &=& \, \dfrac{\delta a}{a} \, f_a +
  \dfrac{\delta \lambda }{\lambda} \, f_\lambda  \, , 
    \label{delfold}
\eea
where the factors $ f_{a , \lambda}$ depend on the model under consideration.

A last comment regards the instantaneous reheating temperature $ T_{ins}$.  This is determined once we know $\rho_{end}$, see Eq. (\ref{tins}) and discussion following it.  With  $g_s^{* (reh)} = 106.75  $, which we have been using,  we have  
\bea
T_{ins} \, = \, 0.411 \, \,  \rho_{end}^{1/4}   \, ,
\label{numt}
\eea
which holds in general. However $\rho_{end} $ depends on the details of the model under consideration. 

The  end of inflation is determined by  $ \epsilon_1 =1$, equivalent to $ \rho + 3 \, p = 0$.  When $L$ - terms are absent this leads to $  \rho_{end} = \sigma \, V_{eff}$, where   $ \sigma = 1.5$.   However, in their presence  a more refined analysis is required.  Still in this case,   the equation 
$ \epsilon_1 =1$  can be trivially solved,  using (\ref{enpre}),  to give $ L \dot{h}^2 / K$ in terms of the potential $V_{eff}$, both evaluated at the end of inflation. That done, it is a fairly easy task to calculate $  \rho_{end} $, 
\bea
 \rho_{end}  = \sigma \, f(c_s) \,  V_{eff}( \bar{h}_{end} )  \quad .
 \label{ro222}
\eea
In this equation, and in order to avoid confusion, we have denoted by $ \bar{h}_{end}  $  the value of the field at the end of inflation. This depends implicitly on $L$ and can be extracted only numerically. The function $f(c_s)$ depends on the speed of sound squared, $c_s^2$, evaluated at the end of inflation,  and is given by $ f(c_s) = 8 c_s^2 / ( 9 c_s^2 -1 )  $.  Due to the fact that $1 / 3 \leq c_s^2 \leq 1 $, as one can see from (\ref{sound2}), it is bounded by $ 1 \leq f(c_s)  \leq 4/3 $.
Had we used the value $h_{end}$, as this is calculated from $ \epsilon_V = 1$, Eq. (\ref{ro222}) would have been expressed as, 
\bea
 \rho_{end}  = \sigma \,  f_\rho \,  V_{eff}( {h}_{end} )  \quad , \quad  
 f_\rho \equiv  f(c_s) \;  \dfrac{V_{eff}( \bar{h}_{end} ) }{ V_{eff}( {h}_{end} )} \quad . 
  \label{ro333}
\eea
This states that the approximate result for  $  \rho_{end}  $ , as given by  $ \sigma \,  V_{eff}( {h}_{end} )    $, is actually dressed by the factor $f_\rho$. In this factor the function   $  f( c_s)$  plays no important role, due to the bounds quoted before,  but  the ratio 
$  {V_{eff}( \bar{h}_{end} ) } / { V_{eff}( {h}_{end} )} $  may deviate substantially from unity. This ratio  can be  calculated only numerically. However in all models considered, and in a wide range of the parameters,  we have found that it lies between 
$ \simeq 0.5$ and $ 0.65$. 
 Taking, also,  into account the bounds on $f(c_s)$,   the factor $f_\rho$   lies in the range $ 0.5 - 0.85 $.
Due to that the  result for $  \rho_{end}  $ derived numerically is reduced  from  the approximate result, $  \rho_{end}  =   \sigma \,  V_{eff}( {h}_{end} )  $, by the  factor $f_\rho$.  For the instantaneous temperature, things are much better since this depends on  the quartic root of 
$  \rho_{end}  $. Therefore  the numerically derived $T_{ins}$ is smaller by a factor in the range $ 0.84 - 0.95  $. Thus the approximate result    $  \rho_{end}  =   \sigma \,  V_{eff}( {h}_{end} )  $, which we can derive analytically,  yields  $T_{ins}$ that are not far from the actual values.

For the models studied in this work, dubbed as  Model I, II as well as the Higgs Model,  using the equation that relates  $ L \dot{h}^2 / K$ to $ V_{eff}$ at the end of inflation, Eq.   (\ref{ro222}) can be further simplified given by, 
\bea
\rho_{end}  = \dfrac{\sigma }{ 2 a}  \, (1 -  c_s^2) \quad ,
  \label{ro444}  
\eea
where  the speed of sound  is meant at the end of inflation. Simple as might be, the value of $c_s^2$ implicitly depends on the parameters of the model under investigation and it can be calculated only numerically.
This is a rather elegant relation, which shows that  only  $c_s$  at the end of inflation is needed,  in order to derive $  \rho_{end}$.  It also shows the prominent  role of the  $ L \dot{h}^2 / K$,  at the end of inflation,  through which $c_s^2$ is determined, see Eq (\ref{sound2}).  Using (\ref{ro444}),  the instantaneous reheating temperature can be cast in the form
\bea
T_{ins} = 0.382 \, a^{-1/4} \, (1 - c_s^2 )^{1/4} \quad .
\label{tinscs}
\eea 
From this, using the fact that $ c_s^2 \geq 1 / 3$, an absolute upper bound can be derived,  $ T_{ins}  \leq 0.345 \;  a^{-1/4} $, valid for any model considered in this work.  From  (\ref{ro444}), one may be mislead to the conclusion that   for large $a$ the instantaneous temperature drops as  
$ a^{-1/4}$.  In fact it may drop much faster,  due to the implicit dependence of $c_s^2$ on the parameters involved. 

Following the previous discussion,  we may derive analytic expressions for the  instantaneous temperature, which are good estimates,   
using the approximate expression $  \rho_{end}  =   \sigma \,  V_{eff}( {h}_{end} )  $.
For the  class of  models studied  in this subsection, the latter  follows from the solution of  (\ref{xx10}) which depends only on the combination $c$. Using the analytic form of the potential it is found, in a straightforward manner, that
\bea
 \rho_{end}   \, =  \, \dfrac{\sigma}{4 a} \left( 1 - \dfrac{2}{n^2} \, h_{end}^2  \right)   \,  .
\label{rendit}
\eea
$   \rho_{end} $, and hence  $T_{ins}$, cannot be quantified further, at this stage, since for this purpose  the value of $h_{end}$ is needed.  
In the following we shall analyze in detail the predictions for this class of models.  As already remarked  at the beginning of this section, when presenting our final results, for each model considered,  we shall solve the pertinent equations numerically using accurate  formulas, without approximations, and take into account  the temperature dependence of the number of e-folds.

  \vspace*{2mm}
  
 \noindent
 {\bf{Model I :}}
 
We first consider the   model ( Model I ) in which the functions $g, M^2$ and $V$  are as given by (\ref{xx1}) with $n=2$, that is the potential $V$ is quadratic in the field $h$, 
\bea
V(h) \, = \, \dfrac{m^2}{2} \, h^2  \, .  
\label{moda}
\eea
For this case we prefer to use $m^2$, instead of $\lambda$,  since it carries dimension of $mass^2$ when $ m_P$ is 
 reinstated.  This models has been discussed in   \cite{Antoniadis:2018ywb} and belongs to the class of the cosmological attractors \cite{Carrasco:2015rva}, which is clearly  seen if one uses the canonically normalized field $\phi$, see (\ref{canon}). However no need to do that as we prefer to   work directly with the  non canonical  field $h$ instead.  Following the previous findings we define, see Eq. (\ref{xx2}),   the constant $c$ as the combination 
\bea
c \, = \,  2\, m^2 \, a   \, .
\label{cccc}
\eea
The value of $h_*$ in this case is given by, using (\ref{xx8}),   
\bea
h_* \, \simeq \, 2 \, \sqrt{N_*}  \, .
\eea
Then from (\ref{xx12}), which arose from  the power spectrum amplitude, we get, for values $N_* = 50 - 60$, 
\bea
m \simeq ( 6.5 \pm 0.5  ) \times 10^{-6} \quad \text{or} \quad \dfrac{c}{a} \simeq ( 8.5 \pm 1.5 )  \times 10^{-11}  \, .
\label{range}
\eea
The lowest ( largest ) limits correspond to $ N_* = 60 \, ( \, N_* = 50 \,)$. Therefore by using reasonable approximations we derived rather tight limits for the parameter $m$.  Recall that $m^2 \equiv \lambda$ and therefore $ \lambda$ is of the order of $ 10^{-11}$. 
From the bound (\ref{xx14}),  which actually arises from the tensor to scalar ratio bound $ r < 0.063$, we get, for $N_* = 50 - 60$, 
a lower bound which is estimated to be in the range, 
 \bea
 a \geq ( 0.65 - 0.75 )\times 10^8    \, .
 \label{abound}
 \eea
In this the lowest value corresponds to  $N_* =  60$ and the largest to $N_* =  50$. Therefore the parameter $a$ cannot be chosen at will. It should be $ \sim 10^8$ or larger.  In the following, due to (\ref{abound}), we shall take the largest value  as the  bound 
set on $a$, i.e. $ a \gtrsim 0.75 \times 10^8 $, which is valid for any $N_*$ in the range of interest.  

Concerning the instantaneous reheating temperature, in this case,  by  solving  analytically (\ref{xx10}), and replacing $ h_{end}$ into 
 (\ref{rendit}), we get
 \bea
 \rho_{end} \, = \, \dfrac{\sigma}{4 a} \, \left(   1 - \dfrac{\sqrt{1+ 8 c} - 1}{4 c}   \right)   \, .
  \label{solrho}
 \eea
 We can consider two separate  regimes, the small $c$ and the large $c$,  for which $\rho_{end}$, and consequently $T_{ins}$, have different dependencies on the parameters involved, as we shall see. Since from (\ref{range}) the ratio $c/a$ should be  of the order of 
 $ \sim 10^{-10}$,  small $c$ values are obtained when $ a < 10^{10}$. On the other hand large $c$ values are obtained when 
 $ a > 10^{10}$. 
 
   For small $c$ - values one  can expand (\ref{solrho}),   and using the fact that $\sigma = 1.5$,  the instantaneous temperature, as   given by (\ref{numt}),  receives the form, 
 \bea
 \rho_{end}  \simeq  \sigma \, \dfrac{c}{2 a} = \sigma  \,  m^2
 \quad \rightarrow \quad
 T_{ins} = 0.455 \times \sqrt{m}   \, .
 \eea
 This,  on account of (\ref{range}), results to a temperature which is 
  $T_{ins} \simeq 2.82  \times 10^{15} \, GeV$, for $ m =  6.5 \times 10^{-6}$.  As we shall see this estimate is not far from the one we get in our numerical treatment. What is more important, perhaps, is the fact that in the regime of small $c$ the power spectrum amplitude, which forces $m$ to be within the limits suggested by (\ref{range}),  also determines the maximum reheating temperature. 
 
 In the case of large $c$, $\rho_{end}$, and hence $ T_{ins}$, have a completely different behavior. In fact in this case,
  from (\ref{solrho}) and (\ref{numt}), we get 
 \bea
  \rho_{end}  \simeq \dfrac{\sigma}{4 a} 
 \quad \rightarrow \quad
 T_{ins} = 0.321  \times a^{-1/4}  \, ,
 \label{highc}  
 \eea 
  that is,  $T_{ins}$ is controlled by the value of $a$, being proportional to $ a^{-1/4}$, and therefore  it  decreases with increasing $a$. 
  Due to the fact  $ a > 10^{10}$, for being within the large $c$ regime,  $T_{ins}$ turns out to have values lower than in the small $c$ case. For instance for $ a= 10^{12}$ we get from (\ref{highc}) a temperature 
 $ T_{ins} \simeq 0.783 \times 10^{15} \, GeV$ and certainly even lower temperatures for larger values of $a$.  
 Therefore for having the largest possible value for the instantaneous temperature, of the order  $  \simeq 10^{15} \, GeV$,  we had better used values  $ a < 10^ {10}$ so that we are within the small $c$ regime. 
      
  As already advertised,  the cosmological predictions of all models considered are based on a numerical analysis in which no approximation is made.    
For the model at hand, predictions for  three different  inputs are presented,  named A, B and C, in the following.  These correspond  to values of the parameters $ a$  and $c$   given by $ ( a, c )  = ( 0.75 \times 10^8, \,  0.006 \,  )   \,  , \, ( 2 \times 10^8 , \,  0.016 \, )$ and $\, ( 2 \times 10^9, \, 0.16  \,)  $.  These have not been randomly chosen. In fact, for  the case A the parameter $a$ touches  its lower bound, discussed before, and $c$ has been taken so that $m $ falls well  within the range suggested by (\ref{range}).  In fact we choose  $m \simeq 6.32 \times 10^{-6}$.  The reasoning behind this particular choice for $m$ will be discussed later.  

For the other cases  larger  values of  $a$'s were chosen  but  the  values of $c$ are tuned so that in  all cases we have  the same value of $m$, i.e.    $m \simeq 6.32 \times 10^{-6}$.  In this way we can check  how predictions vary with changing the parameter  $a$ since we have kept a  fixed  $m$. 
Note that from all cases presented,  the case A has  the lowest allowed  value of $a$ and therefore  the Planck upper bound  on the tensor to scalar ratio parameter $r$ is almost saturated. The other cases $B, C$ are expected to yield smaller values for $r$.

\begin{figure}[t]
\centering
  \centering
  \includegraphics[width=1.07\linewidth]{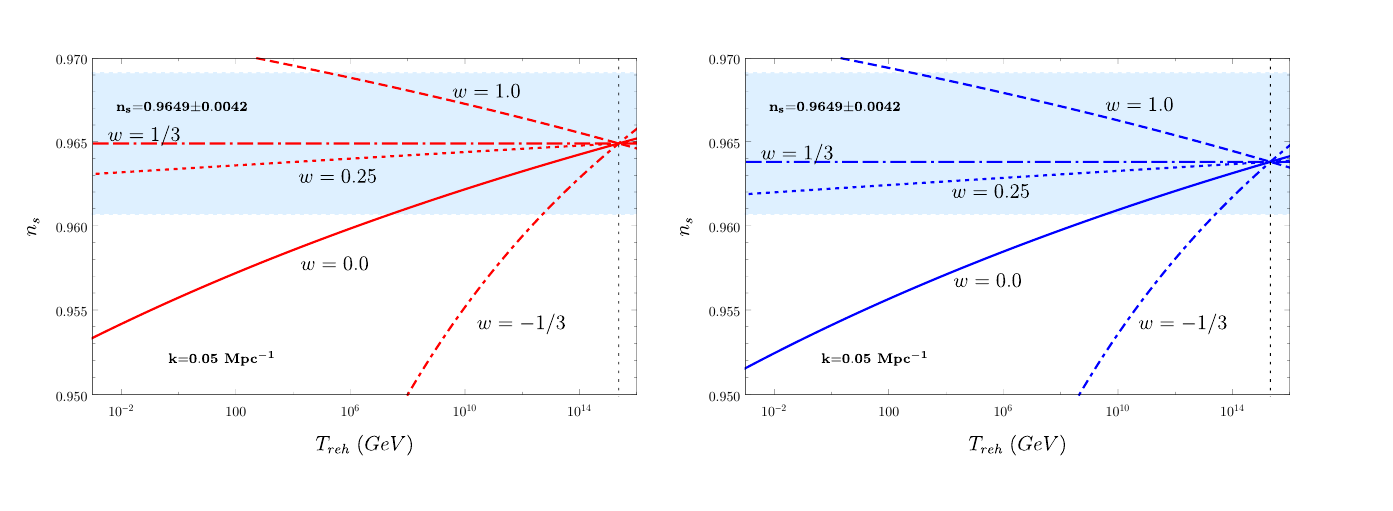}    
   \includegraphics[width=1.07\linewidth]{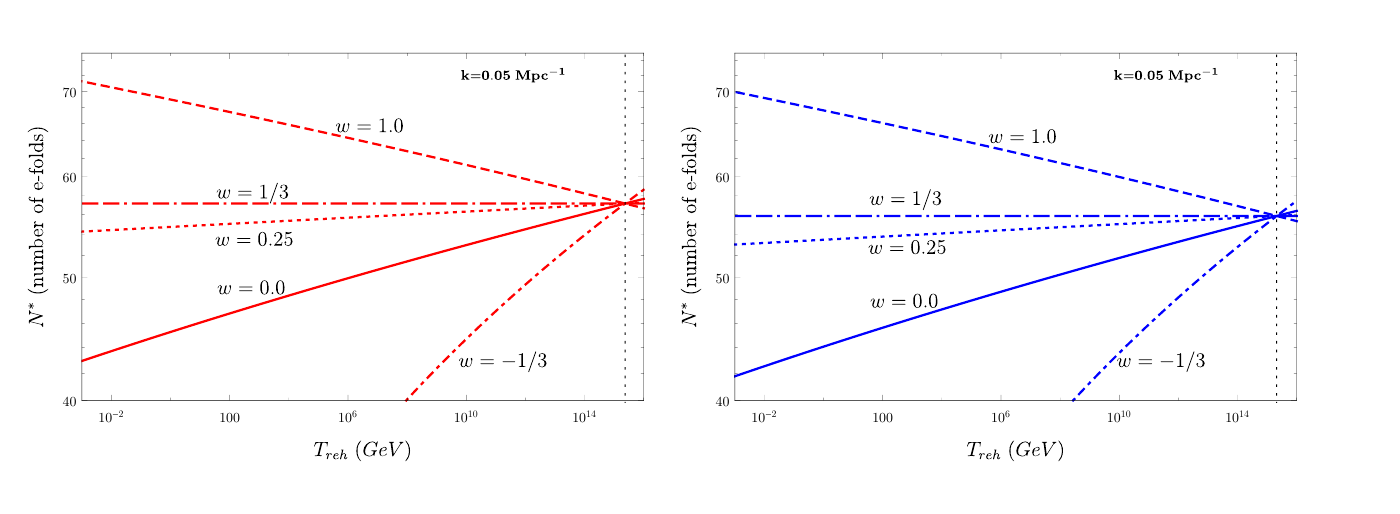}      
\caption{
The spectral index $n_s$ (top)  and the number of e-folds $N_*$  (bottom),  versus the reheating temperature $T_{reh}$,  in 
$GeV$, for a  scale $ k^* = 0.05 \, Mpc^{-1}$,  and for different values of the equation of state parameter, for the cases A ( left ) and C ( right)  of  Model I, discussed in the text. The shaded region marks the allowed values for the spectral index  $ n_s = 0.9649 \pm 0.0042 \; $ while the vertical dotted line the instantaneous reheating temperature.
}
\label{fig9}
\end{figure}

In Figure \ref{fig9}, at the top,  we display,  for the cases A (left) and C (right), the spectral index $n_s$ versus the reheating temperature $T_{reh}$,  for various  values of the equation of state parameter ranging from $w = - 1/3 $ to $w=1.0$. 
The shaded region marks the range  
$ n_s = 0.9649 \pm 0.0042 \; $ allowed by observations.  


\noindent
All lines intersect at a common temperature, the instantaneous reheating temperature $T_{ins}$,  marked by  thin vertical dashed lines, which  for case  A equals to  
$ T_{ins} =  2.337 \times 10^{15} \, GeV$, and for case C to  $ T_{ins} =  2.099 \times 10^{15} \, GeV$.   Values of reheating temperatures beyond that point, although displayed, are not allowed. The data shown correspond to a pivot scale $k^* = 0.05 \, Mpc^{-1}$.   
Note that $n_s$ data by themselves  do not impose any restriction on the reheating temperature,  as long as the equation of state parameter is in the range from  $ 0.25  $ to values slightly lower than $  \simeq 1.0$. 
For these values of $w$ any temperature is allowed. For $ w <  0.25$ a lower reheating temperature is imposed which is larger for smaller values of $w$.  For instance for the canonical reheating scenario, $w = 0$, this is $ \simeq 10^{7} - 10^9 \, GeV$  while for $ w = -1/3$ this is $ \approx   10^{13} \, GeV$.  
At the bottom  of the same figure, and for the same set of inputs,  the corresponding numbers of e-folds, $N_*$, are shown, for the A (left) and the  C (right) cases respectively. 

 Note that  both $n_s$ and $N_*$, shown in the figures, are very similar for the two cases, A and C.   In particular both observables move slightly  downwards in going from A ( left ) to C ( right ), that is by increasing the value of $a$ from $ 0.75 \times 10^8$ to 
 $ 2.0  \times 10^9$, keeping the other  parameter  fixed.  In fact,  varying only the parameter $a$, keeping 
 $\lambda = m^2$ fixed,  which is the case for the inputs we are using,  we get from (\ref{delfold}),
\bea
\delta N_* \, &=& \, \dfrac{\delta a}{a} \, f_a  \, .
\eea
For our input values,   we find that the factor $ f_a$ is of order unity and negative. The result is that  by increasing the value of the parameter $a$, the relative change $  \delta N_* / N_*  $, is negative and therefore, due to (\ref{varns}),   $n_s$  decreases. This decrease is small, as we have already discussed, what is indeed imprinted on this  figure.

\begin{figure}[h]
\centering
  \centering
   \includegraphics[width=1.07\linewidth]{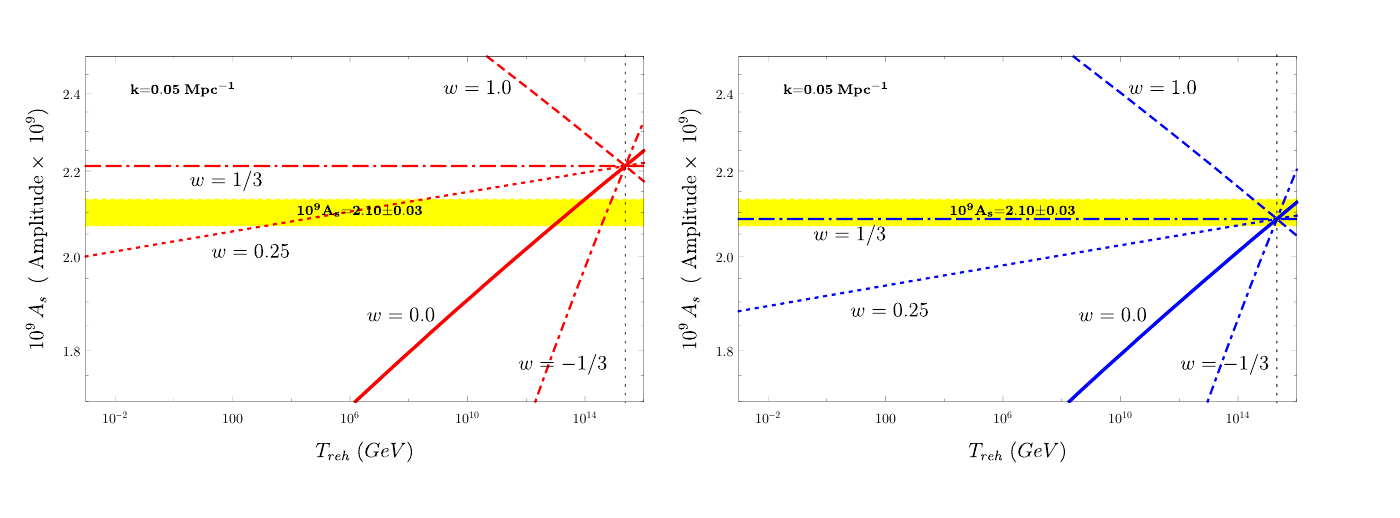}    
\caption{
The amplitude  $10^9 \, A_s$   versus the reheating temperature $T_{reh}$,  in 
$GeV$, for $ k^* = 0.05 \, Mpc^{-1}$,  for different values of the equation of state parameter. The shaded region marks the allowed values   $10^9 \, A_s = 2.10 \pm 0.03 \; $.  On the left the case A is shown and on the right the case C of Model I.
 The instantaneous temperatures, in each case, are marked by   thin dotted vertical lines. }
\label{fig10}
\end{figure}

The power spectrum amplitude imposes more stringent bounds on $ T_{reh}$ than  $n_s$, as shown in Figure \ref{fig10}.  In this figure  we plot the amplitude  $10^9 \times A_s$   versus the reheating temperature $T_{reh}$,  in 
$GeV$, for $ k^* = 0.05 \, Mpc^{-1}$,  and for different values of the equation of state parameter,  as in the previous figure. The shaded region marks the allowed range   $10^9 \times  A_s = 2.10 \pm 0.03 \; $.  On the left the case A is shown and on the right the case C. 
The lines are as in Figure \ref{fig9}.  One notices that for the A-case values $ w \gtrsim 1/3$ are totally excluded by $A_s$ data while for $ w \lesssim 0.25$ limits on the minimum and maximum allowed temperature are imposed. In this case the maximum temperature, for any allowed value of $w$,  can never reach the instantaneous temperature.  For the  C-case, right panel,  one sees, by comparing this figure with the $n_s$ plot, top and right pane of Figure \ref{fig9}, 
  that the bounds set  on the reheating temperature are more constrained. In particular, for values of  $w$, which deviate from  $ w \simeq 1 / 3$,  a lower reheating temperature is established, which is much higher than this imposed by $n_s$  data.

Comparing the two cases, A and C,  we observe that $A_s$  also decreases in accord with (\ref{varas}), and the fact that $ \lambda$, 
or equivalently $m$, is fixed and $  \delta N_* / N_*  $  is negative.  However the change in  $A_s$ is relatively large,   unlike $n_s$,    in the sense that its variation  reaches the order of magnitude of the observational error of $A_s$, as has been  previously discussed.  

It is worth mentioning that given a fixed value for the parameter $a$ there is  a fine-tuned  value of $m$, in the range suggested by (\ref{range}),  for which the case $w = 1/3$ falls within the allowed region by $A_s$ observations
{\footnote{
This requires that the case $ w = 1/3 $ is compatible with $N_*$ in the range $ \approx 50 - 60 $, which is always the case provided the parameter  $a$ does not take extremely high values. 
}}
. In this case the instantaneous reheating temperature is attained for any value of $w$ in the range $ -1/3 \leq w \leq 1$.  However in this case a  lowest temperature is determined, which  is  close to the instantaneous temperature,  for  any $w$ that deviates  from the value $1/3$. This includes the values $ 0.0 \lesssim w \lesssim 0.25 $ which are favored in some reheating scenarios.  This can be clearly seen, for instance,  in the case C where for $ a= 2.0 \times 10^9$ the value $m = 6.32 \times 10^{-6}$ forces the line $w = 1/3$ to be within $A_s$ limits,  as shown on the right panel of Figure   \ref{fig10}. 
Keeping $a$ fixed, any slight change in the value of the parameter $m$, which essentially controls $A_s$,  will move, downwards or upwards,  the line $w=1/3$,   off the allowed range, and in this case the instantaneous reheating scenario is no longer supported. At the same time, depending on the value of $m$, lower and higher limits of  reheating temperatures are imposed, different for each $w$.  However some values of $w$ are totally excluded.  For instance, by increasing $m$, the line 
$w = 1/3$ will be uplifted and move  above the upper observational limit  set on $A_s$.  In this case all values in the range 
$ 1/3 \leq w \leq 1$ are excluded. If, on the other hand $m$ is decreased, the line  $ w = 1/3$ will move below the lower  limit of $A_s$ and  values $ - 1/3  \leq w \leq 1/3$ are excluded.  Increasing, or decreasing, further  the value of $m$ will exclude all possible cases 
$ -1/3 \leq w \leq 1 $.  Therefore, there is a range of $m$ outside of which  agreement with $A_s$ data can not be obtained,  for any value of the equation of state parameter,  in the interval  $ -1/3 \leq w \leq 1 $.  This range  is actually very tight and falls within the suggested range given by (\ref{range}). Within this range there are fine-tuned values for which reheating can be instantaneous. 
Note that the sensitivity of the spectral index $n_s$ on the  value of $m$ is not that dramatic and $n_s$ data leave more ample space for  the observational requirements to be satisfied.  
Therefore, the conclusion is that given $a$, the value of  $m$ should lie in a very narrow range, in order  to comply with power spectrum data. Moreover if reheating is instantaneous it should be fine-tuned accordingly. 
This, as we shall see, holds for other popular models as well, notably the Higgs model that will be discussed later.

 Following the already outlined numerical procedure, in Table \ref{table1} we display sample outputs  of the model under consideration  for the choice of the parameters corresponding to the  inputs A, and C  for  a pivot scale $k^* = 0.05 \,  Mpc^{-1}$. 
The predicted cosmological observables $ n_s, r,  A_s $ are displayed,  for various values of the equation of state  parameter $w$, corresponding to the minimum (upper rows ) and maximum (lower rows)  allowed reheating temperatures $T_{reh}$,  when the limits  $ A_s \simeq ( 2.10 \pm  0.03 ) \times 10^{-9} $,  and  $  n_s = 0.9649 \pm 0.0042   $ are observed. The corresponding  predictions for the number of  e-folds $N_*$ ,  are also shown.  Blanc entries indicate that there are no values  compatible with   observational bounds put on $n_s, A_s$, for the specific value of $w$.  
Note that for the C - case,  the maximum reheating temperature reaches the instantaneous reheating  temperature, $ T_{ins} = 2.099 \times 10^{15} \, GeV $. At this temperature predictions are independent of $w $, due to the fact that $ T_{ins}$  marks the intersection of all $w$-lines. For the same case, the lower limits on  $T_{reh}$ are also shown. For the cases $ w = 0.0,\,  0.25$ and $ w = 1.0$,  these are not very far from the $T_{ins}$, as already discussed, in agreement with  Figure \ref{fig10}, right panel.
For the case A, on the other hand, the minimum and maximum reheating temperatures are both smaller than the corresponding ones of the C - case.  Note in particular the predictions for $ w =0.25$ for which the range of temperatures, allowed by all observations,  is 
$ T_{reh}   \simeq (1.5 \times 10^3  - 1.9 \times 10^8 ) \, GeV  $.  
\begin{table}  
\begin{center}
\begin{tabular}{|cccc|ccc|}  
  \multicolumn{7}{c} {  Model I \quad ( pivot scale $ k^* = 0.05 \, Mpc^{-1}$ ) }    \\      
  \hline \hline   
 & &   A - case  & &  & C - case &      \\ 
  \hline
   \multicolumn{1}{|c} {$w \,$- value } & $\quad w= 0.0 \quad $  & $ \quad w= 0.25 \quad $  &  $\quad w= 1.0 \quad $ & $ \quad w= 0.0 \quad $   
   & $\quad w= 0.25 \quad $ & $\quad w= 1.0 \quad $ \\
   \hline
    \multicolumn{1}{|c|} {$  \quad 10^{9} \,A_s  \quad$ } & 2.07  & 2.07 &   & 2.07  & 2.07 & 2.13 \\
    \hline
    \multicolumn{1}{|c|} {$n_s$ } & 0.9637  & 0.9637 &   & 0.9637  & 0.9637 & 0.9642 \\
    \hline
    \multicolumn{1}{|c|} {$r$ } & 0.0616  & 0.0616 &    & 0.0040  & 0.0040 & 0.0038 \\
    \hline
    \multicolumn{1}{|c|} {$N_*$ } & 55.25  & 55.25 &   & 55.65  & 55.65 & 56.43 \\
   \hline
    \multicolumn{1}{|c|} {$T_{reh}$ } &  $8.542  \times 10^{12}$  &    $ 1.547 \times 10^3 $  &   &   $ 1.138  \times 10^{15}$  &   $ 9.741 \times 10^{13} $ &   $ 3.667 \times 10^{14} $ \\
    \hline \hline
      \multicolumn{1}{|c|} {$ 10^{9} \,A_s $ } & 2.13  & 2.13 &     & 2.08  & 2.08 & 2.08 \\ 
    \hline
    \multicolumn{1}{|c|} {$n_s$ } & 0.9642  & 0.9642 &   & 0.9638 & 0.9638 & 0.9638 \\
    \hline
    \multicolumn{1}{|c|} {$r$ } & 0.0602  & 0.0602 &   & 0.0039  & 0.0039 & 0.0039 \\
    \hline
    \multicolumn{1}{|c|} {$N_*$ } & 56.03  & 56.03 &   &  55.85 &  55.85 & 55.85  \\
    \hline
    \multicolumn{1}{|c|} {$T_{reh}$ } &  $8.861  \times 10^{13}$  &    $ 1.855  \times 10^8 $  &    &    $ 2.099 \times 10^{15}$  &   $ 2.099 \times 10^{15}$ &    $ 2.099 \times 10^{15}$  \\
 \hline
 \end{tabular}
\caption{
Sample outputs for the Model I, for inputs corresponding to  cases A, C (see main text) , for the  cosmological observables 
$ n_s, r,  A_s $ and $N_*$,   for various values of the equation of state  parameter $w$. The  values shown for  the reheating temperature $T_{reh}$, in GeV, correspond to  the minimum (upper rows) and maximum (lower rows) allowed,  when the observational limits  for $ A_s \simeq ( 2.10 \pm  0.03 ) \times 10^{-9} $  and  $  n_s = 0.9649 \pm 0.0042   $ are imposed.  Blank entries indicate that there are no values  compatible with  the observational bounds put on $n_s$ and $A_s$, for the specific value of $w$.  
}
\label{table1} 
\end{center}
\end{table} 

\begin{figure}[t]
\centering
  \centering  
  \includegraphics[width=0.72\linewidth]{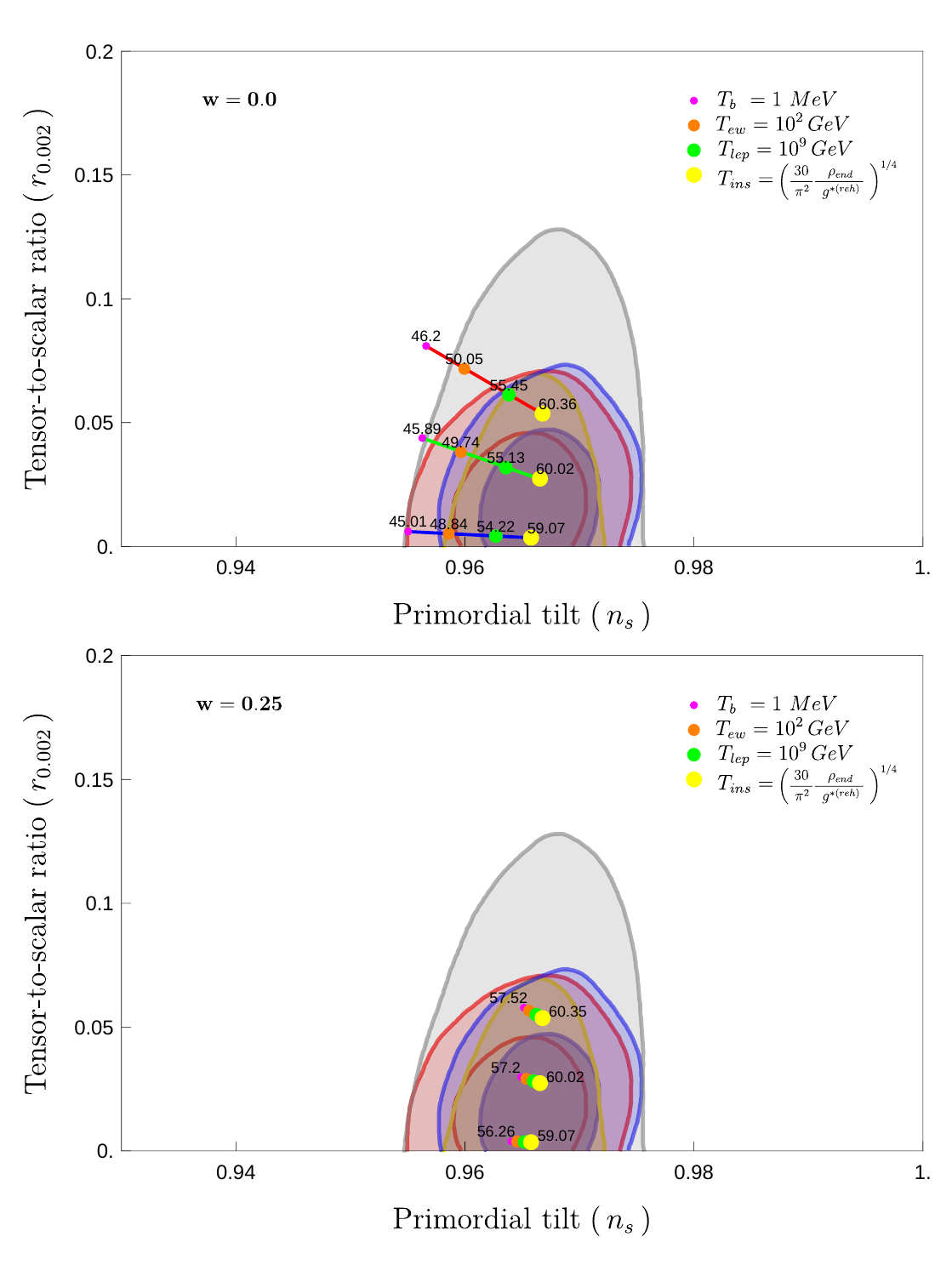} 
\caption{
The tensor-to-scalar ratio $r_{0.002}$ versus the spectral index $n_s$ for the Model I, for the data set A ( red-line), 
B ( green-line) and C ( blue-line) corresponding to different inputs of the parameters ( see main text ). A pivot scale $ k^* = 0.002 \, Mpc^{-1}$ is  used so that a direct comparison with the  corresponding Planck 2018 data is  possible. 
 The value of the equation of state parameter for the figure on top is $w = 0.0 $, while for the figure at the bottom  is $w = 0.25 $. The tiny circle (in magenta), the small (in orange) and  the large (in green)   correspond to reheating temperatures close to BBN, Electroweak and  Leptogenesis scenarios, while the largest  (in yellow) marks the instantaneous reheating temperature ( see main text ).   The numbers indicate the  e-folds,  in each case, when $ k^* = 0.002 \, Mpc^{-1}$.
}
\label{fig11}
\end{figure} 

In Figure \ref{fig11}  the tensor-to-scalar ratio $r_{0.002}$ versus the spectral index $n_s$ is plotted,  for the Model I, for the data set A (red-line),  B (green-line) and C (blue-line).  In drawing this figure a pivot scale $ k^* = 0.002 \, Mpc^{-1}$,  was used so that it can be directly  compared to  the corresponding Planck 2018 bounds  \cite{Akrami:2018odb,Aghanim:2018eyx},  which are also drawn.  
The tiny circle (in magenta), the small (in orange) and  the large (in green)   correspond to reheating temperatures close to BBN, Electroweak and  Leptogenesis scenarios,  given by $T_b = 1 \, MeV,  \, T{ew}=10^2 \, GeV$ and $ T_{lep} = 10^9 \, GeV$, respectively.  The largest circle (in  yellow) marks the instantaneous reheating temperature, see  Eq. (\ref{tins}),  for each case displayed.  The number close to  each circle indicates the corresponding number of e-folds left, at the pivot scale $ k^* = 0.002 \, Mpc^{-1}$. The value of the equation of state parameter for the figure on top is $w = 0.0 $,  while for the one at the bottom
 $w = 0.25 $. In the latter only the e-folds corresponding to $T_b$ and the instantaneous reheating are shown, to be clearly visible.
In both cases shown, $w=0$ and $w=0.25$ ,  the smallest values for the tensor-to-scalar ratio $r$ are obtained in the C-case, that is     for the largest values of the parameters $a, c$. Recall that  the  ratio $c/a$ has been kept fixed.  For smaller values of the parameters,  $r$ gets larger and saturates the Planck upper bound in the A - case,  corresponding  to  the lowest  allowed values of $a, c$, as we have already remarked. 

We point out that in drawing  Figure  \ref{fig11} the $A_s$ constraints have not be taken into account. Including these will shrink considerably  the allowed line segments, displayed on the figure, since $T_{reh}$ is further constrained by $A_s$ data.  For instance,  for the C - case,  which  is well within the region allowed by all observations,  yielding also the smallest value for $r$,  a large portion of the segment, with ends corresponding  to  temperatures  $ T_b$ and  the minimum allowed temperature, as this is   read from Table \ref{table1} for each $w$-case,  will be excised.  Only a tiny part of it,  close to the maximum reheating temperature  $T_{ins}$,  will be left.

\vspace*{2mm}
\noindent
 {\bf{Model II :}}
 
As a  second  model ( Model II ) worth studying, is the one  in which the functions $g, M^2$ are as in (\ref{xx1}), as in the Model I, but the potential is quartic in the scalar field involved, i.e. 
\bea
 V(h) \, = \,   \dfrac{\lambda }{4}  \, h^4 \,  ,
\label{moda2}
\eea
that is $n = 4 $. 
We have already remarked, based on the qualitative arguments presented earlier,  that this model, as well as all with $ n > 4$,  fails to satisfy the observations on the spectral  index unless one has a large number of e-folds, probably larger than $ N_* > 76$, or so.  However a more detailed study is required to get a firm conclusion which also takes into account the reheating temperature. 

Using the general results, given at the beginning of this section, when applied to  this model, we get , 
\bea
h_* \simeq \sqrt{  8 \, N_*}   \, .
\eea
Also on account of  (\ref{xx12}) the coupling $\lambda$  is 
\bea
\lambda \simeq 10^{-8} \; \dfrac{3.11}{N_*^3}   \, ,
\label{zzzw}
\eea
which for e-folds in the  range $N_* = 50 - 60$,   yields 
\bea
\lambda \simeq ( 1.45 - 2.50 ) \times 10^{-13}  \, , 
\label{coupl}
\eea
the lowest value corresponding to $ N_* = 60$.
Therefore the coupling $\lambda$ must be quite small in order to satisfy the constraints put by observations. 
As for the parameter $a$, which sets the inflation scale, employing  (\ref{xx14}), we have   a lower bound  given by
  \bea
 a \gtrsim ( 0.95 - 1.00)  \times 10^8  \, ,
 \label{abound2} 
 \eea
 not much different from the bounds given in  (\ref{abound}).      
 
 For $T_{ins}$ we have to calculate $h_{end}$, as in the $n=2$ case, and use   (\ref{rendit}) adapted to the case $n=4$. Although analytic solution for  $h_{end}$ is feasible, through Eq. (\ref{xx10}), we will not present it. Instead we shall discuss its behavior for small and large $c$-values.  
 For small $c$, omitting  $  {\cal{O}} ( c^2) $  terms,  we find $h_{end}^2 \simeq 8 ( 1 - 64 \, c ) $. Then  
   from (\ref{rendit}) the leading contribution is, 
  \bea
 \rho_{end}  \simeq  \sigma \, \dfrac{16 \, c}{ a} = \,  16 \,  \sigma \,  \lambda
 \quad \rightarrow \quad
 T_{ins} = 0.909  \times \lambda^{1/4}  \, .  
 \label{rho41}
 \eea

\begin{figure}[t]   
\centering
  \centering
  \includegraphics[width=1.07\linewidth]{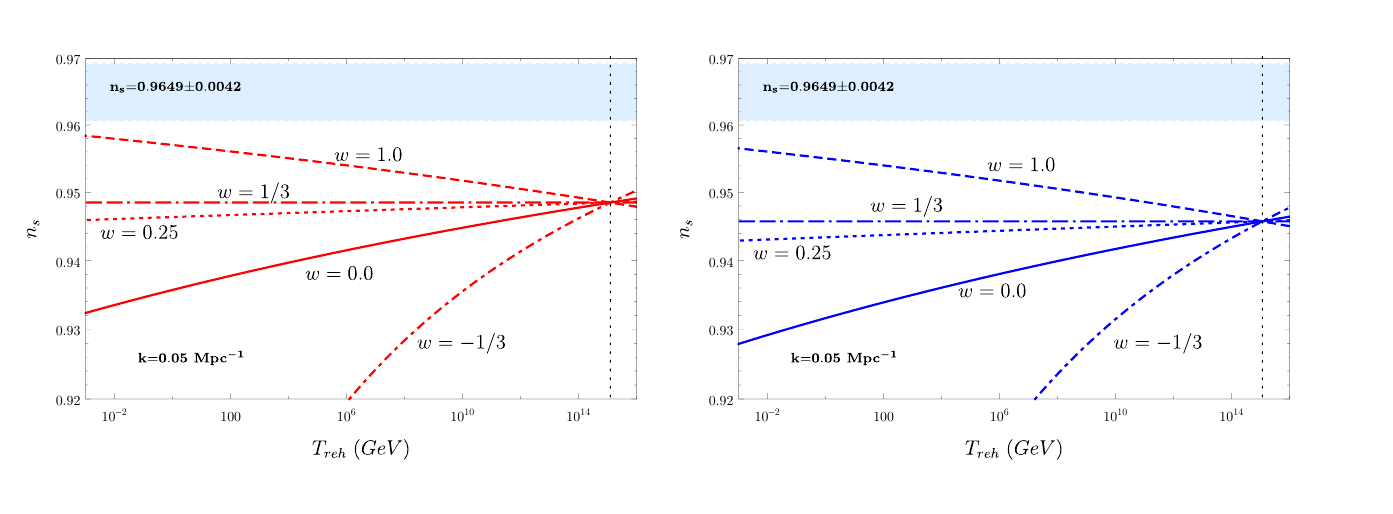}     
\caption{
The spectral index $n_s$,  versus the reheating temperature $T_{reh}$,  in 
$GeV$, for a  scale $ k^* = 0.05 \, Mpc^{-1}$,  and for various values of the equation of state parameter, for the cases A (left) and B (right)  of  Model II  discussed in the text. The shaded regions marks the allowed values for the spectral index  $ n_s = 0.9649 \pm 0.0042 \; $ and  the vertical dotted lines the instantaneous reheating temperatures. 
 }
\label{fig121}
\end{figure}

\begin{figure}[t]    
\centering
  \centering
  \includegraphics[width=1.07\linewidth]{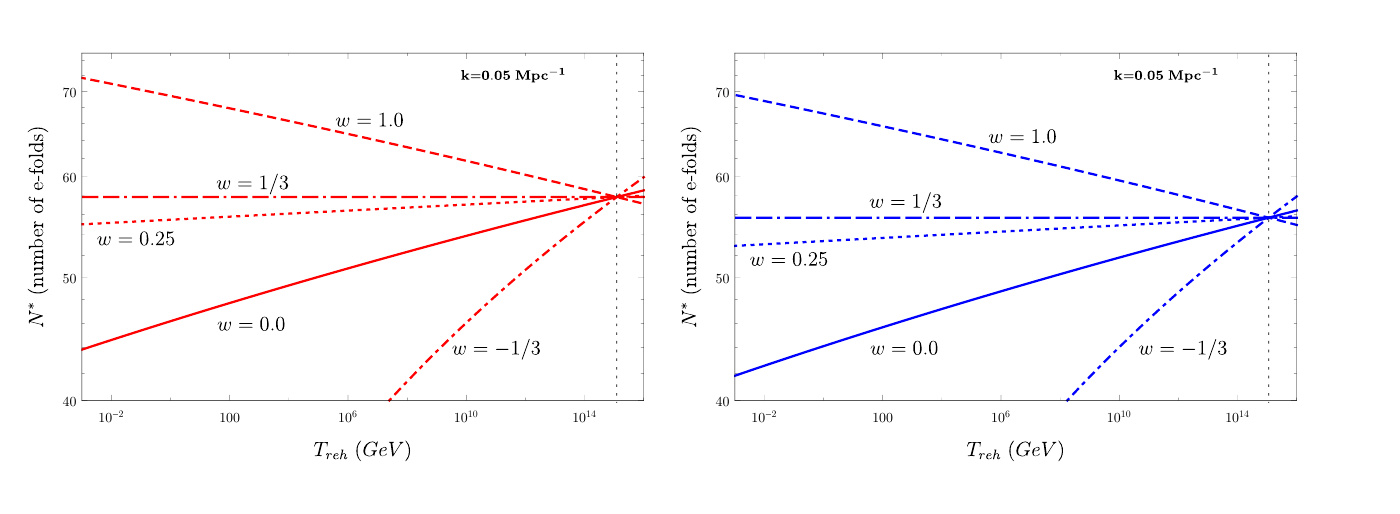}     
\caption{
As in Figure \ref{fig121} for the number of e-folds   $N_*$ . 
   }
   \label{fig122}
\end{figure}
 
 With $\lambda = 2 \times 10^{-13}$, the central value  in the range   (\ref{coupl}),  this yields an instantaneous reheating  temperature around 
  $T_{ins} \simeq 1.48  \times 10^{15} \, GeV$.   As in the previously studied  model, $ n = 2$ case, the power spectrum   determines the maximum  reheating temperature, in the regime of small $c$.
 In the case of large $c$, $h_{end}^2$ behaves as $c^{-1/3}$, and hence little  contributes  to  Eq. (\ref{rendit}). Then keeping only the leading  term in $ \rho_{end}$, we get the same result (\ref{highc}),  as in the previous model,  and $T_{ins}$ is again  proportional to $ a^{-1/4}$.

  For this model we shall also  present sample outputs  of our numerical treatment, considering  
  a fixed value  $\lambda = 2.0  \times 10^{-13}$, in the middle of the range suggested by (\ref{coupl}),  and values of $a$ in the range $ a = 10^8 - 10^{10} \, $, respecting therefore the bound (\ref{abound2}).  The value $a = 10^8$ corresponds to the lowest allowed value, and   for future reference  we name it A - case, while  $ 10^{10} $ is arbitrarily taken to be two orders of magnitude larger, which we name  B - case.  Although, in principle,  one can  consider larger $a$ - values there is no need to do this for reasons that will be shortly  explained.  
  
   On the left panel of  Figure \ref{fig121} the predictions for the spectral index $n_s$, for  the  cases A (left) and B (right), are shown versus the reheating temperature for various values of the equation of state parameter $w$. Note that there is no much difference between the two cases, although the parameter $a$ differs by two orders of magnitude. The explanation is  the same as that discussed for Model I.  Notice that on the right  the lines have been moved  imperceptibly lower. That is,  the tendency is to get lower $n_s$ values as the parameter $a$ increases. Concerning the  instantaneous reheating temperature, for the values taken for $a , \lambda$,  for the A - case it is $T_{ins} = 1.223 \times 10^{15} \, GeV$,  while for for B - case this is  $T_{ins} = 1.129 \times 10^{15} \, GeV$. These are  marked by  vertical thin dotted lines,  as in previous figures. 
  As already remarked for this model agreement with $n_s$  observational data is hard to achieve.  In both cases it is clearly seen from this figure, that  only for very small reheating temperatures, and only for   $w=1$,  values of $n_s$  that are marginally acceptable  can be obtained.  In this case the number of e-folds is large $ N_* > 70$, as is shown in Figure \ref{fig122} where the number of e-folds is displayed. This agrees with the general arguments given before,  see discussion following Eq. (\ref{xx16}).  
  We have not considered larger values of $a$,  since as explained,  they would  predict lower $n_s$, resulting to larger deviations  from the data.    
  
    Although agreement with $n_s$ data cannot be obtained, in this model, for reasons of completeness we shall give a brief account 
  for the predictions for the power spectrum amplitude. 
  Agreement with $A_s$ data requires values of $w$ that are smaller than $ 0.25$, for the A - case, while for the B - case  the value  
  $ w =0.25$ is marginally accepted. Values lower than $w \simeq 0.25$ are allowed.   Whatever the case, such values for the equation of state parameter lead, according to Figure \ref{fig121},  to even smaller   values of $n_s$,  less than  $\simeq 0.945$ or so,  and hence unacceptable.   

 These  results are in agreement with \cite{Tenkanen:2019jiq} where small $n_s$ are also obtained, and indicate that the quartic potential $ ( n = 4 )$ is in tension  with cosmological data. 
 Models with   $ n > 4 $ yield predictions that according to our general arguments are, also,  hard to reconcile with the data. 
 
  Therefore the conclusion   is that,  from the class of the models whose initial potential is  of the monomial form 
  $ V \sim h^n$, and with constant values for the coefficients of $\cal{R}$ and ${\cal{R}}^2$ terms in the Palatini gravity, only the case $n=2$, which belongs to the  class of the cosmological attractors \cite{Carrasco:2015rva},  can lead to successful inflation, if all observational constraints are taken into account.

\subsection{ Non-minimally coupled models }     

A non-minimal coupling arises if in the previously studied models the constants $g$ and/or  $M^2$are promoted to be field dependent.  A particularly interesting case is the model in which
\bea 
g(h) = 1 + \xi h^2 \; ,  \; M^2(h) = \dfrac{1}{3 a} \quad , \quad V(h) \, = \, \dfrac{\lambda }{4} \, h^4   \, .
\label{higg1}
\eea
This belongs to the class of models (\ref{xx1}), with quartic potential,  however  the scalar $h$ is non-minimally coupled to the scalar curvature ${\cal{R}}$,  in the Palatini framework, since  $g$ is field dependent, in the particular way shown above. 
This model arises actually  from the Higgs coupling to Palatini gravity
\bea
\dfrac{m_P^2 + 2 \, \xi H^\dagger H}{2}  \, {\cal{R}} + \dfrac{a}{ 4 } \, {\cal{R}}^2 + | D H |^2 - \lambda \left( |H|^2 - \dfrac{u^2}{2} \right)^2   \, ,
\label{higg2}    
\eea
where $u \simeq 246 \, GeV$ is the Electroweak scale.  In Planck units, $ m_P=1$,  this  is very small $ u \sim 10^{-16} $ and plays no significant  role in inflation. 
Setting therefore $u=0$ and working in the unitary gauge, $ H^{\dagger}  = ( 0, h / \sqrt{2} ) $,   (\ref{higg2}) is actually  the model described by $g, M^2$ and quartic potential  as  given in (\ref{higg1}). 

The Higgs coupling to gravity and its role as the inflaton,  in the metric formulation,  has been proposed in 
\cite{Bezrukov:2007ep,Bezrukov:2008ej}  and it has been  widely studied  since then
 \cite{ Barbon:2009ya,Barvinsky:2009fy,Barvinsky:2009ii,Germani:2010gm,Germani:2010ux,Lerner:2010mq,Bezrukov:2010jz,Kamada:2010qe,Kamada:2012se,Bezrukov:2013fka, Allison:2013uaa,Bezrukov:2014bra,Hamada:2014wna, Salvio:2015kka,Calmet:2016fsr,Jinno:2017lun, Bezrukov:2017dyv, Antoniadis:2018ywb,Antoniadis:2018yfq,Enckell:2018kkc,
 He:2018gyf,Gundhi:2018wyz,Rubio:2018ogq,He:2018mgb,Steinwachs:2019hdr,
 Rubio:2019ypq,Takahashi:2018brt,Tenkanen:2019jiq}  
 both in the context of the metric formulation and  in  Palatini formulation. The importance of the $  {\cal{R}}^2 $ term in (\ref{higg2}) has been discussed in  \cite{Antoniadis:2018ywb,Antoniadis:2018yfq,Takahashi:2018brt,Tenkanen:2019jiq}.  In this work we shall show that the quartic coupling $ \lambda$, as in the minimally coupled quartic model  studied previously, corresponding to $ \xi = 0$, is constrained considerably by cosmological data, especially the power spectrum amplitude $A_s$. 
 This limits the available options especially when the reheating of Universe after inflation is taken into account.  

The functions $K, L$ and $V_{eff}$ in this model are given below, in the limit $ u = 0$,  
\bea
K(h) \, = \,   \dfrac{ 1 + \xi \, h^2 }{(1 +  \xi \, h^2  )^2+ c h^4  }   \; , \;  L(h) \, = \, \dfrac{ a }{(1 +  \xi \, h^2  )^2+ c h^4  }  \, ,
\label{zxx3}
\eea
while the potential $V_{eff}$ receives the form
\bea
V_{eff}(h) \, = \, \dfrac{1 }{4 \, a} \, \dfrac{c h ^4 }{ ( 1 +  \xi \, h^2  )^2+ c h^4 }   \, .
\label{zxx4}
\eea  
As in the simple quartic potential the parameter $c$ is the combination $ c = a \lambda$. Notice however that a non-trivial 
$\xi$-dependence exists and therefore the Higgs model differs from the simple quartic model studied previously. Evidently when 
$\xi =0$ the functions (\ref{zxx3}), (\ref{zxx4}) smoothly go into (\ref{xx3}), (\ref{xx4}) 

For large values of $h$   the potential  (\ref{zxx4})  approaches a plateau $\simeq 1 / 4 ( a  + \xi^2 / \lambda)$, and   therefore  an inflation scale $\mu$ can be set.  In particular,  reinstating units, this is defined by  
 $ \mu \equiv \,  {m_P } / { \sqrt{ 3 (  \xi^2 / \lambda  + a ) }} \, $. Then for large field values the potential approaches $ V_{eff} \simeq 3 \mu^2 \, m_P^2 / 4$.   For comparison, in the Starobinsky model  the inflaton  potential reaches the value   $3 \, \mu_S^2 m_P^2 /4$, where   $\mu_S$ is the scalaron mass, and  in that case cosmological data determine its magnitude, given by $\mu_S \simeq 10^{-5} \, m_P$.  In the model under consideration,  the magnitude of  $\mu$  will be discussed later,  when  imposing  limits  on the parameters $\xi, \lambda$ and $a$.

Proceeding in the same manner, as in the previously studied models, the slow - roll parameters $\epsilon_V, \eta_V$ are given by, as functions of $h$,
\bea
\epsilon_V  = \dfrac{ 8 ( 1 + \xi \, h^2  ) }{ h^2 ( 1 + 2 \xi h^2 + ( \xi^2 + c) h^4  )  } 
\; , \;
\eta_V =  \dfrac{4}{h^2 } \; \left(  - \dfrac{ 3 + 2 \xi h^2}{1 +  \xi h^2 }  + \dfrac{6 ( 1 + \xi \, h^2  )  }{ ( 1 + \xi \, h^2 )^2 + c h^4 }\right)
\, .
\label{epvita}
\eea
Although the parameter $\eta_V$  has a rather complicated form both the spectral index and the power spectrum amplitude have rather simple expressions. In fact they are given by
\bea
n_s = 1 - \dfrac{16}{h_*^2} - \dfrac{ 8}{ h^2_* ( 1 + \xi  h_*^2  ) }   \, 
\label{nshigg}
\eea 
and 
\bea
A_s = \dfrac{\lambda }{ 24 \, \pi^2}  \, \,  \dfrac{ h_*^6}{ 32 ( 1 + \xi h_*^2  ) }   \, ,
\label{ashigg}
\eea 
where we have replaced the field $h$ by its pivot value $h_*$.  These  coincide with (\ref{zz15}) and (\ref{xx5}), respectively, for $n=4$, when 
 $\xi =0$, as they should. However, the presence of the $\xi $ alters the predictions for the cosmological observables, as we shall see.
 
 In order to proceed further we need the pivot value $h_*$. In this case the number of e-folds $N_*$  is given by
  \bea
N_* = \dfrac{1}{8} \, (  h_*^2 - h_{end}^2 )   \, .
\label{zxx7}  
\eea
This does not depend explicitly on the parameter $\xi$ and is identical with (\ref{xx7}) when $n=4$.  Therefore 
\bea
h_*^2 = 8 N_* +  h_{end}^2  \, ,
\label{zxx8}  
\eea
which is functionally the same as  (\ref{xx8}) but the value of $h_{end}$ differs. The latter depends on both $ \xi$ and the combination $ c = a \lambda $, being determined as  solution of the equation
\bea
c \,  h^6_{end} + ( 1 + \xi \,  h^2_{end} )  (    h^2_{end}  ( 1 + \xi \,  h^2_{end} )  - 8  ) \, = \, 0  \, .
\label{eqh2}
\eea
This is a cubic equation in $h_{end}^2$, which we prefer to cast it in the form (\ref{eqh2}) for reasons that will become clear in the following.  Notice that  in the limit $\xi =0$ this equation becomes  (\ref{xx10}), when in the latter we put $n=4$.  
In the form presented by  (\ref{eqh2}) we see that when $c=0$ the solution for $  h^2_{end}  $ is easily obtained since it becomes a quadratic equation for $  h^2_{end}  $.  This observation is useful if we want to study the predictions of the model for small $c$,  and in doing that we expand in powers of $c$   about the zeroth order solution. 

Being a cubic equation for $ h^2_{end}$,  analytic solution can be obtained, and in our case there is  only one  real and  positive solution. The value of this solution, for  $ h^2_{end}$, can never exceed $8$. In fact this value is reached when $\xi , c$ are smaller than $ \sim 10^{-3}$ , or so. For larger  values the root of this equation is smaller. The conclusion is that $h_{end}^2 $ can be neglected in (\ref{zxx8}) and $h_*$ can be approximated by 
\bea
h_* \simeq \sqrt{ 8 N_*}  \, ,
\label{zsol4}
\eea
 as in the simple quartic model.  Replacing this value in (\ref{nshigg}) and (\ref{ashigg}) we get
 \bea
n_s = 1 - \dfrac{2}{N_*} - \dfrac{ 1}{ N_* ( 1 + 8 \xi N_*   ) }   \, ,
\label{xnshigg}
\eea   
and 
\bea
A_s = \dfrac{2 \lambda }{ 3 \, \pi^2}  \, \,  \dfrac{ N_*^3}{  ( 1 + 8 \xi N_*  ) }   \, .
\label{xashigg}
\eea 
As expected in the limit $\xi =0$ these smoothly go to (\ref{xx16}) and (\ref{yy5})   when in the latter we put $n=4$.
 
 However the role of the parameter $\xi$ is very important and can improve the case,  as far as the spectral index $n_s$ is concerned. In the simple quartic model the predictions for $n_s$ are hard to comply with the cosmological observations, unless large values of the e-folds are considered, $ N_* > 70$ or so, as already discussed.  This has  been also pointed out in \cite{Antoniadis:2018ywb,Antoniadis:2018yfq}.  Such large values of e-folds may not be acceptable,  since they demand very low values for the reheating temperature,  at least in the standard reheating scenarios.  
 Accepting large number of e-folds,  $ N_* > 70$,  it may be consistent with alternative reheating scenarios,  which may be interesting per se,  however,  in this work we would prefer to keep a more conservative attitude.   
  
 Concerning $\xi$, we shall assume that it is positive.  Then one sees from (\ref{xnshigg}) that $n_s$ is larger than the one obtained in the quartic potential studied before, which corresponds to $\xi = 0 $. Moreover,  for any $N_*$ the observable  $n_s$ increases as $\xi$ grows and therefore values within limits may be obtained for sufficiently large values of $\xi$. 
 From (\ref{xnshigg}) it can be seen that for values $ \xi \simeq 0.06$ the spectral index can be within observational limits, for  e-folds in the range $ N_* \simeq 52  - 60  $.  That is for this  value of $\xi$ a large portion of e-folds, in the range $  50 - 60$, is covered, which is broadened  for  larger $\xi $ allowing, also,  for  values of $N_*$ lower than $52$. 
 Values of $\xi < 0.06$ are also acceptable,  at the cost of shrinking considerably the  range of the allowed e-folds, that are compatible with the observational limits imposed by $n_s$. For instance for  $\xi \simeq 0.004$ one obtains $n_s = 0.9607$,  at the edge of the lower observational limit, pushing $N_*$ to   $N_* \simeq 60$.  From these arguments it is obvious that a reasonable range to deal with  in our numerical procedure is to focus on  values of $\xi$ of the order of ${\cal{O}}({10^{-2}})$, or larger. In the following  we shall take $ \xi \gtrsim 0.06 $ on the grounds that is likely to cover  a wider range of e-folds, as we explained above.

 From (\ref{xashigg}), and accepting that $A_s \simeq 2.1 \times 10^{-9}$, the quartic coupling is found to be constrained by
 \bea
 \lambda \simeq 3.11  \times 10^{-8} \; \dfrac{1 + 8 \xi N_* }{  N_*^3}  \, .
 \label{aslambda}
 \eea
 In the limit $\xi = 0$ this coincides with (\ref{zzzw}), as it should. From this it is seen that the allowed values for $\lambda$ depend on the parameter $\xi$, and also that larger values of the coupling $\lambda$, as compared to the simple quartic model,  are available in this case.  However even in this case the quartic coupling is small. For $\xi =0.06$ it is of order $\sim 10^{-12}$. In order for $\lambda$  to reach values of order $ \gtrsim 10^{ - 6}$ one needs large values $ \xi \gtrsim 10^4$, when  $N_* \simeq 50 - 60$. 
 
 Concerning the parameter $a$, by the same token, as discussed in previous models, a lower limit on it can be established by  (\ref{boundr}), given by
 \bea  
 a \gtrsim  10^8 \; \left(  1.25 - \, \dfrac{N_* ( 1 + 8 \xi \, N_* )}{200}  \right)  \, .
 \label{alpha}
 \eea
 
 This bound on $a$ depends on $\xi$,  it is quadratic in $N_*$, and  there is a critical value of $\xi$ beyond which  it  becomes negative, signaling that in this case any positive value of $a$ is actually  allowed.  As we prefer to work with values $ \xi > 0.06$  the rhs  of (\ref{alpha}) is  negative, for $N_*  \simeq 50 - 60$, and practically for our purposes  there is no lower  limit  imposed on the parameter $a$. The absence of a lower bound  may be important since in this case $a$ can be chosen either larger, or smaller, than the ratio $\xi^2 / \lambda $.   In the regime 
  \bea
 \xi^2  > a \, \lambda  \, ,
 \label{xia0}
 \eea
 an upper bound on $a$ is imposed, for given  $\xi \, , \lambda$. 
 Of particular interest, within this regime,  is the case where $ \xi^2 \gg a \, \lambda  $. 
 In this limit, it is seen from (\ref{zxx3}) and (\ref{zxx4}) that the functions $K(h)$ and the potential $V_{eff}(h)$ do not depend on the parameter $a$.  In fact $K(h)$ depends only on $\xi$ and $ V_{eff(h)}$ on $ \xi, \lambda$. 
 The function $L(h)$ does depend on $a$, however,  its effect in the equations of motion is small, for the cases of interest,  as we have already  remarked in Section III. Therefore in this case the results are independent of the parameter $a$,  as long as $ \xi^2 \gg a \, \lambda  $  holds. This we have verified in our numerical procedure. 
 In this case the inflation scale $\mu$, as defined before (  see discussion following Eq. (\ref{zxx4}) ),  becomes 
 $ \mu \simeq  \sqrt{ \lambda / 3 \xi^2 } \, m_P $ and lies in the range   $ \sim ( 2 \times  10^{-5} - 5 \times 10^{-7} ) \, m_P  $, for values of $ \xi$ in the range $ 0.06 - 100.0$ and for  $N_*$  between $50 - 60$ ,  the smaller (larger)  scales being  attained for higher (lower)  $\xi$ and $N_*$ values. 
 
 Evidently the arguments given before are no longer valid if  the parameters are chosen in the regime 
  \bea
 \xi^2  < a \, \lambda  \, .
 \label{xia2}
 \eea
 Then we have a lower bound on $a$, for given $\xi , \lambda$. 
 In this case the predictions depend on $a$ and  $\xi, \lambda$ as well. In particular,  when  $ \xi^2 \ll a \, \lambda$ the inflation scale 
 is $ \mu \simeq m_P / \sqrt{ 3 a}$, that is it is determined solely by $a$. 
 
 For facilitating the discussion, in Table \ref{table2} we present order of magnitude estimates of  the quartic coupling $\lambda$,  as these arise from (\ref{aslambda}),  and the corresponding $\xi^2 / \lambda$ for given value of $\xi = 10^\nu$, where $ \nu < 0$ or 
 $ \nu >0$, corresponding to $ \xi < 1$ or $ \xi > 1$, respectively.  We see that the coupling $\lambda$ increases with increasing $\xi$, or same, increasing  $\nu$.  Although not displayed in the table, we remark  that for  $ \nu \geq 0  $ the coupling  $\lambda$  lies within $ ( 0.7 - 1.0 ) \times 10^{-10 + \nu} $.  
 The lower and upper bounds on the parameter $a$, for having  $ a > \xi^2 / \lambda$ and $ a < \xi^2 / \lambda$,  are  shown in the fourth and fifth column,  respectively.  In creating this  table the values of $N_*$ were taken, as usual, in the range  $ N_* \simeq 50 - 60$. 
 
\begin{table}
\begin{center}  
\begin{tabular}{|c|cccc|}  
   \hline
   $\; \xi \, \; $ & $\lambda $  & $ \xi^2 / \lambda $  &  $ a >  \xi^2 / \lambda $ &   $ a < \xi^2 / \lambda $ \\
  \hline
  \quad  $10^{\nu}$ \quad  &  \quad $10^{-10 +\nu}$ \quad  &  \quad $10^{10+\nu}$ \quad  & \quad  \quad $ > 10^{10+\nu} $ \quad  \quad & \quad  \quad $ < 10^{10+\nu}  $  \quad \\
   \hline
 \end{tabular}  
\caption{
Order of magnitude estimates for $\lambda$, and $\xi^2 / \lambda$, as derived from Eq. (\ref{aslambda}), for given value of $\xi$ ( first row ) and value of e-folds $ N_* = 50 - 60 $.  In the fourth (fifth)  column the lower (upper) bound, set on the parameter $a$,  is displayed for having  
$ \; a > \xi^2 / \lambda \;  \; (     a <  \xi^2 / \lambda     )$.  The power $\nu$ is positive, for $ \xi > 1$,  and negative for $ \xi < 1$.  
}
\label{table2} 
\end{center}
\end{table} 

 In order to have an estimate of  the instantaneous reheating temperature, $ T_{ins}$,  which is given by (\ref{numt}),  we need know the energy density at the end of inflation.  Following similar arguments, as for the models studied previously, we find that in this case it is given by 
\bea
 \rho_{end}   \, = \,  \, \dfrac{\sigma}{4 a} \left( 1 - \dfrac{ h_{end}^2 ( 1 + \xi \, h_{end}^2 ) }{8} \,   \right)  
 \, \equiv \dfrac{\sigma}{4 a} \, F(\xi, c)  \, .
 \label{rendit4}
\eea  
Recall that  $\sigma =1.5 $. The function $F(\xi, c)$ is too complicated to be presented, although  analytic expression  for the unique  positive solution $h_{end}^2$ of  Eq. (\ref{eqh2}) does  exist.  This we shall actually   use for the calculation of $\rho_{end}$ through 
(\ref{rendit4}).  
Replacing   $a$ by $c/\lambda$, with $\lambda$ as given by (\ref{aslambda}), we get from  (\ref{numt}),
\bea
T_{ins} \, = \, ( 0.968 \times 10^{-3} ) \, \left( \dfrac{55}{ N_*} \right)^{1/2} \, 
 \left( \xi + 2.27 \times10^{-3} \, \dfrac{55}{ N_*}    \right)^{1/4}   \, R^{1/4}(\xi, c)  \, ,
 \label{tin1}
\eea
where  $ R(\xi, c) = F(\xi, c)/c$. This it gets a very simple form in particular regions, and interestingly enough this includes the region where $T_{ins}$ gets its largest value.   

The first region of interest is when $ c /  \xi^2  < 1 $. 
As we have already remarked,  Eq.  (\ref{eqh2})  is easily solved when $c$ vanishes,  since in that case it is reduced to a quadratic equation for  $h_{end}^2$.  For non-vanishing $c$,  within  the regime  $ c / \xi^2 < 1 $,  we can treat this ratio  as a small parameter,  in order to find the desired  solution as deviation from the zeroth-order solution,  corresponding to $ \, c= 0$.  This is easily implemented, resulting to a function $R(\xi, c)$, which to the lowest order in $c / \xi^2$, is independent of $c$. In particular it is found that, 
\bea
R(\xi, c) \, = \,  \left (  \frac{ 1 + 16 \xi - \sqrt{1+32 \, \xi  }    }{ 16 \, \xi^2} \right)^2 \,   
\equiv \, P(\xi)   \, .
\label{rofx}
\eea 
The function $P(\xi)$ is regular at $\xi = 0$, with limit  $ P(0) = 64$. Using this, we find from (\ref{tin1}) 
\bea
T_{ins} \, = \, ( 0.968 \times 10^{-3} ) \,   
 \left( \xi \, P(\xi)    \right)^{1/4}    \,  \; .
 \label{tin2}
\eea
In this we have set $ 55 / N_* \simeq 1 $,  and, besides,  we assume that $ \xi > 0.01$, which is actually the region we are interested in.  Note that (\ref{tin2}) is valid in the regime $ c / \xi^2 < 1 $ and it is a very handy relation.  
Within the $ c < \xi^2$ regime the  maximum temperature is attained when $ \xi P(\xi) $ reaches its maximum.  This occurs at  
$ \xi  = 3/32$, that is very close to  $ \simeq 0.094$, and for this value $ T_{ins} \simeq 2.47 \times 10^{15} \, GeV$, in natural units. This is independent of $c$ as long as $ c $ is much smaller than $\xi^2$.  Away from this maximum $T_{ins}$ drops, as $\xi$ increases, behaving as $T_{ins} \simeq   ( 0.968 \times 10^{-3} ) \,  \xi^{-1/4}$.  

Another region of interest is when $c$ is large and $ c >> \xi^2$.  In this region the function $F(\xi, c)$, that controls $ \rho_{end}$ in 
(\ref{rendit4}), is very close to unity.  
Note that the largeness of $c$ by itself  is not adequate to have $  F(\xi, c) \simeq 1$, despite the fact that $h_{end}^2$ is small. We must require, in addition,  that $ c >> \xi^2$.
Then $  \rho_{end}$ turns out to be inverse proportional to $a$, and hence the instantaneous reheating temperature is proportional to $ a^{-1/4}$, or same proportional to $( \lambda / c)^{1/4} $.  The latter is proportional to $( \xi / c)^{1/4} $, when (\ref{aslambda}) is used. 
 Then the analytic result for $T_{ins}$, in this case,   is trivially found from (\ref{tin1}),  
\bea
 T_{ins} \, 
\simeq   ( 0.968 \times 10^{-3} ) \, \left(  \dfrac{\xi}{c} \right)^{1/4} \, .
 \label{tin3}
\eea
This holds for large $c$ values, satisfying $ c \gg \xi^2$, and  therefore it cannot be arbitrarily large. The largest value within this regime is about $ \simeq 10^{15} \, GeV$,  which is slightly smaller than the corresponding temperature of the $ c << \xi^2$ region.  
This is obtained for $ c \simeq10^2$, which is relatively large,  and values of  $\xi^2 $ about an order of magnitude smaller than $c$.  Any other pair of values, for these parameters, within this particular regime, results to lower values of $T_{ins}$. 

\begin{figure}[t] 
\centering
  \centering
  \includegraphics[width=0.75\linewidth]{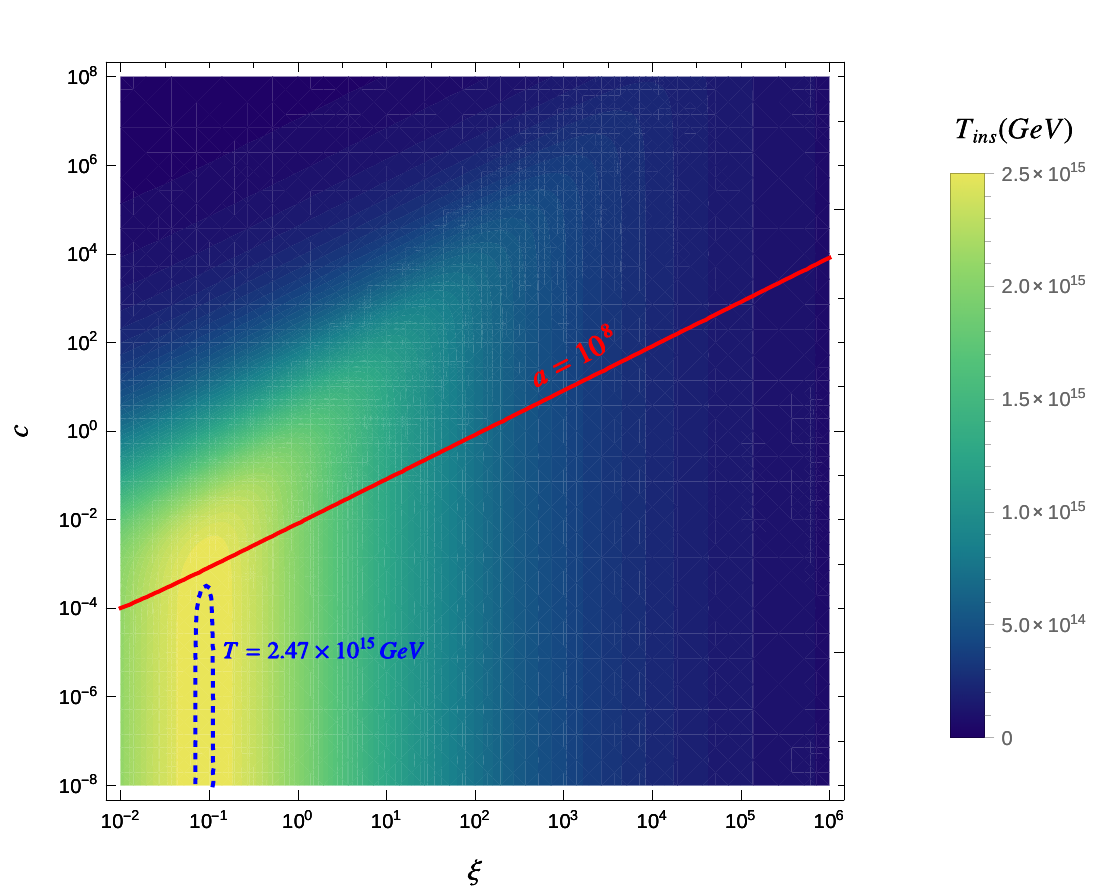}      
\caption{In the $c, \xi$ plane we display the instantaneous reheating temperature, as given by Eq. (\ref{tin1}), for $N_* = 55 $.  
Light colors correspond to larger temperatures. The largest temperature, $ T \simeq 2.47 \times 10^{15} \, GeV$   , is within the  yellow region  near $\xi \simeq 0.1$, with boundary   the blue dashed line. The red line is the locus of  points with  $\, a = 10^8  $.
}
\label{figit1} 
\end{figure}

Unfortunately, outside the aforementioned regions there are not simple mathematical expressions to deal with, and we shall rely on a numerical treatment of  (\ref{tin1}).   In fact scanning the two dimensional parameter space $c, \xi^2$ we found,  that the approximate formulas given before in the appropriate regions, agree to a very good accuracy with the values obtained from (\ref{tin1}).
In Figure \ref{figit1} we display the instantaneous reheating temperature, as given by Eq. (\ref{tin1}), for $N_* = 55 $.  
Light colors correspond to higher temperatures. From this figure it is clearly seen that the larger temperatures are obtained for values of the parameters within the small yellow region, located at the bottom and left. The region with the largest temperature $T_{ins}$  is centered about  $\xi \simeq 0.1$, and values $c \lesssim 10^{-4}$,  having  as boundary   the blue dashed line corresponding to  
$ T_{ins} = 2.47 \times 10^{15} \, GeV$. The maximum temperature attained  is very close to it,  confirming, therefore, our previous arguments. 
Within this region $\xi \simeq 0.1$,  and since Eq. (\ref{aslambda}) is used,  $ \lambda \simeq 10^{-12}$. 
Therefore $ a = c / \lambda \lesssim 10^{8}$ is needed for having the largest possible $T_{ins}$. This is also  seen by drawing  the locus of  points for which the parameter $\, a $ has a constant value,   $\, a = 10^8  $. This lies just above the aforementioned region.  
Lower values, $a < 10^8$, will move this line  downwards, crossing the largest $T_{ins}$ region, and thus the maximum $T_{ins}$ is obtainable.

Note that the analytic expressions  for  $T_{ins}$, given so far,  serve as an estimate of the magnitude of the  instantaneous temperature.  As we have already  pointed out, the actual values are extracted  by solving the pertinent equations of motion numerically. However,  the  numerical analysis reveals that these estimates are  accurate enough.  In fact, the results derived  are lower by less than about $10 \, \%$. Only in a small region,  for  $c \leq 10^{-8} $ and for  $\xi $ values in the vicinity  $\xi \simeq 0.1$,  this difference augments to about  $15 \, \%$, or so.  This is in accord with the discussion following Eq. (\ref{ro333}). 
 As a result the maximum instantaneous reheating temperature  mentioned before, $T_{ins} = 2.47 \times 10^{15} \, GeV$,  drops to $ T_{ins} = 2.07  \times 10^{15} \, GeV$.

 Our numerical study can be summarized  by selecting the following representative  inputs :
 
For the value $\xi = 0.06$, which according to preceding discussion sets the threshold for having sufficient number of e-folds, we choose the quartic coupling $ \lambda =  4.875  \times 10^{-12} $. 
From (\ref{aslambda}) one can see that for $N_* = 50 -60$ the quartic coupling is between  $ 4.29 \times 10^{-12}  $ ( for $N_* = 60$ )  and $ 6.22 \times 10^{-12}  $ ( for $N_* = 50$ ), so that the value chosen is indeed within the appropriate range. 
However,  this fine-tuned value has been chosen so that  the predicted amplitude $A_s$  is within observational limits, in such a way that instantaneous reheating is feasible. 
 It should be remarked that the approximate formula  used  for $A_s$  may differ from the one that the numerical procedure returns. The latter yields more accurate results, since the exact numerical solution for the field $h$ is used, and also, because it incorporates  corrections, see Eq. (\ref{amplitude}),  that although small in some cases  are of the same order of magnitude  with the observational errors.  It is for this reason that fine-tune adjustments are necessary to make the instantaneous reheating mechanism a viable possibility.

  For these inputs $\xi^2 / \lambda \simeq 7.38 \times 10^8$, and  therefore for values $ a \ll 10^8$ we are in the regime 
  $ a \ll \xi^2 / \lambda$ and, as we have discussed,  predictions are insensitive to the choice of $a$. Therefore any value of $a$ yields the same results, provided $ a \ll 10^8$. This we have verified by our numerical code. For definiteness we take $a =10^6$ which is three orders of magnitude smaller than  $ \xi^2 / \lambda$ given above. 
  
\begin{figure}
\centering
  \centering  
  \includegraphics[width=1.07\linewidth]{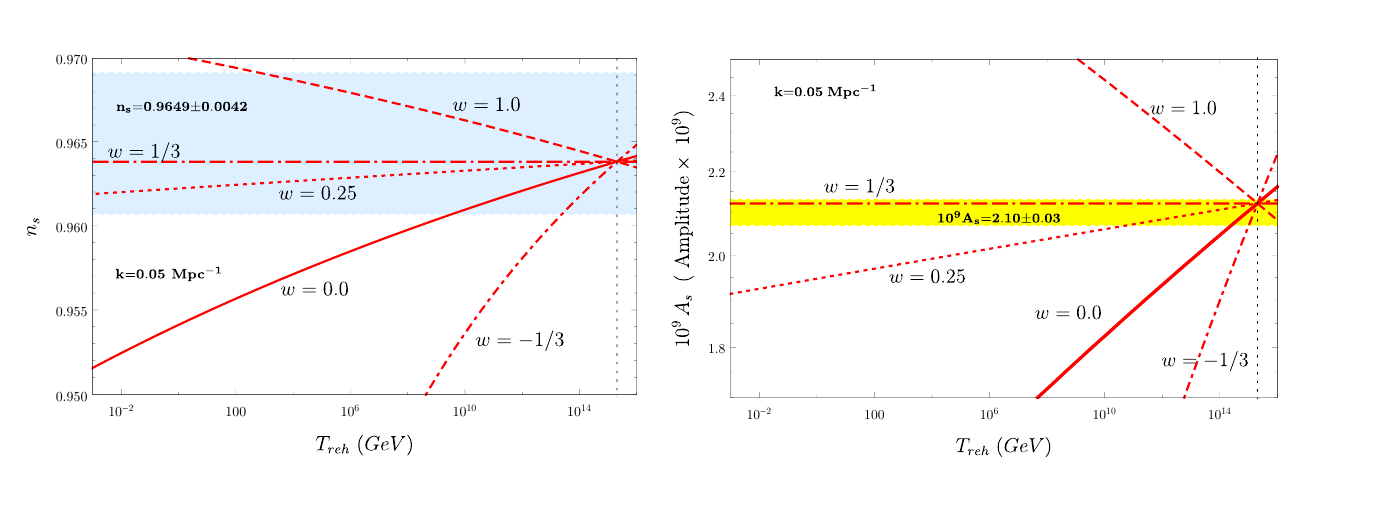}     
   \includegraphics[width=0.48\linewidth]{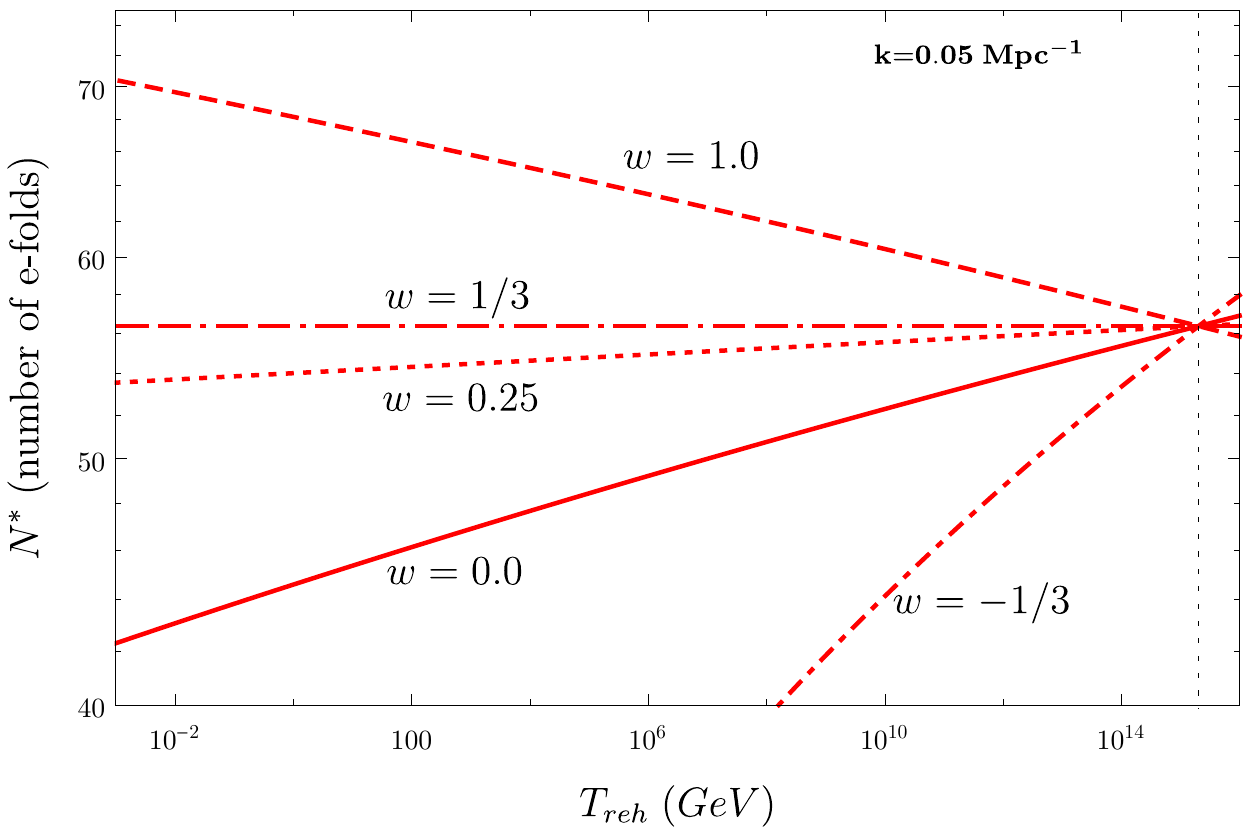} 
\caption{
Top: The spectral index $n_s$, left, and the power spectrum amplitude  $A_s$, right,  versus the reheating temperature $T_{reh}$,  for the Higgs model, for inputs $\xi = 0.06$, $\lambda = 4.875 \times 10^{-12}$ and $ a = 10^6$.  
Bottom:  For the Higgs models and same inputs, the number of e-folds   versus the reheating temperature $T_{reh}$ is displayed. 
   }
   \label{fig124}
\end{figure}

  In Figure \ref{fig124}, at the top, we display the spectral index and the power spectrum amplitude. We see that agreement with $n_s$ data is obtained for any temperature when the parameter $w$ is $\simeq0.25$ or larger,  but smaller than $ 1.0$. For  canonical reheating, $ w =0.0 $,  however a lower bound is imposed $ T_{reh} \gtrsim 10^{10} \, GeV$, while for $w = 1.0$ the lower bound is about $ T_{reh} \gtrsim 100 \, GeV$ . Looking at $A_s$ plot we observe, as advertised, that instantaneous reheating can occur, for the given $\xi , \lambda$  inputs. We also  observe that the constraints are more stringent than those imposed by $n_s$.  In fact values of $ w > 1/3 $, allow for temperatures which are very close to $T_{ins}$. At the same time a lower reheating temperature is imposed for the $ w = 0.25$ case, $ T_{reh} \gtrsim 10^{11} \, GeV$, while  for the canonical scenario the lower limit imposed by $A_s$ is pushed to a much  higher value,  close to $T_{ins}$.  At the bottom of the same figure the number of e-folds is shown. Although values of  e-folds $N_*$ as large as $\simeq 70$ for low $T_{reh}$ are allowed, by $n_s$ data, when $ 1>  w \geq 0.25$, 
 the  $A_s$ measurements restrict the allowed temperature range in such a way that  $N_*$ is forced to be in the range  
 $ \simeq 55.70 - 56.30$,  as shown in  Table \ref{table3}. In this table  the predictions for $A_s, n_s, r, N_*$, corresponding to the minimum ( upper rows ) and maximum ( bottom rows ) reheating temperature, are also shown. The maximum reheating temperature is the instantaneous temperature,  $T_{ins} = 2.027 \times 10^{15} \, GeV $, and for this reason the predictions for the various $w$, in that case, coincide.
  
\begin{table}[]   
\begin{center}
\begin{tabular}{|cccc|}  
  \multicolumn{4}{c} {  Higgs Model  \quad ( pivot scale $ k^* = 0.05 \, Mpc^{-1}$ ) }    \\       
  \hline \hline   
 Input values   \quad   & \quad  \quad $ \xi = 0.06 $ \quad  & \quad $ \lambda = 4.875  \times 10^{-12} $ \quad  & \quad  $ a = 10^6 $    \\ 
  \hline
   \multicolumn{1}{|c} {$\; w \,$- value } \quad  & $\quad w= 0.0 \quad $  & $ \quad w= 0.25 \quad $  &  $\quad w= 1.0 \quad $  \\
   \hline
    \multicolumn{1}{|c|} {$  \quad 10^{9} \,A_s  \quad \quad $ } & 2.07  & 2.07 &  2.13     \\
    \hline
    \multicolumn{1}{|c|} {$n_s$ } & 0.9634  & 0.9633 &   0.9639 \\
    \hline
    \multicolumn{1}{|c|} {$r$ } & 0.0102  & 0.0102 &    0.0100  \\
    \hline
    \multicolumn{1}{|c|} {$N_*$ } & 55.67  & 55.67 &   56.45  \\
   \hline
    \multicolumn{1}{|c|} {$T_{reh}$ } &  $2.562  \times 10^{14}$  &    $ 6.695 \times 10^{10} $  &   $ 1.569 \times 10^{15} $  \\
    \hline \hline
     \multicolumn{1}{|c|} {$  \quad 10^{9} \,A_s  \quad$ } & 2.12  & 2.12 &  2.12    \\
    \hline
    \multicolumn{1}{|c|} {$n_s$ } & 0.9638  & 0.9638 &   0.9638  \\
    \hline
    \multicolumn{1}{|c|} {$r$ } & 0.0100  & 0.0100 &    0.0100  \\
    \hline
    \multicolumn{1}{|c|} {$N_*$ } & 56.36  & 56.36 &   56.36  \\
   \hline
    \multicolumn{1}{|c|} {$T_{reh}$ } &  $2.027  \times 10^{15}$  &   $2.027  \times 10^{15}$  &  $2.027  \times 10^{15}$     \\
    \hline
 \end{tabular}
\caption{
Predictions  of the Higgs  Model , for the input values shown on the top, for the  cosmological observables $ n_s, r,  A_s, N_* $  and for various values of the equation of state  parameter $w$. The  values shown for  the reheating temperature $T_{reh}$, in GeV, correspond to  the minimum (upper rows) and maximum (lower rows) allowed,  when the observational limits  for 
$ A_s  $ and   $n_s$  are imposed.  }
\label{table3} 
\end{center}  
\end{table} 

  As a second sample we consider values of $\xi$ in the range $ \xi = 0.06 - 10.0$ when the parameter $a$ is increased to  $ a = 10^{12}$.  These cases fall in the regime $ a > \xi^2 / \lambda$ when $ \lambda$ is within the range suggested by (\ref{aslambda}).  Following the same reasoning, we may consider  values for the quartic coupling so that agreement with $A_s$ data is obtained, requiring, at the same time,   the maximum reheating temperature can reach the instantaneous temperature $T_{ins}$.  For the lowest value of $\xi$  in this range,  $\xi = 0.06$,  the quartic coupling can be taken $ \lambda = 5.60 \times 10^{-12}$ while for the largest, 
  $ \xi = 10$,  the value    $ \lambda = 8.85 \times 10^{-10}$ suits our needs. For reference, these cases we shall name A and B,  respectively.   

Note that by  changing $a$, from   $ a =10^6$ to $ a = 10^{12}$,  the predicted values for the cosmological parameters change as well, and thus  readjustments of $\lambda$ are necessary, in order to obtain agreement with $A_s$ data,    and have,  at the same time, $T_{ins}$ as the maximum temperature. This is the reason the values of  $\lambda$, for  the case $\xi = 0.06$,  are slightly  different for  $ a =10^6$ and $ a = 10^{12}$.

\begin{figure}  [H]
\centering
  \centering
  \includegraphics[width=1.07\linewidth]{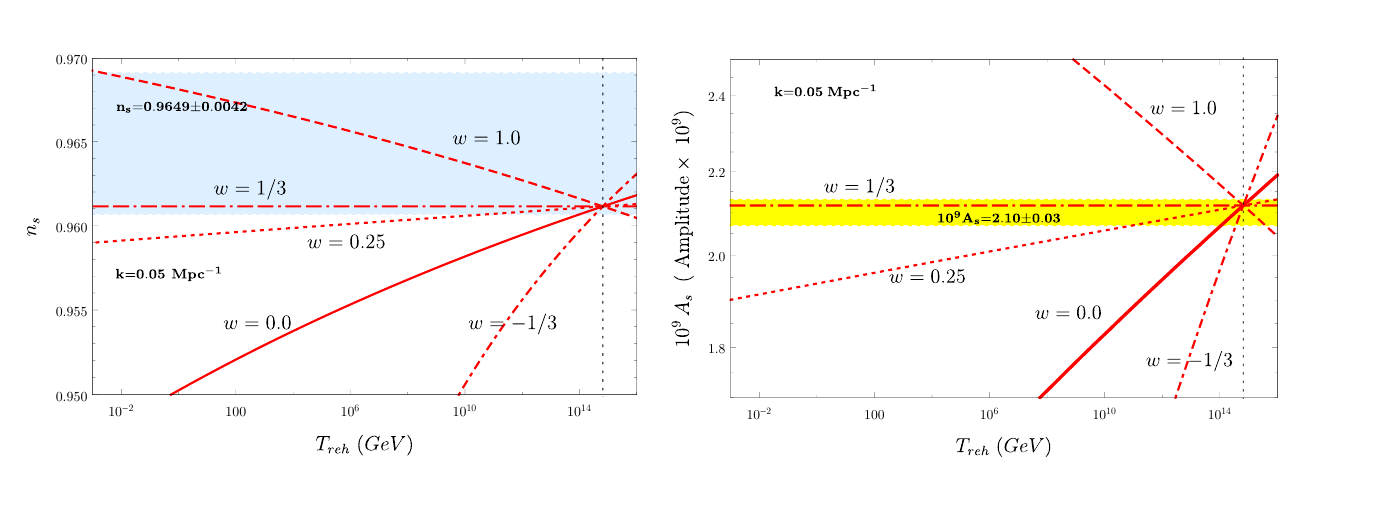}      
\caption{
The spectral index $n_s$, left, and the power spectrum amplitude  $A_s$, right,  versus the reheating temperature $T_{reh}$,  for the Higgs model, for inputs $\xi = 0.06$, $\lambda = 5.6 \times 10^{-12}$ and $ a = 10^{12}$ ( case A ). 
   }
   \label{fig125}
\end{figure}

\begin{figure}  [H]
\centering
  \centering   
    \includegraphics[width=1.07\linewidth]{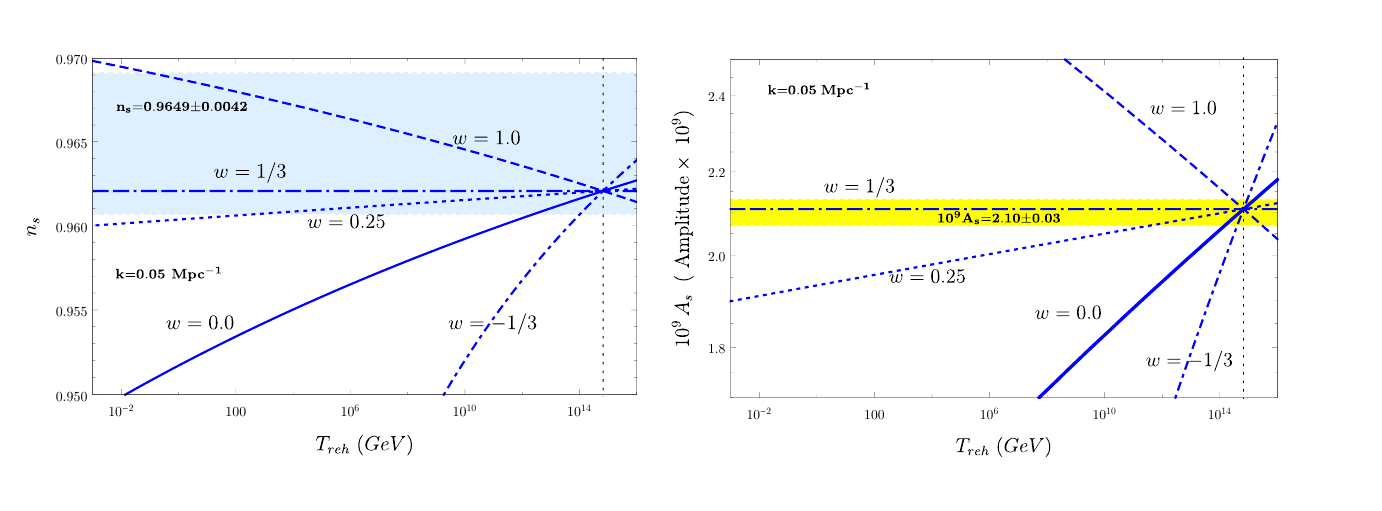}       
\caption{
The same as in Figure \ref{fig125}, for inputs $\xi = 10.0$, $\lambda = 8.85 \times 10^{-10}$ and $ a = 10^{12}$ ( case B ). 
   }
   \label{fig125b}
\end{figure}

In Figures \ref{fig125} and \ref{fig125b} we display the predictions for the spectral index and the power spectrum amplitude for the cases A   and B, respectively,  discussed before. 
Comparing  Figure \ref{fig125}    with Figure  \ref{fig124}  (on top) we first observe that $T_{ins}$ is lowered, in comparison  to the A - case. In fact,   from $T_{ins} = 2.027 \times 10^{15} \, GeV$ it slides down  to $  6.522  \times 10^{14} \, GeV $. 
 Also the lowest reheating temperatures change  a little. For instance  for $w = 0.25$ this   is $ 1.065 \times 10^{11}$, i.e. it has been slightly increased from the corresponding $a = 10^6$ case,  which was $ 6.695 \times 10^{10} $ ( see Table \ref{table3} ).  In  Figure \ref{fig125b} the corresponding  predictions for the B - case are shown.  In this case  $T_{ins} = 6.647 \times 10^{14} \, GeV$. That is,  it is  slightly larger than the case A.  Keeping $a$ fixed the tendency for $T_{ins}$ is to decrease, with increasing the parameter $\xi$, 
 as long as $  a  > \xi^2 / \lambda $,  while tuning the  quartic coupling to have  agreement with $A_s$ data.

\begin{figure}[]    
\centering
  \centering
  \includegraphics[width=0.8\linewidth]{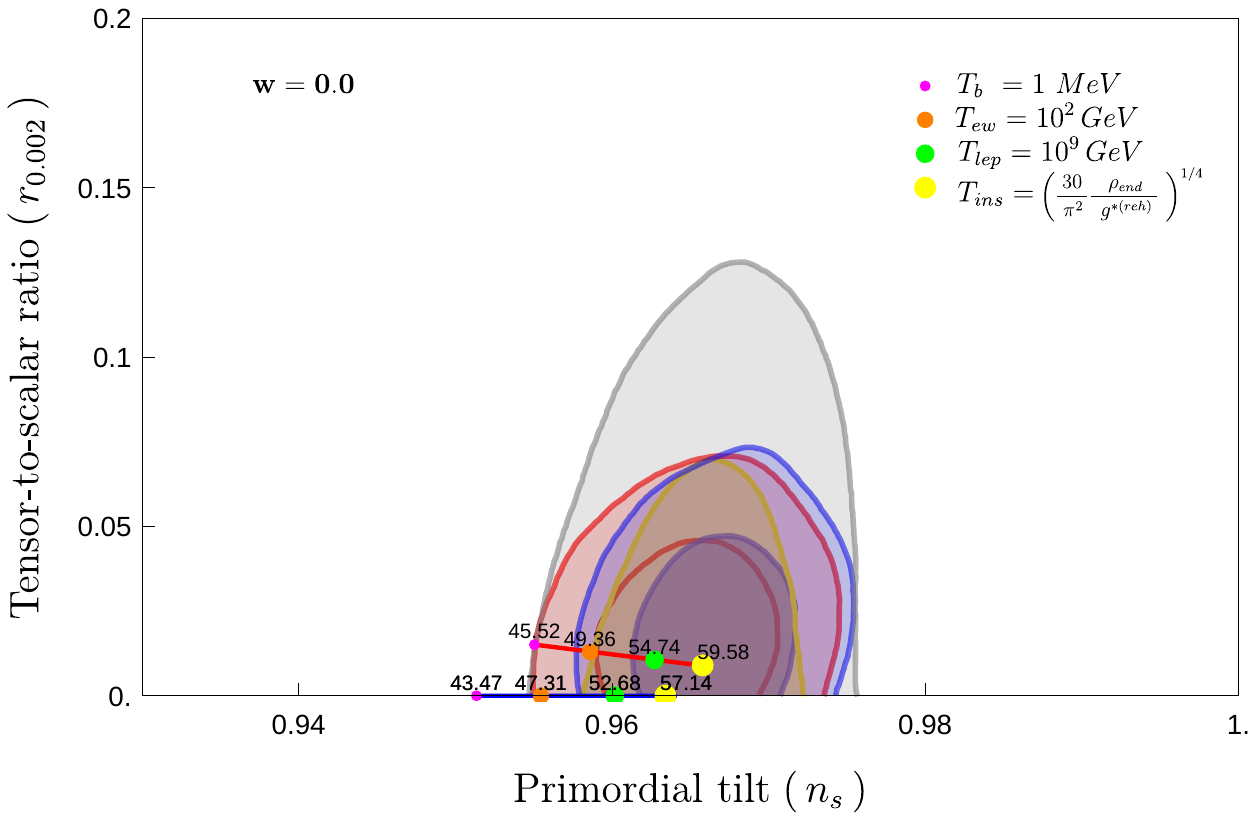}     
\caption{
The tensor-to-scalar ratio $r_{0.002}$ versus the spectral index $n_s$ for the Higgs. As in Figure \ref{fig11} the numbers shown correspond to  the e-folds and the circles designate different reheating temperatures. For the line on top ( in red )  the parameters are 
$ a= 10^6$ , $\xi = 0.06$  and $\lambda = 4.875 \times 10^{-12}$ while for the one at the bottom  ( in blue )
$ a= 10^{12}$  , $\xi = 0.06$  and  $\lambda = 5.60 \times 10^{-12}$.  Only the cases for the canonical scenario are shown, 
$ w =0$.
   }
   \label{fig126}
\end{figure}

In Figure \ref{fig126}, we show the tensor-to-scalar ratio $r_{0.002}$ versus the spectral index $n_s$ for the Higgs model.  The numbers of  the e-folds are shown, and the circles designate different reheating temperatures, exactly as  in Figure \ref{fig11}. The upper line ( in red )   corresponds to parameters  $ a= 10^6$ , $\xi = 0.06$  and $\lambda = 4.875 \times 10^{-12}$ while for the one at the  bottom ( in blue )  the parameters are 
$ a= 10^{12}$  , $\xi = 0.06$  and  $\lambda = 5.60 \times 10^{-12}$.  Only the cases for the canonical reheating are shown, i.e. 
$ w =0$.  Note that in drawing  this figure  the constraints arising from $A_s$ have not been taken into account.  When they are a small segment including the $ T_{ins} $ temperature is left.  In any case, we observe from these figures that by increasing the parameter $a$  the tensor-to-scalar ratio gets smaller and the predictions move lower and the instantaneous reheating temperature  mechanism is  in  full agreement with Planck 2018 cosmological constraints. 

  \section{Conclusions}
  
  In this work we have considered ${\cal{R}}^2$  theories in the framework of the Palatini formulation.  Although this is not new, we have presented a general setup within which inflation models can be studied. The actions, in the Einstein frame, resemble $K$ - inflation models, however additional terms, that are quartic in the derivatives of the fields involved,  emerge.  
  For  the models studied in this work, these are small during inflation,  in accord with the findings of other authors, as we have verified by our numerical approach, which duly takes into account these terms.   
   However,  their role towards the end of inflation is essential for the determination of  the instantaneous reheating temperature.
  
  This formulation is model independent and  can be applied to any inflationary model.  These theories are described by three arbitrary functions. Two of them are associated with the coupling of  the scalars to the linear and the quadratic terms,  with respect the Palatini curvature ${\cal{R}}$,  and the third is a scalar potential. Inflation can be studied in this framework without the need of using  canonically normalized fields. 
  
  We have applied this for the study of popular inflationary models that are minimally coupled to gravity, with monomial potentials of the form $ V \sim h^{n}$, with the power  $n$ a positive and even integer. We also considered the Higgs potential  non-minimally coupled to gravity. These models have been put under scrutiny over the years, in the metric formalism,  and  recently have been extensively studied in the non-metric, or Palatini, formalism.  However the stringent constraints arising from the scalar power spectrum measurements have not been duly taken into account, in most of the studies, in conjunction with the reheating temperature of the Universe. In \cite{Tenkanen:2019jiq}  such a study has been undertaken, in the context of  the quartic Higgs model that is minimally coupled to gravity.  
  
   In this work, without invoking any particular reheating temperature mechanism, we have undertaken this study,  and show that the measurements of the primordial power spectrum amplitude imposes very stringent constraints.    These, 
   in combination with the restrictions arising from the measurements of other  cosmological observables, in particular the primordial tilt $ n_s$ and the tensor to scalar ratio $r$,  restrict considerably these models. 
   
   For the quadratic model $ V = \frac{m^2}{2} h^2$ we have seen that the scalar  power spectrum amplitude $A_s$ puts constraints on the parameter $m$, and agreement with data is obtained for values of it that lie  in a tight range. The maximum reheating, or instantaneous ,  temperature $T_{ins}$,  is of  order  $ \, \sim 10^{15} \,  GeV$, and this  is attained for  fine-tuned values of $m$, within this range.  For these fine-tuned values, the range of the allowed temperatures is rather narrow, and depends  on the effective equation of state parameter $w$, with a lowest  temperature  not  far from the instantaneous temperature. For the canonical scenario, although smaller,  this is of the same order of magnitude with $T_{ins}$.
   If we allow for small  deviations, from this fine-tuned values, agreement with data is still feasible. However these deviations, although they do not disturb substantially the observable $n_s$, should lie in a narrow range, outside which agreement with $A_s$ data is hard to achieve.  In these cases the allowed temperatures are well below $T_{ins}$  and rapid thermalization is not possible. Besides, depending on the value of $m$, not any  value of $w$ in the range $ - 1/3 < w < 1 $ is allowed.   The conclusion, concerning this model, is that, agreement with all cosmological data is possible for values of the potential coupling $m$ that lie in a narrow range. 
 Instantaneous reheating is possible at the cost of a very fine-tuned values of $m$.

   The model with the quartic potential $ V \sim h^4 $ is in conflict with the spectral index $n_s$ data. Only marginal agreement with the primordial tilt can be obtained, with $n_s \simeq 0.960 $,  but this occurs for  very low reheating temperatures close to Nucleosynthesis $ T_{reh} \sim MeV$, and for values $ w$ close to $ w = 1.0$. On the other hand, the amplitude $A_s$  prefers smaller values of the equation of state parameter $ w \lesssim 0.25$. 
   The conclusion is that, this model is hard to reconcile with $n_s$, the scalar power spectrum measurements and reheating temperatures that are reasonably larger than $ T_{reh} \sim MeV$ so that we do not run into problems with Big Bang Nucleosynthesis.  As our qualitative arguments have shown, for the descendant models,  $ V \sim h^{n}$ with $ n > 4$, the situation is even worse.  
   
   The situation with the quartic potential  is rescued in the Higgs model when  the scalar field couples in a non-minimal manner to gravity, specified by a parameter $ \xi$.  This helps in that, as we have explicitly shown,  the value of $n_s$ depends on $\xi$ allowing for larger values of $n_s$.  Agreement with $n_s$ observations demands that  $\xi$ is not smaller than about $\sim 0.06$.  Given 
   $\xi$, the primordial spectrum measurements in the Higgs model restricts severely the quartic coupling $ \lambda$.  The larger the value of $\xi$ is  the largest  the values of the allowed $\lambda$ are. The quartic coupling is small, smaller than $ \sim 10^{-6}$,  for values  $\xi$ that do not exceed  $ \sim 10^4$.   Higher   $\lambda$ 
    values  are in principle allowed but these require very large values of $\xi$ leading to  instantaneous reheating temperatures,  lower than $\sim 10^{15} \, GeV$.  
   Note that in the Higgs  case there is no bound on the parameter $a$ specifying the coupling of the scalar field to the gravity term 
   ${\cal{R}}^2$, which, unlike the previous models is unrestricted.   Thus both large and low values of $a$ are allowed. Due to that, and for  given $\xi$ and $\lambda$ in the appropriate range,  two regimes can be distinguished. The small-$a$, when $a < \xi^2 / \lambda$, and the large $a > \xi^2 / \lambda$ regime.  In the small $a$-regime, and particularly when $ a \ll \xi^2 / \lambda $, the predictions are independent of the parameter $a$, provided that $a$ stays much smaller than  $  \xi^2 / \lambda $.  The inflationary scale in this case is $ \mu \sim \sqrt{( \lambda / \xi^2 ) }$ and lies in the range $10^{-5} -  10^{-7} $ Planck masses, for $\xi$ between $ 0.06$ and $10^2$.  The instantaneous reheating temperature $T_{ins}$, in this case, is larger for smaller values of the parameter 
   $\xi$ and receives its largest possible value,     $\simeq 2.1 \times  10^{15} \, GeV$, when $\xi$ is in the vicinity of $\xi\simeq 0.1$. 
   
 In the large-$a$ regime, on the other hand, the inflationary scale is $ \mu \sim a^{-1/2}$.  
      At the same time $T_{ins}$ behaves as $  \sim a^{-1/4} $.  Unless $a$ is not exceedingly  large, $T_{ins}$ can be as large as 
 $\simeq 10^{15} \, GeV$,  and this requires values of $\xi $ of order unity or so.  In both regimes there are values of the parameters for which all cosmological data can be satisfied. However for given $\xi$, as in the   models discussed previously, the quartic coupling $\lambda$ should lie in a tight range, as the  power spectrum observations dictate. Moreover for instantaneous reheating $\lambda$ should be fine-tuned.  In that case the allowed temperatures are close to $T_{ins}$ for the canonical scenario, $w = 0.0$, while a broader range of $T_{reh}$ is allowed,  bracketing values $  T_{reh} \sim 10^9 \, GeV$ or so, for values of the equation of state parameter $w$ in the vicinity of $ \simeq 0.25$.

  \vspace*{4mm}  
{\textbf{Acknowledgments}}   
 A.B.L.  wishes to thank V. C. Spanos and K. Tamvakis  for discussions.   I.D.G.  thanks A. Karam for  illuminating discussions. 
\textquote{This research is co-financed by Greece and the European   Union (European Social Fund- ESF) through the Operational Programme \textquote{Human Resources Development, Education and Lifelong Learning} in the context of the project \textquote{Strengthening Human Resources Research Potential via Doctorate Research}  (MIS-5000432), implemented by the State Scholarships Foundation (IKY).}



\begin{thebibliography}{99}   

\bibitem{Sotiriou:2006hs} 
  T.~P.~Sotiriou,
  ``f(R) gravity and scalar-tensor theory,''
\href{https://iopscience.iop.org/article/10.1088/0264-9381/23/17/003}{Class.\ Quant.\ Grav.\  {\bf 23}, 5117 (2006),  [gr-qc/0604028]}.
    
  
\bibitem{Sotiriou:2006qn} 
  T.~P.~Sotiriou and S.~Liberati,
  ``Metric-affine f(R) theories of gravity,''
  \href{https://www.sciencedirect.com/science/article/pii/S0003491606001321?via%3Dihub}{Annals Phys.\  {\bf 322}, 935 (2007), [gr-qc/0604006]}.
  
     
\bibitem{Sotiriou:2008rp} 
  T.~P.~Sotiriou and V.~Faraoni,
  ``f(R) Theories of Gravity,''
  \href{https://journals.aps.org/rmp/abstract/10.1103/RevModPhys.82.451}{Rev.\ Mod.\ Phys.\  {\bf 82}, 451 (2010),arXiv:0805.1726 [gr-qc]}.
  
  
\bibitem{Borunda:2008kf} 
  M.~Borunda, B.~Janssen and M.~Bastero-Gil,
  ``Palatini versus metric formulation in higher curvature gravity,''
  \href{https://iopscience.iop.org/article/10.1088/1475-7516/2008/11/008}{JCAP {\bf 0811}, 008 (2008),arXiv:0804.4440 [hep-th]}.


\bibitem{DeFelice:2010aj} 
  A.~De Felice and S.~Tsujikawa,
  ``f(R) theories,''
  \href{https://link.springer.com/article/10.12942%2Flrr-2010-3}{Living Rev.\ Rel.\  {\bf 13}, 3 (2010),arXiv:1002.4928 [gr-qc]}.
  
 
\bibitem{Olmo:2011uz} 
  G.~J.~Olmo,
  ``Palatini Approach to Modified Gravity: f(R) Theories and Beyond,''
  \href{https://www.worldscientific.com/doi/abs/10.1142/S0218271811018925}{Int.\ J.\ Mod.\ Phys.\ D {\bf 20}, 413 (2011),arXiv:1101.3864 [gr-qc]}.
  
   
\bibitem{Capozziello:2011et} 
  S.~Capozziello and M.~De Laurentis, 
  ``Extended Theories of Gravity,''
 \href{https://www.sciencedirect.com/science/article/pii/S0370157311002432?via%3Dihub}{  Phys.\ Rept.\  {\bf 509}, 167 (2011), arXiv:1108.6266 [gr-qc]}. 
  
\bibitem{Clifton:2011jh} 
  T.~Clifton, P.~G.~Ferreira, A.~Padilla and C.~Skordis,
  ``Modified Gravity and Cosmology,''
 \href{https://www.sciencedirect.com/science/article/pii/S0370157312000105?via%3Dihub}{ Phys.\ Rept.\  {\bf 513}, 1 (2012), arXiv:1106.2476 [astro-ph.CO]}.
  
\bibitem{Nojiri:2017ncd} 
  S.~Nojiri, S.~D.~Odintsov and V.~K.~Oikonomou,
  ``Modified Gravity Theories on a Nutshell: Inflation, Bounce and Late-time Evolution,''
  \href{https://www.sciencedirect.com/science/article/pii/S0370157317301527?via%3Dihub}{ Phys.\ Rept.\  {\bf 692}, 1 (2017),   arXiv:1705.11098 [gr-qc]}.


  
\bibitem{Starobinsky:1980te} 
  A.~A.~Starobinsky,
  ``A New Type of Isotropic Cosmological Models Without Singularity,''
  \href{https://www.sciencedirect.com/science/article/pii/037026938090670X?via%3Dihub}{  Phys.\ Lett.\  {\bf 91B}, 99 (1980)
  [Adv.\ Ser.\ Astrophys.\ Cosmol.\  {\bf 3}, 130 (1987)].} 
  
  
\bibitem{Mukhanov:1981xt} 
  V.~F.~Mukhanov and G.~V.~Chibisov,
  ``Quantum Fluctuations and a Nonsingular Universe,''
  JETP Lett.\  {\bf 33}, 532 (1981)
  [Pisma Zh.\ Eksp.\ Teor.\ Fiz.\  {\bf 33}, 549 (1981)].
  
\bibitem{Starobinsky:1983zz} 
  A.~A.~Starobinsky,
  ``The Perturbation Spectrum Evolving from a Nonsingular Initially De-Sitter Cosmology and the Microwave Background Anisotropy,''
  Sov.\ Astron.\ Lett.\  {\bf 9}, 302 (1983).
  
\bibitem{Bauer:2008zj} 
  F.~Bauer and D.~A.~Demir,
  ``Inflation with Non-Minimal Coupling: Metric versus Palatini Formulations,''
  \href{https://www.sciencedirect.com/science/article/abs/pii/S0370269308007351?via%3Dihub}{Phys.\ Lett.\ B {\bf 665}, 222 (2008),arXiv:0803.2664 [hep-ph]}.
  
  
\bibitem{Koivisto:2005yc} 
  T.~Koivisto and H.~Kurki-Suonio,
  ``Cosmological perturbations in the palatini formulation of modified gravity,''
 \href{https://iopscience.iop.org/article/10.1088/0264-9381/23/7/009}{Class.\ Quant.\ Grav.\  {\bf 23}, 2355 (2006),  [astro-ph/0509422]}.
  
  
\bibitem{Tamanini:2010uq} 
  N.~Tamanini and C.~R.~Contaldi,
  ``Inflationary Perturbations in Palatini Generalised Gravity,''
  \href{https://journals.aps.org/prd/abstract/10.1103/PhysRevD.83.044018}{Phys.\ Rev.\ D {\bf 83}, 044018 (2011), arXiv:1010.0689 [gr-qc]}.
  
         
\bibitem{Bauer:2010jg} 
  F.~Bauer and D.~A.~Demir,
  ``Higgs-Palatini Inflation and Unitarity,''
   \href{https://www.sciencedirect.com/science/article/abs/pii/S0370269311003157?via%3Dihub}{Phys.\ Lett.\ B {\bf 698}, 425 (2011), arXiv:1012.2900 [hep-ph]}.
  
  
\bibitem{Enqvist:2011qm} 
  K.~Enqvist, T.~Koivisto and G.~Rigopoulos,
  ``Non-metric chaotic inflation,''
   \href{https://iopscience.iop.org/article/10.1088/1475-7516/2012/05/023}{JCAP {\bf 1205}, 023 (2012),arXiv:1107.3739, [astro-ph.CO]}.

\bibitem{Borowiec:2011wd} 
  A.~Borowiec, M.~Kamionka, A.~Kurek and M.~Szydlowski,
  ``Cosmic acceleration from modified gravity with Palatini formalism,''
  \href{https://iopscience.iop.org/article/10.1088/1475-7516/2012/02/027}{JCAP {\bf 1202}, 027 (2012), arXiv:1109.3420 [gr-qc]}.


\bibitem{Stachowski:2016zio} 
  A.~Stachowski, M.~Szydlowski and A.~Borowiec,
  ``Starobinsky cosmological model in Palatini formalism,''
   \href{https://link.springer.com/article/10.1140%2Fepjc%2Fs10052-017-4981-8}{Eur.\ Phys.\ J.\ C {\bf 77}, no. 6, 406 (2017), arXiv:1608.03196 [gr-qc]}.
  

\bibitem{Fu:2017iqg} 
  C.~Fu, P.~Wu and H.~Yu,
  ``Inflationary dynamics and preheating of the nonminimally coupled inflaton field in the metric and Palatini formalisms,''
  \href{https://journals.aps.org/prd/abstract/10.1103/PhysRevD.96.103542}{Phys.\ Rev.\ D {\bf 96}, no. 10, 103542 (2017), arXiv:1801.04089 [gr-qc]}.

  
\bibitem{Rasanen:2017ivk} 
  S.~Rasanen and P.~Wahlman,
  ``Higgs inflation with loop corrections in the Palatini formulation,''
   \href{https://iopscience.iop.org/article/10.1088/1475-7516/2017/11/047}{JCAP {\bf 1711}, no. 11, 047 (2017), arXiv:1709.07853 [astro-ph.CO]}.
  
  
\bibitem{Tenkanen:2017jih} 
  T.~Tenkanen,
  ``Resurrecting Quadratic Inflation with a non-minimal coupling to gravity,''
   \href{https://iopscience.iop.org/article/10.1088/1475-7516/2017/12/001}{JCAP {\bf 1712}, no. 12, 001 (2017), arXiv:1710.02758 [astro-ph.CO]}.
  

\bibitem{Racioppi:2017spw} 
  A.~Racioppi,
  ``Coleman-Weinberg linear inflation: metric vs. Palatini formulation,''
  \href{https://iopscience.iop.org/article/10.1088/1475-7516/2017/12/041}{JCAP {\bf 1712}, no. 12, 041 (2017), arXiv:1710.04853 [astro-ph.CO]}.
    
  
\bibitem{Markkanen:2017tun} 
  T.~Markkanen, T.~Tenkanen, V.~Vaskonen and H.~Veerm\"ae,
  ``Quantum corrections to quartic inflation with a non-minimal coupling: metric vs. Palatini,''
  \href{https://iopscience.iop.org/article/10.1088/1475-7516/2018/03/029}{JCAP {\bf 1803}, no. 03, 029 (2018), arXiv:1712.04874 [gr-qc]}.

  
\bibitem{Jarv:2017azx} 
  L.~J\"arv, A.~Racioppi and T.~Tenkanen,
  ``Palatini side of inflationary attractors,''
  \href{https://journals.aps.org/prd/abstract/10.1103/PhysRevD.97.083513}{Phys.\ Rev.\ D {\bf 97}, no. 8, 083513 (2018), arXiv:1712.08471 [gr-qc]}.


 
\bibitem{Rasanen:2018ihz} 
  S.~Rasanen,
  ``Higgs inflation in the Palatini formulation with kinetic terms for the metric,''
  \href{https://astro.theoj.org/article/7243-higgs-inflation-in-the-palatini-formulation-with-kinetic-terms-for-the-metric}{The Open Journal of Astrophysics, 2018, arXiv:1811.09514 [gr-qc]}.
  

\bibitem{Racioppi:2018zoy} 
  A.~Racioppi,
  ``New universal attractor in nonminimally coupled gravity: Linear inflation,'' 
  \href{https://journals.aps.org/prd/abstract/10.1103/PhysRevD.97.123514}{Phys.\ Rev.\ D {\bf 97}, no. 12, 123514 (2018), arXiv:1801.08810 [astro-ph.CO]}. 
  
  
\bibitem{Carrilho:2018ffi} 
  P.~Carrilho, D.~Mulryne, J.~Ronayne and T.~Tenkanen,
  ``Attractor Behaviour in Multifield Inflation,''
  \href{https://iopscience.iop.org/article/10.1088/1475-7516/2018/06/032}{JCAP {\bf 1806}, no. 06, 032 (2018), arXiv:1804.10489 [astro-ph.CO]}.
  
  
\bibitem{Enckell:2018kkc} 
  V.~M.~Enckell, K.~Enqvist, S.~Rasanen and E.~Tomberg,
  ``Higgs inflation at the hilltop,''
  \href{https://iopscience.iop.org/article/10.1088/1475-7516/2018/06/005}{JCAP {\bf 1806}, no. 06, 005 (2018), arXiv:1802.09299 [astro-ph.CO]}.


\bibitem{Bombacigno:2018tyw} 
  F.~Bombacigno and G.~Montani,
  ``Big bounce cosmology for Palatini $R^2$ gravity with a Nieh–Yan term,''
  \href{https://link.springer.com/article/10.1140%2Fepjc%2Fs10052-019-6918-x}{Eur.\ Phys.\ J.\ C {\bf 79}, no. 5, 405 (2019), arXiv:1809.07563 [gr-qc]}.


\bibitem{Enckell:2018hmo} 
  V.~M.~Enckell, K.~Enqvist, S.~Rasanen and L.~P.~Wahlman,
  ``Inflation with $R^2$ term in the Palatini formalism,''
  \href{https://iopscience.iop.org/article/10.1088/1475-7516/2019/02/022}{JCAP {\bf 1902}, 022 (2019), arXiv:1810.05536 [gr-qc]}.
  
  
\bibitem{Antoniadis:2018ywb} 
  I.~Antoniadis, A.~Karam, A.~Lykkas and K.~Tamvakis,
  ``Palatini inflation in models with an $R^2$ term,''
  \href{https://iopscience.iop.org/article/10.1088/1475-7516/2018/11/028}{JCAP {\bf 1811}, 028 (2018), arXiv:1810.10418 [gr-qc].}
  
  
\bibitem{Antoniadis:2018yfq} 
  I.~Antoniadis, A.~Karam, A.~Lykkas, T.~Pappas and K.~Tamvakis,
  ``Rescuing Quartic and Natural Inflation in the Palatini Formalism,''
   \href{https://iopscience.iop.org/article/10.1088/1475-7516/2019/03/005}{JCAP {\bf 1903}, 005 (2019), arXiv:1812.00847 [gr-qc]}.
  

\bibitem{Rasanen:2018fom} 
  S.~Rasanen and E.~Tomberg,
  ``Planck scale black hole dark matter from Higgs inflation,''
   \href{https://iopscience.iop.org/article/10.1088/1475-7516/2019/01/038}{JCAP {\bf 1901}, 038 (2019), arXiv:1810.12608 [astro-ph.CO]}.
  
    
  
\bibitem{Almeida:2018oid} 
  J.~P.~B.~Almeida, N.~Bernal, J.~Rubio and T.~Tenkanen,
  ``Hidden Inflaton Dark Matter,''
  \href{https://iopscience.iop.org/article/10.1088/1475-7516/2019/03/012}{ JCAP {\bf 1903}, 012 (2019), arXiv:1811.09640 [hep-ph]}. 
  

  
\bibitem{Takahashi:2018brt} 
  T.~Takahashi and T.~Tenkanen,
  ``Towards distinguishing variants of non-minimal inflation,''
   \href{https://iopscience.iop.org/article/10.1088/1475-7516/2019/04/035}{JCAP {\bf 1904}, no. 04, 035 (2019), arXiv:1812.08492 [astro-ph.CO]}.
    
  
\bibitem{Kannike:2018zwn} 
  K.~Kannike, A.~Kubarski, L.~Marzola and A.~Racioppi,
  ``A minimal model of inflation and dark radiation,''
 \href{https://www.sciencedirect.com/science/article/pii/S0370269319301807?via%3Dihub}{ Phys.\ Lett.\ B {\bf 792}, 74 (2019), arXiv:1810.12689 [hep-ph]}.
  
  
    
\bibitem{Tenkanen:2019jiq} 
  T.~Tenkanen,
  ``Minimal Higgs inflation with an $R^2$ term in Palatini gravity,''
  \href{https://journals.aps.org/prd/abstract/10.1103/PhysRevD.99.063528}{Phys.\ Rev.\ D {\bf 99}, no. 6, 063528 (2019), arXiv:1901.01794 [astro-ph.CO]]}.
  

 
\bibitem{Shimada:2018lnm} 
  K.~Shimada, K.~Aoki and K.~i.~Maeda,
  ``Metric-affine Gravity and Inflation,''
 \href{https://journals.aps.org/prd/abstract/10.1103/PhysRevD.99.104020}{ Phys.\ Rev.\ D {\bf 99}, no. 10, 104020 (2019), arXiv:1812.03420 [gr-qc]}. 
  
  
     
\bibitem{Wu:2018idg} 
  J.~Wu, G.~Li, T.~Harko and S.~D.~Liang,
  ``Palatini formulation of $f(R,T)$ gravity theory, and its cosmological implications,''
  \href{https://link.springer.com/article/10.1140/epjc/s10052-018-5923-9}{Eur.\ Phys.\ J.\ C {\bf 78}, no. 5, 430 (2018), arXiv:1805.07419 [gr-qc]}.
  
  
\bibitem{Kozak:2018vlp} 
  A.~Kozak and A.~Borowiec,
  ``Palatini frames in scalar tensor theories of gravity,''
  \href{https://link.springer.com/article/10.1140/epjc/s10052-019-6836-y}{Eur.\ Phys.\ J.\ C {\bf 79}, no. 4, 335 (2019),  arXiv:1808.05598 [hep-th].}
  
  
\bibitem{Jinno:2018jei} 
  R.~Jinno, K.~Kaneta, K.~y.~Oda and S.~C.~Park,
  ``Hillclimbing inflation in metric and Palatini formulations,''
  \href{https://www.sciencedirect.com/science/article/pii/S0370269319301662?via%3Dihub}{Phys.\ Lett.\ B {\bf 791}, 396 (2019), arXiv:1812.11077 [gr-qc]}. 
  
\bibitem{Edery:2019txq} 
  A.~Edery and Y.~Nakayama,
  ``Palatini formulation of pure $R^2$ gravity yields Einstein gravity with no massless scalar,''
  \href{https://journals.aps.org/prd/abstract/10.1103/PhysRevD.99.124018}{Phys.\ Rev.\ D {\bf 99}, no. 12, 124018 (2019),  arXiv:1902.07876 [hep-th] }. 

  
\bibitem{Rubio:2019ypq} 
  J.~Rubio and E.~S.~Tomberg,
  ``Preheating in Palatini Higgs inflation,''
 \href{https://iopscience.iop.org/article/10.1088/1475-7516/2019/04/021}{ JCAP {\bf 1904}, 021 (2019), arXiv:1902.10148 [hep-ph]}.
  

\bibitem{Jinno:2019und} 
  R.~Jinno, M.~Kubota, K.~y.~Oda and S.~C.~Park,
  ``Higgs inflation in metric and Palatini formalisms: Required suppression of higher dimensional operators,''
  \href{https://arxiv.org/abs/1904.05699}{arXiv:1904.05699 [hep-ph]}.
  
  
\bibitem{Giovannini:2019mgk} 
  M.~Giovannini,
  ``Post-inflationary phases stiffer than radiation and Palatini formulation,''
  \href{https://arxiv.org/abs/1905.06182}{arXiv:1905.06182 [gr-qc]}.


\bibitem{Tenkanen:2019xzn} 
  T.~Tenkanen and L.~Visinelli,
  ``Axion dark matter from Higgs inflation with an intermediate $H_*$,''
  \href{https://iopscience.iop.org/article/10.1088/1475-7516/2019/08/033}{JCAP {\bf 1908}, 033 (2019), arXiv:1906.11837 [astro-ph.CO]}.

\bibitem{Bostan:2019wsd} 
  N.~Bostan,
  ``Quadratic, Higgs and hilltop potentials in the Palatini gravity,''
  \href{https://arxiv.org/abs/1908.09674}{arXiv:1908.09674 [astro-ph.CO]}.



\bibitem{Tenkanen:2019wsd} 
  T.~Tenkanen,
  ``Trans-Planckian Censorship, Inflation and Dark Matter,''
  \href{https://arxiv.org/abs/1910.00521}{arXiv:1910.00521 [astro-ph.CO].}



\bibitem{Aghanim:2018eyx}
  N.~Aghanim {\it et al.} [Planck Collaboration],
  ``Planck 2018 results. VI. Cosmological parameters,''
  \href{https://arxiv.org/abs/1807.06209}{arXiv:1807.06209 [astro-ph.CO]}.
  
\bibitem{Akrami:2018odb} 
  Y.~Akrami {\it et al.} [Planck Collaboration],
  ``Planck 2018 results. X. Constraints on inflation,''
   \href{https://arxiv.org/abs/1807.06211}{arXiv:1807.06211v2 [astro-ph.CO]}.
  
\bibitem{Ade:2018gkx} 
  P.~A.~R.~Ade {\it et al.} [BICEP2 and Keck Array Collaborations],
  ``BICEP2 / Keck Array x: Constraints on Primordial Gravitational Waves using Planck, WMAP, and New BICEP2/Keck Observations through the 2015 Season,''
\href{https://journals.aps.org/prl/abstract/10.1103/PhysRevLett.121.221301}{Phys.\ Rev.\ Lett.\  {\bf 121}, 221301 (2018), arXiv:1810.05216 [astro-ph.CO]}.
  
  
\bibitem{Gerbino:2016sgw}
  M.~Gerbino, K.~Freese, S.~Vagnozzi, M.~Lattanzi, O.~Mena, E.~Giusarma and S.~Ho,
  ``Impact of neutrino properties on the estimation of inflationary parameters from current and future observations,''
  \href{https://journals.aps.org/prd/abstract/10.1103/PhysRevD.95.043512}{Phys.\ Rev.\ D {\bf 95} (2017) no.4,  043512  , arXiv:1610.08830 [astro-ph.CO]}.

  

\bibitem{ArmendarizPicon:1999rj} 
  C.~Armendariz-Picon, T.~Damour and V.~F.~Mukhanov,
  ``k - inflation,''
  \href{https://www.sciencedirect.com/science/article/abs/pii/S0370269399006036?via%3Dihub}{Phys.\ Lett.\ B {\bf 458}, 209 (1999), [hep-th/9904075]}.
  
  
\bibitem{Garriga:1999vw} 
  J.~Garriga and V.~F.~Mukhanov,
  ``Perturbations in k-inflation,''
   \href{https://linkinghub.elsevier.com/retrieve/pii/S0370269399006024}{Phys.\ Lett.\ B {\bf 458}, 219 (1999), [hep-th/9904176]}.
  
  
\bibitem{Lorenz:2008je} 
  L.~Lorenz, J.~Martin and C.~Ringeval,
  ``Constraints on Kinetically Modified Inflation from WMAP5,''
  \href{https://journals.aps.org/prd/abstract/10.1103/PhysRevD.78.063543}{ Phys.\ Rev.\ D {\bf 78}, 063543 (2008),  [arXiv:0807.2414 [astro-ph]]}.

  

  
\bibitem{Li:2012vta} 
  S.~Li and A.~R.~Liddle,
  ``Observational constraints on K-inflation models,''
   \href{https://iopscience.iop.org/article/10.1088/1475-7516/2012/10/011}{JCAP {\bf 1210}, 011 (2012), arXiv:1204.6214 [astro-ph.CO]}.

  
  
  
\bibitem{Starobinsky:1979ty} 
  A.~A.~Starobinsky,
  ``Spectrum of relict gravitational radiation and the early state of the universe,''
 \href{http://www.jetpletters.ac.ru/ps/1370/article_20738.shtml}{ JETP Lett.\  {\bf 30}, 682 (1979)
  [Pisma Zh.\ Eksp.\ Teor.\ Fiz.\  {\bf 30}, 719 (1979)]}. 


\bibitem{Mukhanov:1985rz}   
  V.~F.~Mukhanov,
  ``Gravitational Instability of the Universe Filled with a Scalar Field,''
   \href{http://www.jetpletters.ac.ru/ps/1467/article_22380.shtml}{JETP Lett.\  {\bf 41}, 493 (1985) [Pisma Zh.\ Eksp.\ Teor.\ Fiz.\  {\bf 41}, 402 (1985)]}.
  
  
\bibitem{Mukhanov:1988jd} 
  V.~F.~Mukhanov,
  Sov.\ Phys.\ JETP {\bf 67}, 1297 (1988)
  [Zh.\ Eksp.\ Teor.\ Fiz.\  {\bf 94N7}, 1 (1988)].
  
  
\bibitem{Lucchin:1984yf} 
  F.~Lucchin and S.~Matarrese,
  ``Power Law Inflation,''
  \href{https://journals.aps.org/prd/abstract/10.1103/PhysRevD.32.1316}{Phys.\ Rev.\ D {\bf 32}, 1316 (1985).}
  


\bibitem{Stewart:1993bc} 
  E.~D.~Stewart and D.~H.~Lyth,
  ``A More accurate analytic calculation of the spectrum of cosmological perturbations produced during inflation,''
  \href{https://www.sciencedirect.com/science/article/pii/037026939390379V?via%3Dihub}{Phys.\ Lett.\ B {\bf 302}, 171 (1993),[gr-qc/9302019]}.
  
  
  
  
\bibitem{Gong:2001he} 
  J.~O.~Gong and E.~D.~Stewart,
  ``The Density perturbation power spectrum to second order corrections in the slow roll expansion,''
  \href{https://www.sciencedirect.com/science/article/pii/S0370269301006165?via%3Dihub}{Phys.\ Lett.\ B {\bf 510}, 1 (2001),  [astro-ph/0101225]}.
  
  
\bibitem{Schwarz:2001vv} 
  D.~J.~Schwarz, C.~A.~Terrero-Escalante and A.~A.~Garcia,
  ``Higher order corrections to primordial spectra from cosmological inflation,''
  \href{https://www.sciencedirect.com/science/article/pii/S037026930101036X?via%3Dihub}{Phys.\ Lett.\ B {\bf 517}, 243 (2001),   [astro-ph/0106020]}.


\bibitem{Leach:2002ar} 
  S.~M.~Leach, A.~R.~Liddle, J.~Martin and D.~J.~Schwarz,
  ``Cosmological parameter estimation and the inflationary cosmology,''
  \href{https://journals.aps.org/prd/abstract/10.1103/PhysRevD.66.023515}{ Phys.\ Rev.\ D {\bf 66}, 023515 (2002), [astro-ph/0202094]}.


\bibitem{Habib:2002yi} 
  S.~Habib, K.~Heitmann, G.~Jungman and C.~Molina-Paris,
  ``The Inflationary perturbation spectrum,''
  \href{https://journals.aps.org/prl/abstract/10.1103/PhysRevLett.89.281301}{ Phys.\ Rev.\ Lett.\  {\bf 89}, 281301 (2002),  [astro-ph/0208443]}.
  
  
\bibitem{Martin:2002vn} 
  J.~Martin and D.~J.~Schwarz,
  ``WKB approximation for inflationary cosmological perturbations,''
  \href{https://journals.aps.org/prd/abstract/10.1103/PhysRevD.67.083512}{ Phys.\ Rev.\ D {\bf 67}, 083512 (2003),  [astro-ph/0210090]}.
  
  
\bibitem{Habib:2004kc} 
  S.~Habib, A.~Heinen, K.~Heitmann, G.~Jungman and C.~Molina-Paris,
  ``Characterizing inflationary perturbations: The Uniform approximation,''
  \href{https://journals.aps.org/prd/abstract/10.1103/PhysRevD.70.083507}{ Phys.\ Rev.\ D {\bf 70}, 083507 (2004), [astro-ph/0406134]}.
  
  
\bibitem{Wei:2004xx} 
  H.~Wei, R.~G.~Cai and A.~Wang,
  ``Second-order corrections to the power spectrum in the slow-roll expansion with a time-dependent sound speed,''
  \href{https://www.sciencedirect.com/science/article/pii/S0370269304014832?via%3Dihub}{Phys.\ Lett.\ B {\bf 603}, 95 (2004),   [hep-th/0409130]}.
  
\bibitem{Casadio:2004ru} 
  R.~Casadio, F.~Finelli, M.~Luzzi and G.~Venturi,
 ``Improved WKB analysis of cosmological perturbations,''
  \href{https://journals.aps.org/prd/abstract/10.1103/PhysRevD.71.043517}{Phys.\ Rev.\ D {\bf 71}, 043517 (2005), [gr-qc/0410092].}.
  
\bibitem{Casadio:2005xv} 
  R.~Casadio, F.~Finelli, M.~Luzzi and G.~Venturi,
  ``Higher order slow-roll predictions for inflation,'' 
  \href{https://www.sciencedirect.com/science/article/pii/S0370269305011640?via%3Dihub}{Phys.\ Lett.\ B {\bf 625}, 1 (2005), [gr-qc/0506043] }.
  

\bibitem{Kinney:2007ag} 
  W.~H.~Kinney and K.~Tzirakis,
  ``Quantum modes in DBI inflation: exact solutions and constraints from vacuum selection,''  
  \href{https://journals.aps.org/prd/abstract/10.1103/PhysRevD.77.103517}{Phys.\ Rev.\ D {\bf 77}, 103517 (2008),   [arXiv:0712.2043 [astro-ph]].}.
  
    
 
\bibitem{Lorenz:2008et} 
  L.~Lorenz, J.~Martin and C.~Ringeval,
  ``K-inflationary Power Spectra in the Uniform Approximation,''
  \href{https://journals.aps.org/prd/abstract/10.1103/PhysRevD.78.083513}{Phys.\ Rev.\ D {\bf 78}, 083513 (2008),  [arXiv:0807.3037 [astro-ph]]}.
  
  
\bibitem{Agarwal:2008ah} 
  N.~Agarwal and R.~Bean,
  ``Cosmological constraints on general, single field inflation,''
  \href{https://journals.aps.org/prd/abstract/10.1103/PhysRevD.79.023503}{ Phys.\ Rev.\ D {\bf 79}, 023503 (2009),  [arXiv:0809.2798 [astro-ph]] }.
  

\bibitem{Martin:2013uma} 
  J.~Martin, C.~Ringeval and V.~Vennin,
  ``K-inflationary Power Spectra at Second Order,''
  \href{https://iopscience.iop.org/article/10.1088/1475-7516/2013/06/021}{ JCAP {\bf 1306}, 021 (2013), [arXiv:1303.2120 [astro-ph.CO]]}.
  
  
\bibitem{Jimenez:2013xwa} 
  J.~Beltran Jimenez, M.~Musso and C.~Ringeval,
  ``Exact Mapping between Tensor and Most General Scalar Power Spectra,''
  \href{https://journals.aps.org/prd/abstract/10.1103/PhysRevD.88.043524}{Phys.\ Rev.\ D {\bf 88}, 043524 (2013),  [arXiv:1303.2788 [astro-ph.CO]]}.
  
  
\bibitem{Alinea:2015gpa} 
  A.~L.~Alinea, T.~Kubota and W.~Naylor,
  ``Logarithmic divergences in the $k$-inflationary power spectra computed through the uniform approximation,''
  \href{https://iopscience.iop.org/article/10.1088/1475-7516/2016/02/028}{ JCAP {\bf 1602}, 028 (2016), [arXiv:1506.08344 [gr-qc]]}.
  

    
\bibitem{Liddle:2003as} 
  A.~R.~Liddle and S.~M.~Leach,
  ``How long before the end of inflation were observable perturbations produced?,''
  \href{https://journals.aps.org/prd/abstract/10.1103/PhysRevD.68.103503}{ Phys.\ Rev.\ D {\bf 68}, 103503 (2003), [astro-ph/0305263]}.
    
  
  
   
\bibitem{Dodelson:2003vq} 
  S.~Dodelson and L.~Hui,
  ``A Horizon ratio bound for inflationary fluctuations,''
  \href{https://journals.aps.org/prl/abstract/10.1103/PhysRevLett.91.131301}{ Phys.\ Rev.\ Lett.\  {\bf 91}, 131301 (2003), [astro-ph/0305113]}.
  
  
  
\bibitem{Martin:2010kz} 
  J.~Martin and C.~Ringeval,
  ``First CMB Constraints on the Inflationary Reheating Temperature,''
   \href{https://journals.aps.org/prd/abstract/10.1103/PhysRevD.82.023511}{Phys.\ Rev.\ D {\bf 82}, 023511 (2010), arXiv:1004.5525 [astro-ph.CO]}.
  
  
\bibitem{Lozanov:2017hjm} 
  K.~D.~Lozanov and M.~A.~Amin,
  ``Self-resonance after inflation: oscillons, transients and radiation domination,''
 \href{https://journals.aps.org/prd/abstract/10.1103/PhysRevD.97.023533}{Phys.\ Rev.\ D {\bf 97}, no. 2, 023533 (2018),  arXiv:1710.06851 [astro-ph.CO]}.



 
    
\bibitem{Allahverdi:2010xz} 
  R.~Allahverdi, R.~Brandenberger, F.~Y.~Cyr-Racine and A.~Mazumdar,
  ``Reheating in Inflationary Cosmology: Theory and Applications,''
   \href{https://www.annualreviews.org/doi/10.1146/annurev.nucl.012809.104511}{Ann.\ Rev.\ Nucl.\ Part.\ Sci.\  {\bf 60}, 27 (2010), arXiv:1001.2600 [hep-th]} . 
  

 
    
\bibitem{Podolsky:2005bw} 
  D.~I.~Podolsky, G.~N.~Felder, L.~Kofman and M.~Peloso,
  ``Equation of state and beginning of thermalization after preheating,''
  \href{https://journals.aps.org/prd/abstract/10.1103/PhysRevD.73.023501}{Phys.\ Rev.\ D {\bf 73}, 023501 (2006), [hep-ph/0507096]}.
  
     
 
  
\bibitem{Adshead:2010mc} 
  P.~Adshead, R.~Easther, J.~Pritchard and A.~Loeb,
  ``Inflation and the Scale Dependent Spectral Index: Prospects and Strategies,''
 \href{https://iopscience.iop.org/article/10.1088/1475-7516/2011/02/021}{JCAP {\bf 1102}, 021 (2011),  arXiv:1007.3748 [astro-ph.CO]}. 
  
  
\bibitem{Mielczarek:2010ag} 
  J.~Mielczarek,
  ``Reheating temperature from the CMB,''
 \href{https://journals.aps.org/prd/abstract/10.1103/PhysRevD.83.023502}{Phys.\ Rev.\ D {\bf 83}, 023502 (2011), arXiv:1009.2359 [astro-ph.CO]}.
  
  
\bibitem{Easther:2011yq} 
  R.~Easther and H.~V.~Peiris,
  ``Bayesian Analysis of Inflation II: Model Selection and Constraints on Reheating,''
  \href{https://journals.aps.org/prd/abstract/10.1103/PhysRevD.85.103533}{Phys.\ Rev.\ D {\bf 85}, 103533 (2012), arXiv:1112.0326 [astro-ph.CO]}.
  
\bibitem{Dai:2014jja} 
  L.~Dai, M.~Kamionkowski and J.~Wang,
  ``Reheating constraints to inflationary models,''
\href{https://journals.aps.org/prl/abstract/10.1103/PhysRevLett.113.041302}{Phys.\ Rev.\ Lett.\  {\bf 113}, 041302 (2014),  arXiv:1404.6704 [astro-ph.CO]}.
  
  
\bibitem{Munoz:2014eqa} 
  J.~B.~Munoz and M.~Kamionkowski,
  ``Equation-of-State Parameter for Reheating,''
   \href{https://journals.aps.org/prd/abstract/10.1103/PhysRevD.91.043521}{Phys.\ Rev.\ D {\bf 91}, no. 4, 043521 (2015), arXiv:1412.0656 [astro-ph.CO]}.
  
  
\bibitem{Cook:2015vqa} 
  J.~L.~Cook, E.~Dimastrogiovanni, D.~A.~Easson and L.~M.~Krauss,
  ``Reheating predictions in single field inflation,''
  \href{https://iopscience.iop.org/article/10.1088/1475-7516/2015/04/047}{JCAP {\bf 1504}, 047 (2015),  arXiv:1502.04673 [astro-ph.CO]}.
  
  
\bibitem{Gong:2015qha} 
  J.~O.~Gong, S.~Pi and G.~Leung,
  ``Probing reheating with primordial spectrum,''
 \href{https://iopscience.iop.org/article/10.1088/1475-7516/2015/05/027}{JCAP {\bf 1505}, 027 (2015), arXiv:1501.03604 [hep-ph]}.


\bibitem{Rehagen:2015zma} 
  T.~Rehagen and G.~B.~Gelmini,
  ``Low reheating temperatures in monomial and binomial inflationary potentials,''
  \href{https://iopscience.iop.org/article/10.1088/1475-7516/2015/06/039}{ JCAP {\bf 1506}, 039 (2015),  arXiv:1504.03768 [hep-ph]}. 
  
  
\bibitem{Lozanov:2016hid} 
  K.~D.~Lozanov and M.~A.~Amin,
  ``Equation of State and Duration to Radiation Domination after Inflation,''
  \href{https://journals.aps.org/prl/abstract/10.1103/PhysRevLett.119.061301}{Phys.\ Rev.\ Lett.\  {\bf 119}, no. 6, 061301 (2017),  arXiv:1608.01213 [astro-ph.CO]}.
  
  
  
\bibitem{Kawasaki:1999na}
  M.~Kawasaki, K.~Kohri and N.~Sugiyama,
  ``Cosmological constraints on late time entropy production,''
 \href{https://journals.aps.org/prl/abstract/10.1103/PhysRevLett.82.4168}{Phys.\ Rev.\ Lett.\  {\bf 82} (1999) 4168,  astro-ph/9811437}.
  
  
  
\bibitem{Hasegawa:2019jsa} 
  T.~Hasegawa, N.~Hiroshima, K.~Kohri, R.~S.~L.~Hansen, T.~Tram and S.~Hannestad,
  ``MeV-scale reheating temperature and thermalization of oscillating neutrinos by radiative and hadronic decays of massive particles,''
  \href{https://arxiv.org/abs/1908.10189}{arXiv:1908.10189 [hep-ph]}.

    
  
\bibitem{Carrasco:2015rva} 
  J.~J.~M.~Carrasco, R.~Kallosh and A.~Linde,
  ``Cosmological Attractors and Initial Conditions for Inflation,''
  \href{https://journals.aps.org/prd/abstract/10.1103/PhysRevD.92.063519}{Phys.\ Rev.\ D {\bf 92}, no. 6, 063519 (2015), arXiv:1506.00936 [hep-th], cea-01690086}.

    
\bibitem{Bezrukov:2007ep} 
  F.~L.~Bezrukov and M.~Shaposhnikov,
  ``The Standard Model Higgs boson as the inflaton,''
  \href{https://www.sciencedirect.com/science/article/abs/pii/S0370269307014128?via%3Dihub}{ Phys.\ Lett.\ B {\bf 659}, 703 (2008), arXiv:0710.3755 [hep-th]. }
  
  
  
  
\bibitem{Bezrukov:2008ej} 
  F.~L.~Bezrukov, A.~Magnin and M.~Shaposhnikov,
  ``Standard Model Higgs boson mass from inflation,''
   \href{https://www.sciencedirect.com/science/article/abs/pii/S0370269309003220?via%3Dihub}{ Phys.\ Lett.\ B {\bf 675}, 88 (2009), arXiv:0812.4950 [hep-ph]. }
  
   
   
\bibitem{Barbon:2009ya} 
  J.~L.~F.~Barbon and J.~R.~Espinosa,
  ``On the Naturalness of Higgs Inflation,''  
\href{https://journals.aps.org/prd/abstract/10.1103/PhysRevD.79.081302}{Phys.\ Rev.\ D {\bf 79}, 081302 (2009), arXiv:0903.0355 [hep-ph]}.
  
\bibitem{Barvinsky:2009fy} 
  A.~O.~Barvinsky, A.~Y.~Kamenshchik, C.~Kiefer, A.~A.~Starobinsky and C.~Steinwachs,
 ``Asymptotic freedom in inflationary cosmology with a non-minimally coupled Higgs field,''
 \href{https://iopscience.iop.org/article/10.1088/1475-7516/2009/12/003}{ JCAP {\bf 0912}, 003 (2009), arXiv:0904.1698 [hep-ph]}.
  
\bibitem{Barvinsky:2009ii} 
  A.~O.~Barvinsky, A.~Y.~Kamenshchik, C.~Kiefer, A.~A.~Starobinsky and C.~F.~Steinwachs,
  ``Higgs boson, renormalization group, and naturalness in cosmology,''
 \href{https://link.springer.com/article/10.1140%2Fepjc%2Fs10052-012-2219-3}{ Eur.\ Phys.\ J.\ C {\bf 72}, 2219 (2012), arXiv:0910.1041 [hep-ph] }.
  
  
\bibitem{Germani:2010gm} 
  C.~Germani and A.~Kehagias,
  ``New Model of Inflation with Non-minimal Derivative Coupling of Standard Model Higgs Boson to Gravity,''
\href{https://journals.aps.org/prl/abstract/10.1103/PhysRevLett.105.011302}{ Phys.\ Rev.\ Lett.\  {\bf 105}, 011302 (2010), arXiv:1003.2635 [hep-ph]}.
  
  
\bibitem{Germani:2010ux} 
  C.~Germani and A.~Kehagias,
  ``Cosmological Perturbations in the New Higgs Inflation,''
\href{https://iopscience.iop.org/article/10.1088/1475-7516/2010/05/019}{JCAP {\bf 1005}, 019 (2010)}, 
 \href{https://iopscience.iop.org/article/10.1088/1475-7516/2010/06/E01}{ Erratum: [JCAP {\bf 1006}, E01 (2010)],   arXiv:1003.4285 [astro-ph.CO]}.
  
  
  
  
\bibitem{Lerner:2010mq} 
  R.~N.~Lerner and J.~McDonald,
  ``A Unitarity-Conserving Higgs Inflation Model,''
  \href{https://journals.aps.org/prd/abstract/10.1103/PhysRevD.82.103525}{Phys.\ Rev.\ D {\bf 82}, 103525 (2010) , arXiv:1005.2978 [hep-ph]}.
  
  
    
  
\bibitem{Bezrukov:2010jz} 
  F.~Bezrukov, A.~Magnin, M.~Shaposhnikov and S.~Sibiryakov,
  ``Higgs inflation: consistency and generalisations,''
  \href{https://link.springer.com/article/10.1007%2FJHEP01%282011%29016}{JHEP {\bf 1101}, 016 (2011), arXiv:1008.5157 [hep-ph] }.
  
  
\bibitem{Kamada:2010qe} 
  K.~Kamada, T.~Kobayashi, M.~Yamaguchi and J.~Yokoyama,
  ``Higgs G-inflation,''
 \href{https://journals.aps.org/prd/abstract/10.1103/PhysRevD.83.083515}{Phys.\ Rev.\ D {\bf 83}, 083515 (2011), arXiv:1012.4238 [astro-ph.CO]}.
  
    
\bibitem{Kamada:2012se} 
  K.~Kamada, T.~Kobayashi, T.~Takahashi, M.~Yamaguchi and J.~Yokoyama,
  ``Generalized Higgs inflation,''
  \href{https://journals.aps.org/prd/abstract/10.1103/PhysRevD.86.023504}{Phys.\ Rev.\ D {\bf 86}, 023504 (2012), arXiv:1203.4059 [hep-ph]}.
  
  
  
\bibitem{Bezrukov:2013fka} 
  F.~Bezrukov,
  ``The Higgs field as an inflaton,''
 \href{https://iopscience.iop.org/article/10.1088/0264-9381/30/21/214001}{  Class.\ Quant.\ Grav.\  {\bf 30}, 214001 (2013),  arXiv:1307.0708 [hep-ph]}. 
  

  
\bibitem{Allison:2013uaa} 
  K.~Allison,
  ``Higgs xi-inflation for the 125-126 GeV Higgs: a two-loop analysis,''
  \href{https://link.springer.com/article/10.1007%2FJHEP02%282014%29040}{JHEP {\bf 1402}, 040 (2014),  arXiv:1306.6931 [hep-ph]}.
  
  
\bibitem{Bezrukov:2014bra} 
  F.~Bezrukov and M.~Shaposhnikov,
  ``Higgs inflation at the critical point,''
  \href{https://www.sciencedirect.com/science/article/pii/S0370269314003840?via%3Dihub}{Phys.\ Lett.\ B {\bf 734}, 249 (2014),  arXiv:1403.6078 [hep-ph]}.
  
\bibitem{Hamada:2014wna} 
  Y.~Hamada, H.~Kawai, K.~y.~Oda and S.~C.~Park,
  ``Higgs inflation from Standard Model criticality,''
  \href{https://journals.aps.org/prd/abstract/10.1103/PhysRevD.91.053008}{ Phys.\ Rev.\ D {\bf 91}, 053008 (2015),  arXiv:1408.4864 [hep-ph]}. 
  
  
\bibitem{Salvio:2015kka} 
  A.~Salvio and A.~Mazumdar,
  ``Classical and Quantum Initial Conditions for Higgs Inflation,''
 \href{https://www.sciencedirect.com/science/article/pii/S0370269315006954?via%3Dihub}{Phys.\ Lett.\ B {\bf 750}, 194 (2015),  arXiv:1506.07520 [hep-ph]} .
  
\bibitem{Calmet:2016fsr} 
  X.~Calmet and I.~Kuntz,
  ``Higgs Starobinsky Inflation,''
  \href{https://link.springer.com/article/10.1140%2Fepjc%2Fs10052-016-4136-3}{Eur.\ Phys.\ J.\ C {\bf 76}, no. 5, 289 (2016),  arXiv:1605.02236 [hep-th]}.
  
  
  
\bibitem{Jinno:2017lun} 
  R.~Jinno, K.~Kaneta and K.~y.~Oda,
  ``Hill-climbing Higgs inflation,''
 \href{https://journals.aps.org/prd/abstract/10.1103/PhysRevD.97.023523}{ Phys.\ Rev.\ D {\bf 97}, no. 2, 023523 (2018), arXiv:1705.03696 [hep-ph]}. 
  
    
\bibitem{Bezrukov:2017dyv} 
  F.~Bezrukov, M.~Pauly and J.~Rubio,
  ``On the robustness of the primordial power spectrum in renormalized Higgs inflation,''
 \href{https://iopscience.iop.org/article/10.1088/1475-7516/2018/02/040}{ JCAP {\bf 1802}, 040 (2018), arXiv:1706.05007 [hep-ph]}. 

\bibitem{He:2018gyf} 
  M.~He, A.~A.~Starobinsky and J.~Yokoyama,
  ``Inflation in the mixed Higgs-$R^2$ model,''
 \href{https://iopscience.iop.org/article/10.1088/1475-7516/2018/05/064}{ JCAP {\bf 1805}, 064 (2018), arXiv:1804.00409 [astro-ph.CO]}.
  
\bibitem{Gundhi:2018wyz} 
  A.~Gundhi and C.~F.~Steinwachs,
  ``Scalaron-Higgs inflation,''
  \href{https://arxiv.org/abs/1810.10546}{arXiv:1810.10546 [hep-th]}.
  
\bibitem{Rubio:2018ogq} 
  J.~Rubio,
  ``Higgs inflation,''
  \href{https://www.frontiersin.org/articles/10.3389/fspas.2018.00050/full}{ Front.\ Astron.\ Space Sci.\  {\bf 5}, 50 (2019), arXiv:1807.02376 [hep-ph]}. 
  
  
\bibitem{He:2018mgb} 
  M.~He, R.~Jinno, K.~Kamada, S.~C.~Park, A.~A.~Starobinsky and J.~Yokoyama,
  ``On the violent preheating in the mixed Higgs-$R^2$ inflationary model,''
  \href{https://www.sciencedirect.com/science/article/pii/S0370269319300991?via%3Dihub}{ Phys.\ Lett.\ B {\bf 791}, 36 (2019) , arXiv:1812.10099 [hep-ph]}.
  
  
\bibitem{Steinwachs:2019hdr} 
  C.~F.~Steinwachs,
  ``Higgs field in cosmology,''
  \href{https://arxiv.org/abs/1909.10528}{arXiv:1909.10528 [hep-ph]}.
  
  



\end{thebibliography}
\end{document}